\documentclass[fleqn,usenatbib]{mnras}

\usepackage{newtxtext,newtxmath}

\usepackage[T1]{fontenc}
\usepackage{ae,aecompl}

\usepackage{graphicx}	
\usepackage{amsmath}	
\usepackage{multicol}   
\usepackage{bm}		    
\usepackage{pdflscape}	
\usepackage{epsfig,color,pifont}
\usepackage{times}
\usepackage{subfig}
\usepackage{url}
\usepackage{booktabs}   

\usepackage{graphicx}	

\def\Msun{\mbox{~M$_\odot$}}

\def\Msunpc2{\mbox{~M$_\odot$~pc$^{-2}$}}
\def\kms{\mbox{~km~s$^{-1}$}}
\def\kmsns{\mbox{km~s$^{-1}$}}

\def\kpc{\mbox{~kpc}}
\def\Mpc{\mbox{~Mpc}}

\def\Gyr{\mbox{~Gyr}}

\def\vr{v_{r}}
\def\vphi{v_{\phi}}
\def\vtheta{v_{\theta}}
\def\vt{v_{\rm t}}
\def\Ekin{E_{\rm k}}
\def\Etot{E}
\def\Epot{E_{\rm pot}}
\def\Lz{L_{\rm z}}

\def\rperi{r_{\rm peri}}
\def\rapo{r_{\rm apo}}

\def\Nmajor{N_{>1:4}}
\def\Nminor{N_{1:100-1:4}}
\def\Nsmall{N_{1:100-1:20}}
\def\Nmedium{N_{1:20-1:4}}
\def\Ntiny{N_{<1:100}}
\def\rmm{r_{\rm mm}}
\def\Nbr{N_{\rm br}}
\def\Nbrz{N_{{\rm br}, z>2}}
\def\Nleaf{N_{\rm leaf}}
\def\rbl{r_{\rm bl}}
\def\fexgcs{f_{\rm ex, GCs}}
\def\fexstars{f_{\rm ex, stars}}
\def\tmax{\tau_{\rm max}}
\def\ta{\tau_{\rm a}}
\def\tf{\tau_{\rm f}}
\def\t25{\tau_{25}}
\def\t50{\tau_{50}}
\def\tmm{\tau_{\rm mm}}
\def\zmm{z_{\rm mm}}
\def\tam{\tau_{\rm am}}
\def\tH{\tau_{\rm H}}
\def\rt{r_{t}}
\def\Rvmax{R_{V_{\rm max}}}
\def\cnfw{c_{\rm NFW}}
\def\feh{[\rm{Fe}/\rm{H}]}

\def\Ncorr{N_{\rm corr}}
\def\pref{p_{\rm ref}}
\def\peff{p_{\rm eff}}

\def\vmax{V_{\rm max}}

\def\M200{M_{\rm 200}}

\def\Mstar{M_{\rm star}}
\def\Mhalo{M_{\rm halo}}

\def\apj{\rm Astrophys. J.}
\def\aap{\rm Astron. \&Astrophys. }
\def\mnras{\rm Mon.\,Not.\,RAS. }

\def\LCDM{$\Lambda$CDM }

\def\mathnew{\mathsurround=0pt}
\def\simov#1#2{\lower .5pt\vbox{\baselineskip0pt
    \lineskip-.5pt\ialign{$\mathnew#1\hfil##\hfil$\crcr#2\crcr\sim\crcr}}}

\def\apj{\rm ApJ}
\def\apjl{\rm ApJL}
\def\apjs{\rm ApJS}
\def\aj{\rm AJ}
\def\mnras{\rm MNRAS}
\def\nat{\rm Nature}

\def\aap{\rm A\&A}
\def\araa{\rm ARAA}

\defcitealias{Reina-Campos17}{RK17}
\defcitealias{Kruijssen12b}{K12}
\defcitealias{Trujillo-Gomez19}{TRK19}

\title[GC kinematics and the assembly of the MW]{The kinematics of globular cluster populations in the E-MOSAICS simulations and their implications for the assembly history of the Milky Way}

\author[S. Trujillo-Gomez et al.]
{Sebastian Trujillo-Gomez$^{1}$\thanks{E-mail: strujill@gmail.com}, 
J.~M.~Diederik Kruijssen$^{1}$, 
Marta Reina-Campos$^{1}$,
\newauthor
Joel L.~Pfeffer$^{2}$, Benjamin W.~Keller$^1$, Robert A.~Crain$^2$, Nate Bastian$^2$, 
\newauthor
and Meghan E.~Hughes$^2$
\\
$^{1}$Astronomisches Rechen-Institut, Zentrum f{\"u}r Astronomie der Universit{\"a}t Heidelberg, Monchhofstra{\ss}e 12-14, D-69120 Heidelberg, Germany \\
$^{2}$Astrophysics Research Institute, Liverpool John Moores University, 146 Brownlow Hill, Liverpool L3 5RF, UK
}

\date{Accepted 2021 February 1. Received 2020 December 22; in original form 2020 May 5}

\pubyear{2020}

\begin{document}
\label{firstpage}
\pagerange{\pageref{firstpage}--\pageref{lastpage}}
\maketitle

\begin{abstract}
We present a detailed comparison of the Milky Way (MW) globular cluster (GC) kinematics with the 25 Milky Way-mass cosmological simulations from the E-MOSAICS project. While the MW falls within the kinematic distribution of GCs spanned by the simulations, the relative kinematics of its metal-rich ($\feh>-1.2$) versus metal-poor ($\feh<-1.2$), and inner ($r<8\kpc$) versus outer ($r>8\kpc$) populations are atypical for its mass. To understand the origins of these features, we perform a comprehensive statistical analysis of the simulations, and find 18 correlations describing the assembly of $L^*$ galaxies and their dark matter haloes based on their GC population kinematics. The correlations arise because the orbital distributions of accreted and in-situ GCs depend on the masses and accretion redshifts of accreted satellites, driven by the combined effects of dynamical fraction, tidal stripping, and dynamical heating. Because the kinematics of in-situ/accreted GCs are broadly traced by the metal-rich/metal-poor and inner/outer populations, the observed GC kinematics are a sensitive probe of galaxy assembly. We predict that relative to the population of $L^*$ galaxies, the MW assembled its dark matter and stellar mass rapidly through a combination of in-situ star formation, more than a dozen low-mass mergers, and $1.4\pm1.2$ early ($z=3.1\pm1.3$) major mergers. The rapid assembly period ended early, limiting the fraction of accreted stars. We conclude by providing detailed  quantitative predictions for the assembly history of the MW. 
\end{abstract}




\section{Introduction}
\label{sec:intro}

Understanding the origin of galaxies, and in particular the MW, remains one of the most important goals of astrophysics. It has been known for several decades that the main components of the Galaxy, namely its disc and stellar halo, have distinct origins. This has been established using studies of the spatial distribution, abundance patterns, and dynamics of stars \citep{Eggen62,Ibata94,Majewski96,Helmi99,ChibaBeers00,Bullock01b,Gilmore02,Yanny03,Crane03,Belokurov06}. Due to their brightness and ubiquity, globular clusters (GCs) have also been used as tracers to study the origin of the Galaxy. Using the chemical composition of stars in galactic GCs, \citet{SearleZinn78} showed that the galactic halo GCs formed over a longer time-scale than GCs associated with the galactic bulge. They concluded that the halo GCs must have formed in independent galactic fragments and accreted into the MW after its initial collapse. Decades later, proper motion measurements facilitated the study of the 3D kinematics of many galactic GCs \citep{CudworthHanson93,Dinescu99,Dinescu03,Massari13}, lending further support to the idea of two-phase buildup of the MW and its GC population. Over the past 50 years, many observational studies established that the disc was mostly formed in-situ, while the stellar halo was at least in part formed through accretion of lower mass galaxies \citep[see][for recent reviews]{Helmi08, DeLucia12, Belokurov13, Helmi20}. 

With the first measurements of the cosmic microwave background fluctuations \citep{Smoot92}, the advent of the era of precision cosmology firmly established the framework for understanding the formation and evolution of galaxies. In the current paradigm, galaxies began their life as intergalactic gas was accreted onto gravitationally collapsing dark matter (DM) haloes, allowing it to cool, condense and form stars. These proto-galaxies then grew rapidly as the hierarchical assembly of their host DM haloes continued through accretion of lower mass galaxies with their own stellar and cluster populations \citep[e.g.,][]{Press74, ReesOstriker77, WhiteRees78, FallEfstathiou80, Blumenthal84, WhiteFrenk91, cole94, Navarro95, NavarroSteinmetz00, Cole00}. This hierarchical assembly paradigm leads to the prediction that stars and GCs that formed in satellites and were later accreted will have distinct properties (such as chemical abundances and kinematics) from those that formed within the main progenitor.

Following the second data release of the \emph{Gaia} astrometry mission \citep{GaiaDR2}, the last two years have witnessed a deluge of studies aiming to characterise the precise details of the assembly history of the MW using the precise 6D phase-space distribution of stars and GCs. These studies have improved our knowledge of the history of the MW with unprecedented detail, including the discovery of at least six new galactic progenitors which had major contributions to the buildup of its stellar halo and GC system \citep[e.g.,][]{Belokurov18, splash, Deason18, Deason19, Haywood18, enceladus, Myeong18a, Myeong18c, Myeong18b, Myeong18d, sequoia, Iorio19, Koppelman19, thamnos, Mackereth19, Massari19, Necib19a, Necib19b, Vasiliev19, Gallart19, Kruijssen20, Pfeffer20} 

The complexity of the processes involved in hierarchical galaxy assembly within the $\Lambda$ cold dark matter ($\Lambda$CDM) paradigm make the task of reconstructing the formation and merger history of a galaxy using only the present-day phase space distribution of its stars extremely difficult. Over the past two decades, however, collisionless $N$-body simulations of galaxy assembly have increased the amount of information that can be derived from dynamical studies \citep[e.g.,][]{Helmi03,BullockJohnston05,Johnston08,Bell08,Cooper10}. Unfortunately, the predictive power of these approaches is often limited by three main factors: the simulations are often idealised and do not include the cosmological environment, they do not include gas-dynamics and the physics of star formation \citep[see][for their effect on the radial halo profile]{Font11}, and/or they do not sample statistically the large variety of galaxy assembly histories which result from evolution within different cosmological environments (i.e., cosmic variance). Moreover, because stars are generally used as tracers, and the mass-to-light ratio of sub-$L^*$ galaxies increases steeply with decreasing halo mass \citep{Moster13,Behroozi19}, the signatures of accretion are generally dominated by the few most massive accretion events. More recently, large hydrodynamical simulations of cosmologically representative volumes aimed to reproduce the general properties of present-day galaxy populations have become available \citep[e.g.,][]{Illustris, Dubois14, Schaye15, IllustrisTNG}. These simulations overcome the earlier shortcomings and present a unique opportunity to piece together the detailed history of the Galaxy using the present phase-space distribution of its stars. 

Decades after the pioneering work of \citet{SearleZinn78} demonstrated the potential of GCs as tracers of galaxy formation, new studies began to exploit it \citep[e.g.][]{Cote98,Bekki05,Rhode05,MuratovGnedin10,Arnold11,Tonini13,Choksi18,Beasley18,Fahrion20,Ramos-Almendares20}. Theoretical studies of the formation and co-evolution of galaxies and GCs have shown that GCs trace the buildup of $L^*$ galaxies across cosmic time \citep{Reina-Campos19}, and that their abundances and ages contain a record of the assembly history of their host \citep{emosaicsII,kraken,Massari19}. GCs are intrinsically bright, ubiquitous \citep{Harris16}, and can be studied at distances beyond the Local Group \citep[e.g.,][]{Norris12, Zhu14, Alabi17}, making them a promising tool for tracing the formation and assembly of galaxies. Most importantly, because the number of GCs per unit host stellar mass increases with decreasing galaxy mass \citep{Peng08, Georgiev10}, and their phase-mixing time is much longer than for stars, GCs should be more sensitive tracers of early and low-mass accretion events than field stars. 

In this work we use cosmological hydrodynamical simulations that include the physics of star cluster formation and evolution to study the kinematics of GCs in an unbiased sample of 25 MW-mass galaxies. We compare the kinematics of GCs in the MW with the E-MOSAICS\footnote{This is an acronym for `MOdelling Star cluster population Assembly In Cosmological Simulations within EAGLE'} simulations \citep{emosaicsI, emosaicsII} and use unique features in the MW system to identify GC kinematic tracers of the formation and assembly history of galaxies. Then, by statistically modelling the relationship between the GC kinematics and the assembly of the simulations, we combine it with the precise \emph{Gaia} measurements and obtain detailed quantitative predictions for the assembly history of the MW. 

This paper is organised as follows.  Section~\ref{sec:simsanddata} describes the E-MOSAICS simulations and the \emph{Gaia} GC kinematics data. In Section~\ref{sec:comparison} we present the comparison of the distributions of median GC 3D velocities, orbits, and integrals of motion, as well as the relative differences between metallicity and galactocentric radius subpopulations. Section~\ref{sec:origin} compares the properties of accreted and in-situ GC populations in the simulations. Section~\ref{sec:tracing_assembly_with_GCs} describes the statistical method to search for GC kinematic tracers of galaxy assembly,  presents detailed predictions for the formation and assembly of the MW, and compares them to existing constraints within the context of the $L^*$ galaxy population. Section~\ref{sec:limitations} discusses the limitations and caveats of the simulations and the analysis. We discuss the results and summarise our conclusions in Section~\ref{sec:conclusions}.

\section{Description of the simulations and observations}
\label{sec:simsanddata}

\subsection{Simulating galaxies and their star cluster populations}
\label{sec:simulations}

The E-MOSAICS simulations combine the subgrid modelling of the formation and evolution of star cluster populations using MOSAICS \citep{Kruijssen11,emosaicsI} with the EAGLE model for galaxy formation simulations \citep{Schaye15,Crain15}. EAGLE uses a modified version of the $N$-body TreePM smoothed particle hydrodynamics code {\sc Gadget} 3 \citep{springel05g2}. It implements subgrid models for several relevant physical processes including radiative cooling \citep{Wiersma09b} in the presence of a spatially uniform and time-dependent extragalactic UV background \citep{HaardtMadau01}, star formation in gas with a density above a metallicity-dependent  threshold \citep{SchayeDallaVecchia08}, stellar feedback \citep{DallaVecchiaSchaye12}, the time-dependent return of mass and metals due to stellar evolution \citep{Wiersma09b}, the formation and growth of supermassive black holes (BH) due to gas accretion and BH-BH mergers \citep{Springel05,Rosas-Guevara15,Schaye15}, and feedback from active galactic nuclei \citep[AGN;][]{BoothSchaye09,Schaye15}. The efficiency of feedback processes was calibrated to reproduce the present-day stellar mass function, the sizes of galaxies, and the $M_{\rm BH} - M_*$ relation. In addition, the EAGLE model has been shown to reproduce several other galaxy observables including the redshift evolution of the stellar mass function, star formation rates \citep{Furlong15}, and galaxy sizes \citep{Furlong17}, present-day galaxy luminosities and colours \citep{Trayford15}, cold gas distribution \citep{Lagos15,Lagos16,Bahe16,Marasco16,Crain17}, the properties of circumgalactic and intergalactic gas \citep{Rahmati15,Rahmati16,Turner16,Turner17,Oppenheimer16,Oppenheimer18}, and the abundance patterns of stars in the MW \citep{Mackereth18}

E-MOSAICS adds a subgrid treatment of the formation and evolution of star clusters to the EAGLE model. Cluster populations are formed as a subgrid component within newly formed star particles using a model for the fraction of star formation in bound clusters \citep{Kruijssen12}, and a Schechter initial cluster mass function with a $-2$ power-law slope and a maximum truncation mass \citep{Reina-Campos17}. Both the bound fraction and the maximum truncation mass are environmentally dependent and increase with gas pressure, resulting in more efficient formation of massive clusters at high redshift and in galaxy  mergers \citep{Reina-Campos19, Keller20}. Cluster evolution is also environmentally dependent and is modelled by following four different physical processes. First, clusters lose mass due to tidal shocks from the interstellar medium. Second, clusters predominantly lose mass in low density environments due to two-body relaxation \citep{Kruijssen11}. For both mechanisms the mass loss is calculated using the instantaneous local tidal field at the position of each particle. Third, mass loss due to stellar evolution is followed according to the standard EAGLE stellar evolution model \citep{Wiersma09b}. Last, the contribution of dynamical friction to the destruction of clusters (which is particularly important for the most massive GCs) is calculated in post-processing \citep{emosaicsI}. 

The E-MOSAICS simulations broadly reproduce several properties of observed GC populations, including the high-mass end of the GC mass function \citep{emosaicsI}, specific frequencies, and age-metallicity relations \citep{emosaicsII}, radial density profiles (Reina-Campos et al. in prep.), as well as their colour-magnitude relation  \citep{Usher18}. The same physics that gives rise to present-day GCs in the simulations also produces young cluster populations in agreement with observations of nearby galaxies \citep{Pfeffer19b}. The fact that E-MOSAICS generally reproduces many of the properties of galaxies and their young and old stellar cluster populations makes it a valuable tool for tracing the formation and assembly of galaxies using their observed GC populations. Following this approach, \citet{emosaicsII} show that the age-metallicity relation of GCs is an excellent probe of the details of the galaxy assembly process. \citet{kraken} apply the method to the MW to reconstruct a detailed picture of the merger tree of the Galaxy, and predict the existence of the `Kraken' satellite progenitor, which was one of the most massive accretion events in the MW's history. \citet{Kruijssen20} and \citet{Pfeffer20}, used the GC orbits in the simulations to infer the mass and accretion redshift of known MW progenitors. Due to the limitations of the EAGLE model, the cold and dense ISM is not resolved in the simulations. This leads to an underestimation of the disruption rate of clusters while they remain in their natal galaxies. \citet{emosaicsII} shows that this results in an excess of metal-rich GCs with $\feh>-1.0$ with respect to the combined distribution in the MW and M31 (their figure D1), and that this is due to metal-rich GCs remaining in their natal galaxy for much longer periods compared to metal-poor GCs (their figure D2). This issue reduces the applicability of E-MOSAICS to GCs with $\feh<-0.5$. There is a remaining excess of a factor of $\sim 2.5$ for GCs with $-1.0<\feh<-0.5$, which corresponds to $34$ per cent of all the GCs considered for this work. In Section~\ref{sec:limitations} we show that the effect on our analysis is minimal. Figure 2 of \citet{emosaicsII} shows the GC metallicity distribution of each of the 25 simulations compared to both the MW and M31. While on average E-MOSAICS contains about twice as many metal-rich ($\feh>-1.2$) as metal-poor GCs, there is large scatter in the metal-rich end of the distribution. As a result, some of the simulations resemble the MW (e.g. MW18), while others have a peak at $\feh>-0.5$, similar to M31 (e.g. MW09). Tests of the impact of other potential systematics in the simulations are discussed in Section~\ref{sec:limitations}. 

For the analysis in this paper, we first transform the coordinates and velocities of each of the 25 simulated galaxies at $z=0$ to a coordinate frame where the $z$-axis corresponds to the direction of the total angular momentum vector of the star particles bound to the central galaxy and located within a galactocentric radius of $30\kpc$. This value is chosen to align the disc with the $x$-$y$ plane while avoiding spurious alignments with satellites at large radii due to their high orbital angular momenta. We define as GCs in the simulations all the clusters with masses $M > 10^5\Msun$ and metallicities in the range $-2.5 < \feh < -0.5$ which are bound to the central galaxy, regardless of cluster age. The use of a metallicity criterion was chosen to mitigate the underestimated disruption rate of clusters in E-MOSAICS due to the lack of a resolved cold interstellar medium in EAGLE \citep[for details see][and appendix D of \citealt{emosaicsII}]{emosaicsI}. For each simulation, we consider only the clusters that are bound to the central galaxy at $z=0$. When comparing to the kinematics of the stars we include all the field stars bound to the central galaxies at $z=0$, as identified by the {\sc subfind} algorithm \citep{subfind01,subfind09}. 

Throughout the analysis the simulated GC sample is divided into distinct metal-rich ($\feh>-1.2$), and metal-poor ($\feh<-1.2$) subpopulations. The threshold value $\feh = -1.2$ approximately bisects the range of metallicities spanned by the MW GC population. According to this definition, across the 25 simulated galaxies there are a total of 2474 metal-rich, and 1247 metal-poor GCs (or 100.0 metal-rich and 49.9 metal-poor on average per galaxy). The sample is also divided into distinct subpopulations based on GC radial distribution, with `inner' GCs at galactocentric radii $r<8\kpc$, and `outer' GCs at $r>8\kpc$. Following this definition, across the 25 simulations there are 2231 inner, and 1490 outer GCs (or 89.2 inner and 59.6 outer GCs on average per galaxy), which matches the relative numbers of inner and outer Galactic GCs.

\subsection{Observational data: MW GC kinematics}
\label{sec:observations}

Using a combination of \emph{Gaia} DR2 \citep{GaiaDR2} proper motions and line-of-sight velocities from the literature, \citet{Baumgardt19} obtained the 3D positions and velocities of 154 GCs, or nearly the entire MW GC population. Their derived kinematics are consistent with those found by \citet{Helmi18} as well as \citet{Vasiliev19}. Using the metallicities from the \citet[][2010 edition]{Harris96} catalogue we selected the subsample of GCs in \citet{Baumgardt19} with $-2.5 < \feh < -0.5$. This metallicity range matches the selection of the GCs in the E-MOSAICS simulations where the effects of underdisruption are not important (see Section~\ref{sec:simulations}), and should prevent any bias in the comparison with observations. A lower GC mass limit is not imposed on the observational sample, because the cut is meant to correct for underdisruption in the simulations, which is only significant for GC masses below $10^5\Msun$. Because the Galactic GC population exhibits no relation between GC mass and kinematics \citep[as verified using the dynamical masses estimates from][]{Baumgardt18}, this correction is not relevant for the observed clusters. The selection criteria above result in an observational sample of 132 GCs that we use from here on when referring to the kinematics of the MW GC system. Within this sample of 132 GCs, subpopulations are defined as follows: metal-poor GCs have metallicities $\feh < -1.2$ (91 objects), while metal-rich GCs have $\feh > -1.2$ (41 objects). `Inner' GCs are those located at galactocentric distances $r < 8\kpc$ (78 objects), whilst `outer' GCs have distances $r > 8\kpc$ (54 objects).

\section{Comparison of observed and simulated GC kinematics}
\label{sec:comparison}

\subsection{3D velocities}
\label{sec:velocities}

\begin{figure*}
    \includegraphics[width=0.70\textwidth]{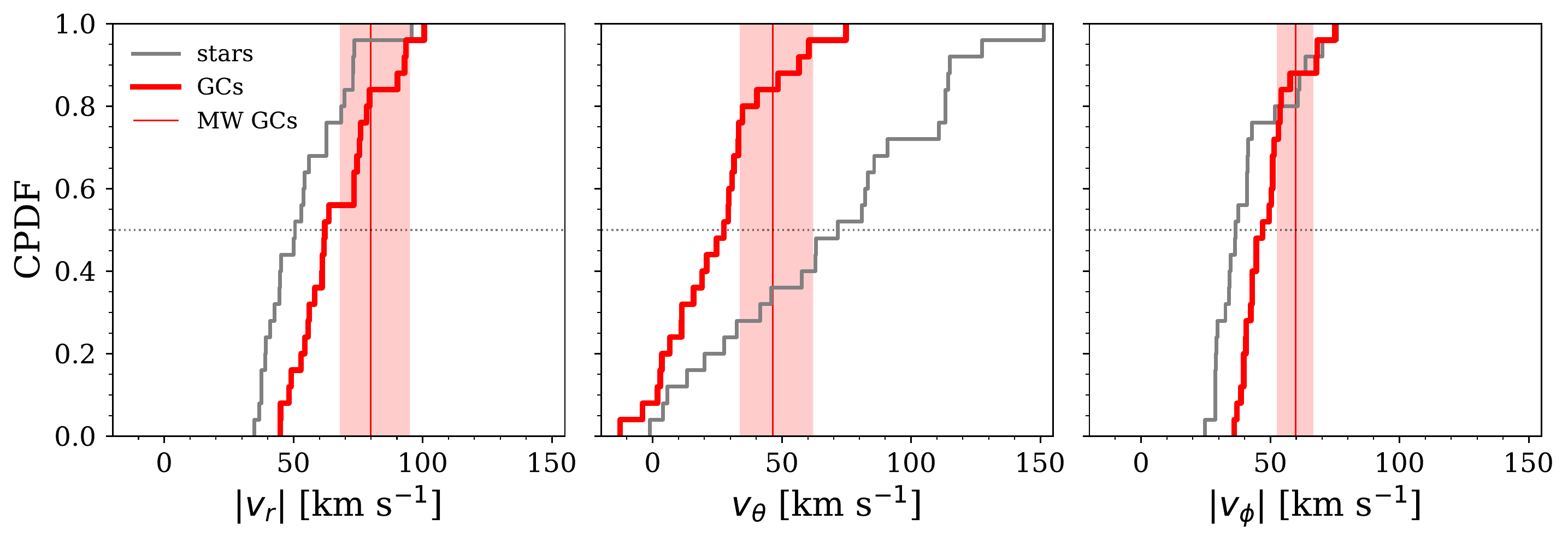}
    \includegraphics[width=0.70\textwidth]{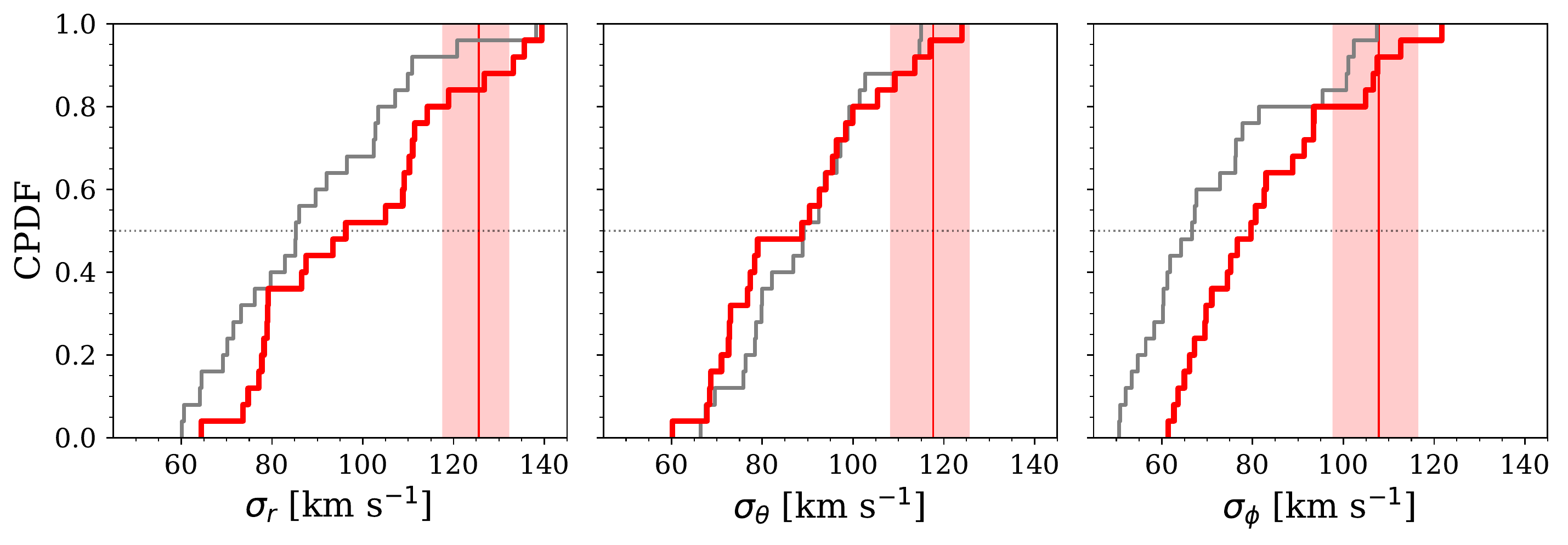}
    \caption{Cumulative PDF of the median 3D velocity components (top row) and velocity dispersions (bottom row) of the GC systems of the 25 simulated galaxies. From left to right each row shows the radial, azimuthal, and polar components. In these probability distributions each data point represents the median velocity or dispersion of the GC population of one galaxy. For comparison, the grey line shows the distribution of the stars. The observed values for the MW GC system and their uncertainties are shown by the vertical line and shading in each panel. Nearly all MW-mass galaxies, including the MW, have GC systems with average prograde rotation. The MW fits well within the distributions but the median rotation and high velocity dispersion of its GC system are unusual, and larger than in $\sim 84$ per cent of the simulated galaxies.}
\label{fig:velocities}
\end{figure*}

We begin by comparing the simulated GC system kinematics with the MW GC distribution in phase-space. The velocity vectors are expressed using their components in spherical coordinates, where $\theta$ is the azimuthal angle (in the $x$-$y$ plane), and $\phi$ is the polar angle. The top row of Figure~\ref{fig:velocities} shows the cumulative distribution of the median 3D spherical velocity components across the 25 galaxies compared to the median of the MW GCs and its uncertainty. To estimate the uncertainties conservatively we use the bootstrapping method \citep{Efron79}. Since the GC samples are sparse, we do not expect them to fully sample the distribution function. However, Figure~\ref{fig:velocities} shows that the distribution of all three components of the median GC velocities across the simulations enclose those observed in the MW. Note that we use the absolute values of the radial and polar velocities, because the direction of motion is not relevant in these cases. For the azimuthal component, we keep the true value, because the sign indicates the direction parallel ($+$) or opposite ($-$) to the galactic rotation.

In the simulations, the median GC radial and polar velocity are shifted to systematically larger velocities compared to the stars. The azimuthal component shows a broader distribution, and indicates that almost all the simulations have GC systems with prograde rotation with respect to the disc. This is to be expected if a significant fraction of GCs formed within the disc and their orbits did not evolve significantly until $z=0$. The MW GC median velocities fit very well within the distribution of the simulations, including its prograde rotation velocity, which exceeds the value for about 80 per cent of the simulated galaxies. Note that comparisons of instantaneous velocities should be treated with caution, as even equilibrium systems should show stochastic fluctuations in the median when using only a small number of tracers. However, physical effects also cause deviations from equilibrium. For instance, recent accretion events may skew the velocity distribution away from this expectation in a way that could enable tracing the assembly history of the galaxy.

The distribution of each of the components of the velocity dispersion across the GC systems of the 25 simulations is shown in the bottom row of Figure~\ref{fig:velocities}. The GC velocity dispersions in the simulations are typically larger than for the stars. The MW GC system has a larger dispersion than about 84 per cent of the E-MOSAICS galaxies across all components. This likely indicates that a significant fraction of the MW GCs were accreted during many small mergers with diverse infall trajectories. To ensure that the lower dispersions in the simulations compared to the MW are not due to the under-massive stellar components of $L^*$ galaxies in the EAGLE model \citep{Schaye15}, we also computed the distributions for the most massive half of the galaxy sample (with a median $\log M_*/\Msun = 10.46$, or $0.17$ dex above the median of the full sample). The MW dispersions are still significantly larger in each component compared to this massive galaxy subsample, confirming the atypical location of the MW in the high dispersion tail of the $L^*$ galaxy distribution. We will demonstrate in Section~\ref{sec:tracing_assembly_with_GCs} that the MW seems to have had an atypically large number of low-mass mergers. 

\begin{figure}
    \includegraphics[width=0.85\columnwidth]{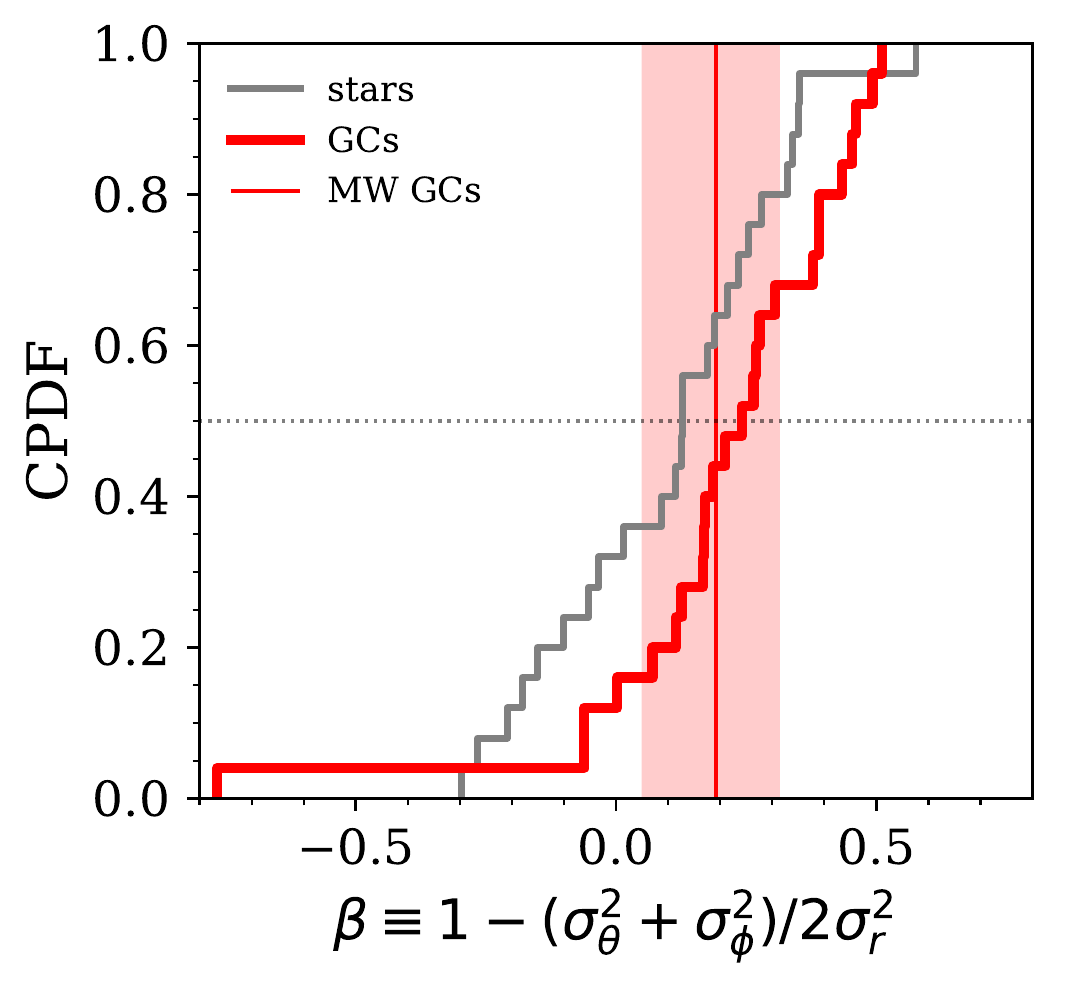}
    \caption{Cumulative PDF of the velocity anisotropy parameter $\beta$ of the GC systems of the 25 MW-mass simulations. Each data point corresponds to the median over the population of one galaxy. The observed values for the MW GC system and their uncertainties are shown by the vertical line and shading. In the simulations GCs typically have more radial orbits than field stars. The MW anisotropy of the MW GC system is typical among the simulations.} 
\label{fig:anisotropy}
\end{figure}

Figure~\ref{fig:anisotropy} shows the distribution of the velocity anisotropy parameter, $\beta \equiv 1 - (\sigma_\theta^2 + \sigma_\phi^2)/2\sigma_r^2$. This parameter is zero in the case of isotropic orbits (the tangential and radial dispersions are comparable), and becomes positive for radially dominated orbits, or negative for tangentially dominated orbits. Overall, both the stars and the GCs in the simulations have on average radially dominated orbits. Stars are offset towards slightly more tangential motions due to the higher degree of rotational support in the disc \citep[the stellar anisotropy of EAGLE galaxies was examined by][]{Thob19}. The MW GCs seem to have a typical degree of rotational support with respect to the simulations. Hence, although the dispersions are larger in the MW's GC population, its distribution of tangential versus radial orbits is common. In Section~\ref{sec:subpopulations} we will investigate which GC subpopulations are responsible for these trends.

\subsubsection{Radial profiles}

\begin{figure*}
    \includegraphics[width=0.38\textwidth]{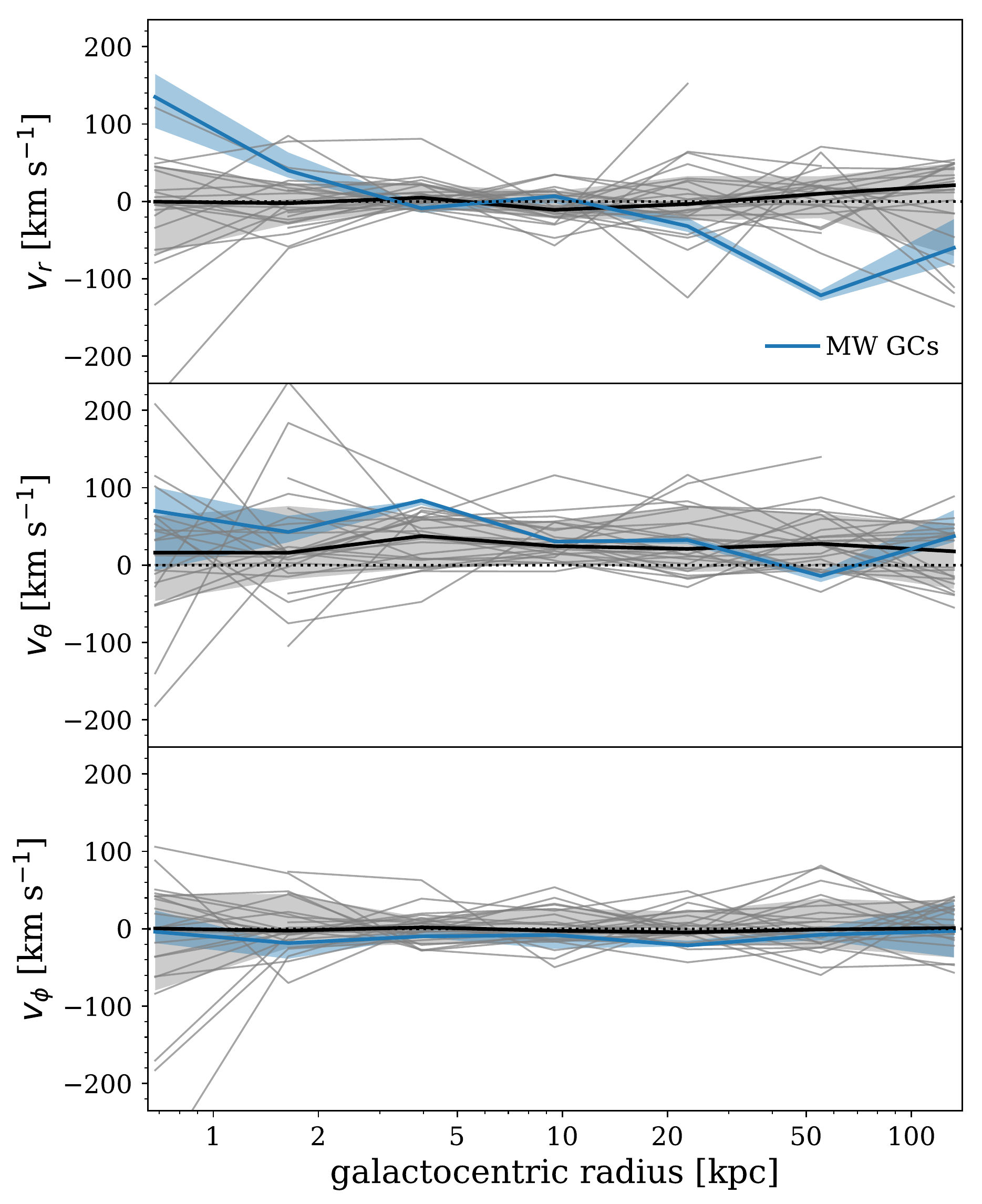}
    \includegraphics[width=0.37\textwidth]{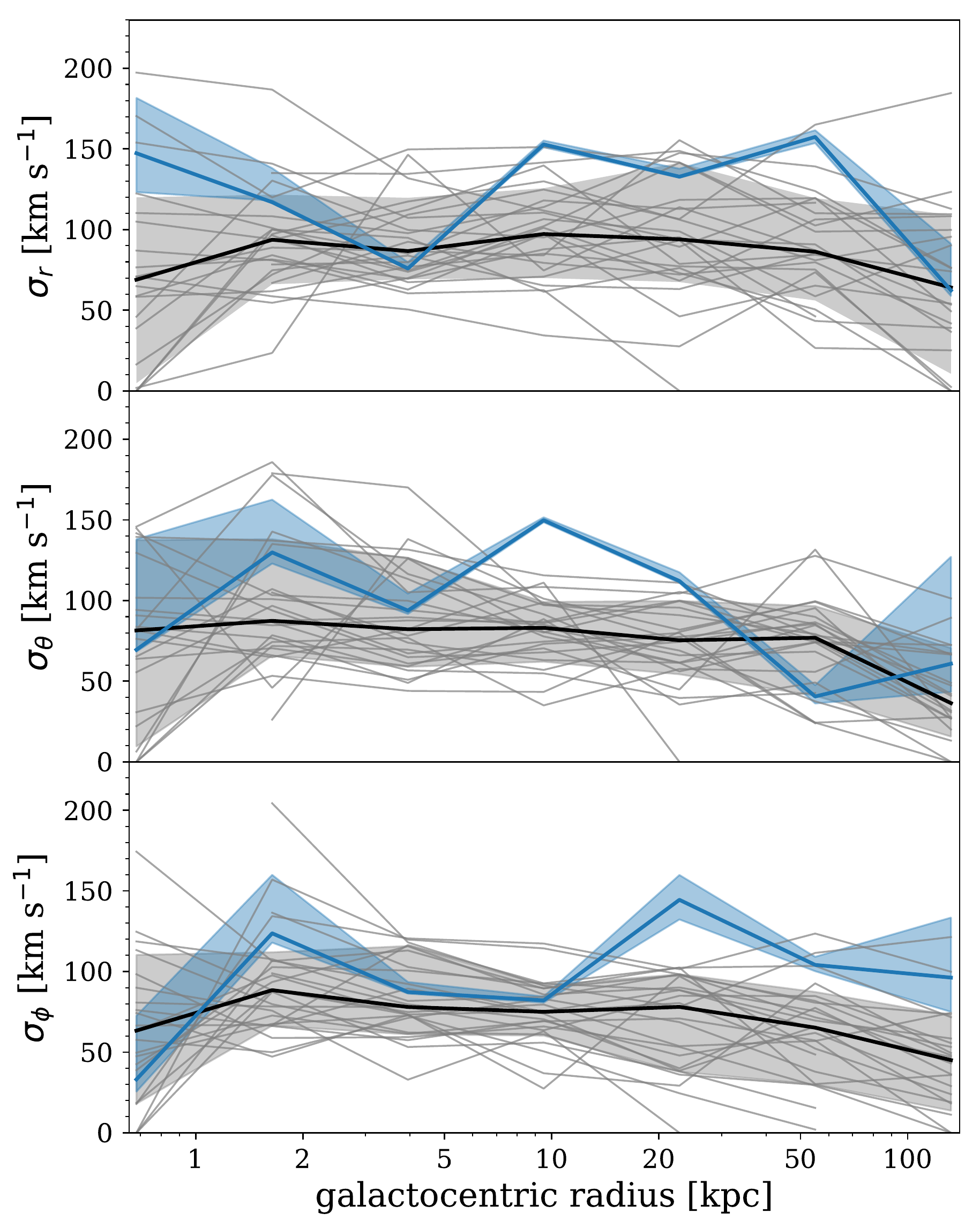}
    \caption{Radial profiles of median GC velocities and velocity dispersions across the GC systems of the 25 MW-mass simulations. Left: Radial profiles of median velocity components. Right: Radial profiles of velocity dispersion. Across all panels the black lines and shading show the median and 16th to 84th percentile envelopes across the 25 simulations respectively, while the thin grey lines show the individual profiles for each galaxy. The observed values for the MW GC system and their uncertainties (estimated using Monte Carlo sampling) are shown by the coloured lines and shading in each panel. The MW fits well within the distributions of the simulations, but shows larger prograde rotation in the inner galaxy ($r \la 8\kpc$) compared to the median simulation. Moreover, the MW GCs have larger dispersions throughout the galaxy relative to the median simulation.}
\label{fig:profiles}
\end{figure*}

To examine how the velocity components vary with galactocentric radius, Figure~\ref{fig:profiles} shows the binned radial profiles of the 3D velocities and velocity dispersions of the MW system compared to the median profiles in E-MOSAICS. In addition to the median and 16-84th percentile range across the 25 simulations, we show the individual median velocity and dispersion profiles for each galaxy. Figure~\ref{fig:profiles} shows that clusters in MW-mass galaxies have on average a prograde rotation, $\vtheta \approx 20-40\kms$, which extends all the way from the inner disc into the outer halo. While the simulations show a broad spread in radial and polar velocity and dispersion profiles, the median velocity across all 25 galaxies is consistent with zero, as expected for dynamical equilibrium. 

In general, the MW fits well within the range of velocity profiles spanned by the 25 simulations. However, its GC population is atypical in three aspects. First, it has a radial velocity gradient, with the median bulge GC at positive radial velocity, and the median outer halo GC moving towards the centre. As discussed in Section~\ref{sec:velocities}, for small numbers of tracers the median GC radial and polar velocities are time dependent, such that the radial profiles may fluctuate stochastically in time even in an equilibrium system, and any trends should not be over-interpreted. On the other hand, accretion of massive satellites causes out-of-equilibrium fluctuations in the velocity distributions, shifting the median. This effect seems to be dominant even for large numbers of tracers, as seen in the deviation from zero of the median radial and polar velocities of star particles in many of the simulated galaxies.  Second, the MW inner GCs (those with $r \la 8\kpc$) show atypically fast prograde rotation ($\vtheta \approx 40-80\kms$), while in the outer galaxy ($r\ga 10\kpc$) they show little rotation. This is a potential signature of the lack of disruptive mergers in the MW's recent history. The analysis in Section~\ref{sec:tracing_assembly_with_GCs} confirms this hypothesis. Third, the velocity dispersions, especially in the radial component, seem to be atypically high in the MW outer halo GC populations. The magnitude of the effect is larger than what is expected from the underpredicted stellar masses of $L^*$ galaxies in EAGLE, possibly indicating that the MW GCs originated from many incoherent accretion events, each bringing a few GCs along a very different infall orbit.

\subsubsection{Metallicity and radial GC subpopulations}
\label{sec:subpopulations}

In this section, we explore which cluster subpopulations are responsible for the overall trends found in the velocities and dispersions. First we split the sample into two metallicity bins, the metal-poor GCs with $[\rm{Fe/H}] < -1.2$, and the metal-rich population with $[\rm{Fe/H}] > -1.2$.  To compare their relative kinematics, we calculate the distribution of the ratios of the median velocities and dispersions of the two populations, respectively (except for the azimuthal velocity, where the difference is used instead). 

\begin{figure*}
    \includegraphics[width=0.63\textwidth]{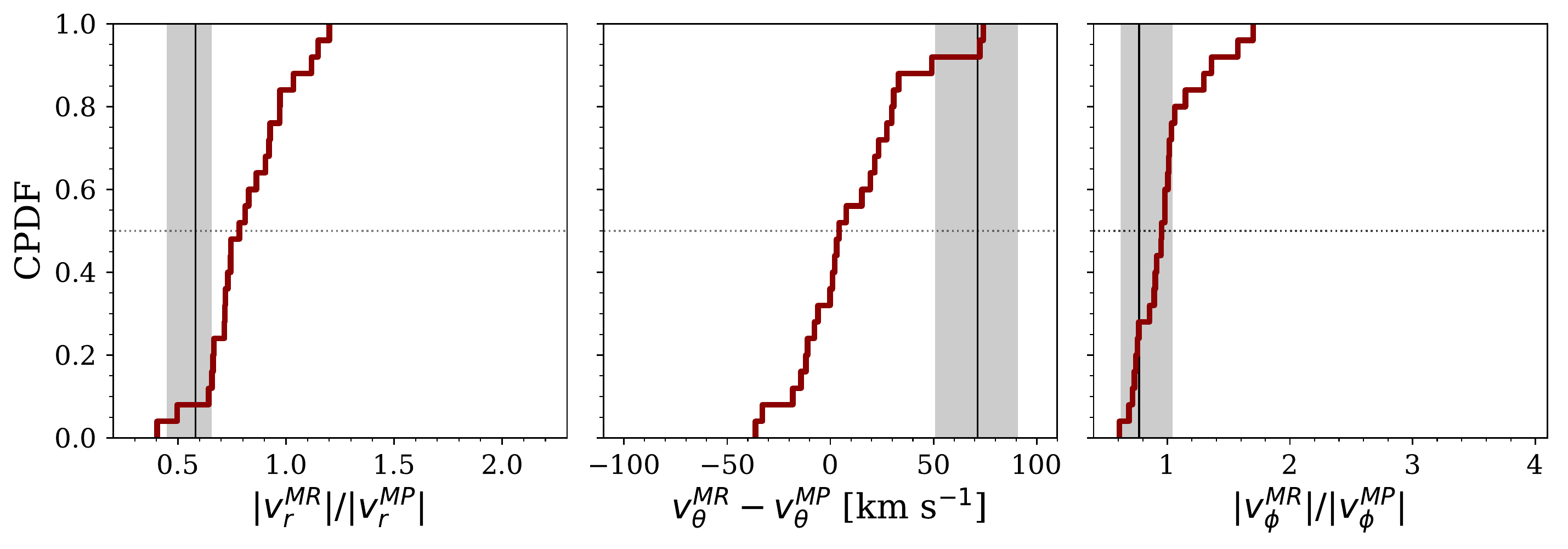}
    \includegraphics[width=0.63\textwidth]{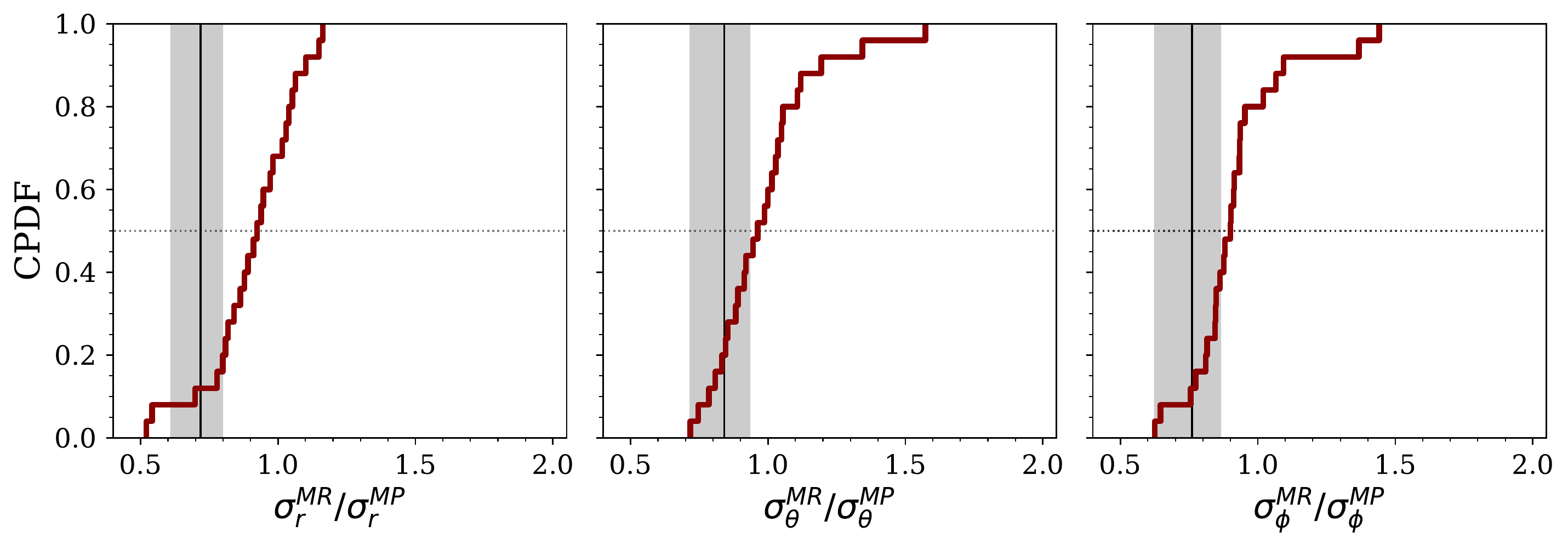}
    \includegraphics[width=0.63\textwidth]{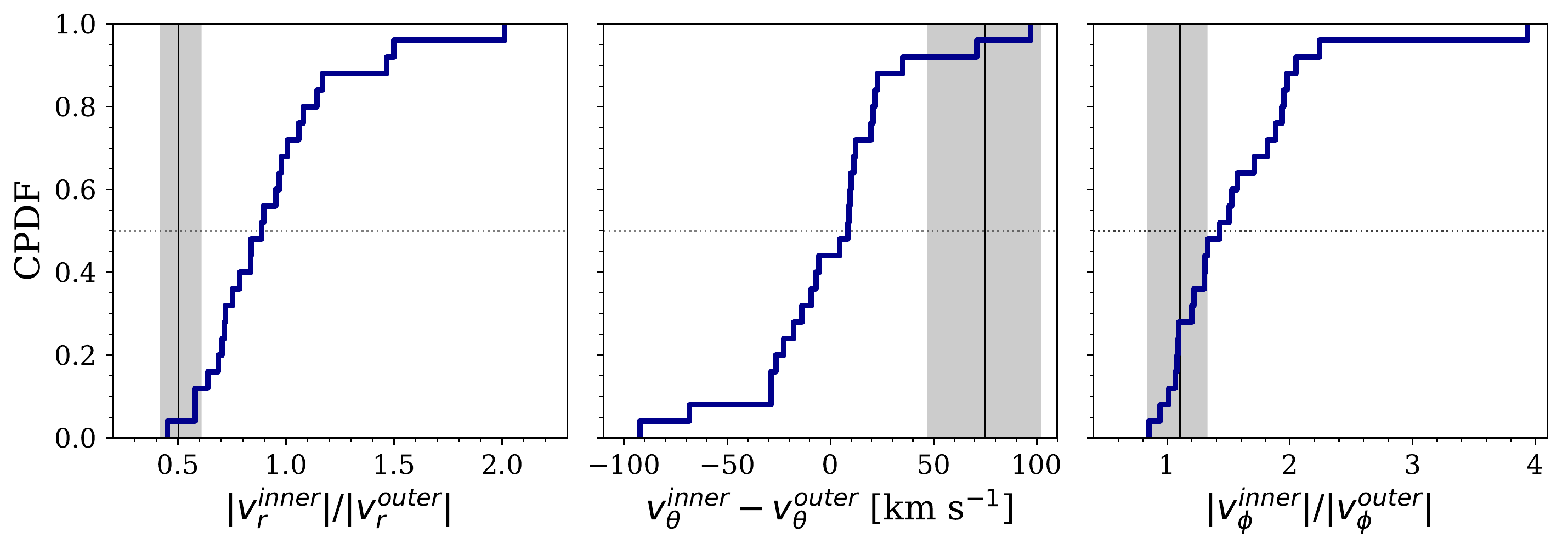}
    \includegraphics[width=0.63\textwidth]{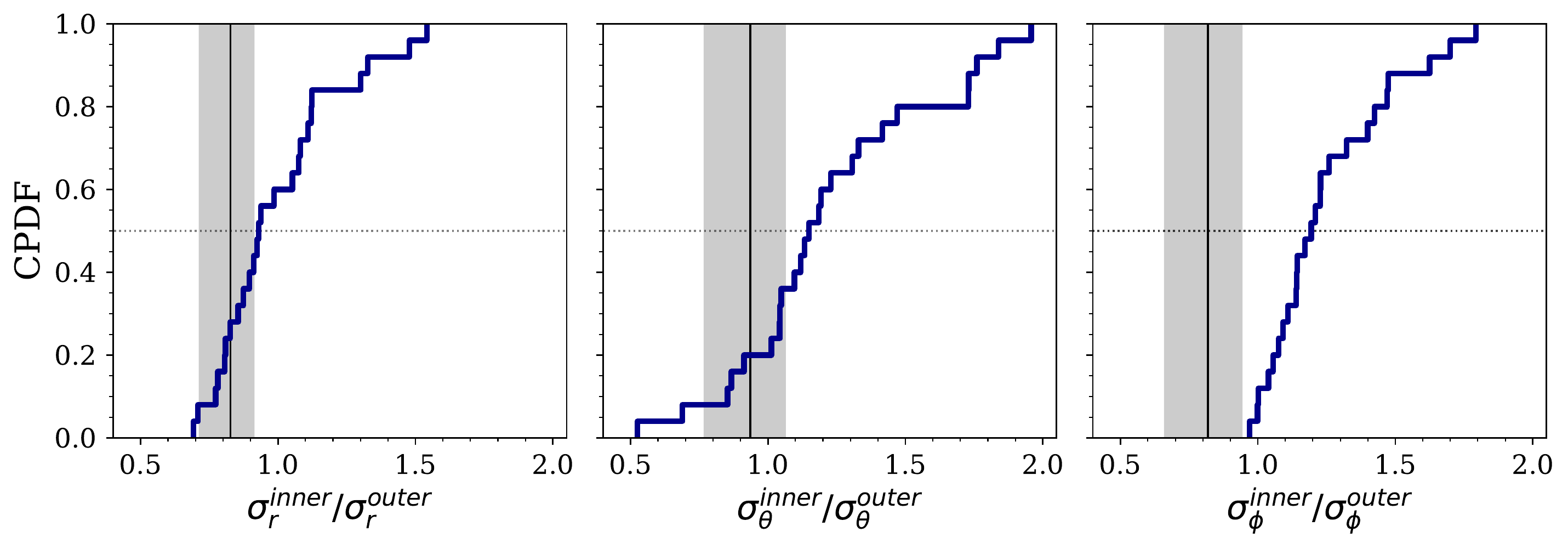}
    \caption{Comparison of the relative kinematics of the metal-rich/metal-poor and inner/outer GC system subpopulations. First row: cumulative PDF of the ratio (or difference for the azimuthal component) between the median velocities of the metal-rich ($[\rm{Fe/H}] > -1.2$) and the metal-poor ($[\rm{Fe/H}] < -1.2$) clusters. Second row: ratios of the velocity dispersions of the metallicity subpopulations. Third row: cumulative PDF of the ratio (between the median velocities of the inner ($r<8\kpc$) and outer ($r>8\kpc$) clusters. Fourth row: ratios of the velocity dispersions of the radial subpopulations. The MW values and uncertainties are shown by the vertical line and shading. Metal-rich GCs in the simulations are on average slightly kinematically colder than metal-poor GCs. The MW lies consistently in the tail of the distributions, with its metal-rich and inner clusters rotating faster than its metal-poor and outer GCs. The MW also has atypically low dispersions of its metal-rich and inner GCs relative to its metal-poor and outer GCs.}
\label{fig:velocities_metrad}
\end{figure*}

The first row of Figure~\ref{fig:velocities_metrad} shows the results for the cumulative distribution of the relative median velocity components of the metal-rich and metal-poor subpopulations, and compares them to the MW. The relative velocities of the MW's subpopulations are not typical compared to the simulations: metal-rich GCs in the MW have distinctly low radial velocities and faster prograde rotation relative to the metal-poor population. This suggests that the strong prograde rotation of the entire GC population is dominated by the metal-rich GCs. The second row of Figure~\ref{fig:velocities_metrad} shows the ratios of the velocity dispersions. Significant differences between the velocity dispersions of the two populations are uncommon in the simulations. However, both the fast rotation and the low dispersion of metal-rich MW GCs relative to the metal-poor population lie in the 80th-90th percentile tail of the simulations. Since metal-rich clusters in the simulations are found preferentially at smaller galactocentric radii \citep{Keller20},  this is likely evidence of relatively weak dynamical heating of the MW disc in comparison to similarly massive galaxies. \footnote{The absence of a resolved cold ISM in EAGLE could contribute to this by artificially thickening the discs and increasing the vertical dispersion. However, the relatively low disc dispersions that result from the slightly undermassive stellar components of MW-mass haloes in EAGLE \citep[see][]{Schaye15} dominate the systematics. This is at least partially compensated by taking the ratio of the dispersions for the two subpopulations. } 

Next, we divide the GC samples into two radially distinct populations, inner GCs at $r<8\kpc$, and outer GCs at $r>8\kpc$.  The third row of Figure~\ref{fig:velocities_metrad} shows the relative velocity distributions of the two populations. On average, the velocities of the inner and outer subpopulations in E-MOSAICS do not differ significantly \footnote{The radial gradient in the MW radial velocity is not evident in Figure \ref{fig:velocities_metrad} because the inner and outer GCs have similar radial velocity magnitudes.}, except for the magnitude of the polar component, which is larger in the inner population across most of the simulations. In the MW, the inner GCs rotate on average significantly faster than the outer GCs.\footnote{We verified that this feature is not due to incompleteness in the MW bulge GC population. Excluding GCs with $r<3\kpc$ in the simulations has little effect on the distribution of relative velocities of inner and outer clusters.} 

In terms of the dispersions, the bottom row of Figure~\ref{fig:velocities_metrad} shows that in the simulations, inner clusters stand out due to their larger azimuthal and polar velocity dispersions compared to the outer GCs. In equilibrium dispersion-supported systems, this results from the drop in the rotation curve at large galactocentric radii. Compared to the simulations, the ratio of all three components of velocity dispersion in the MW inner and outer populations is relatively low. In addition to the similar trend found in the metal-rich/metal-poor populations above, this is an indication of the coherence or dynamical coldness of inner GCs, which are predominantly in-situ, due to an absence of late major mergers. We will expand on this statement more quantitatively in Section~\ref{sec:tracing_assembly_with_GCs}.

\subsection{Orbits}
\label{sec:orbits}

Using the 3D velocities, if the potential of the galaxy is known a priori, the orbits of GCs can be fully characterised by integrating the equations of motion. Here we use  the pericentre $\rperi$ and apocentre $\rapo$ radii, and the eccentricity $e$ to describe the GC orbits in the simulations. These orbital characteristics are commonly used in dynamical studies of the Galaxy because they remain constant in slowly varying potentials.

\begin{figure*}
    \includegraphics[width=0.75\textwidth]{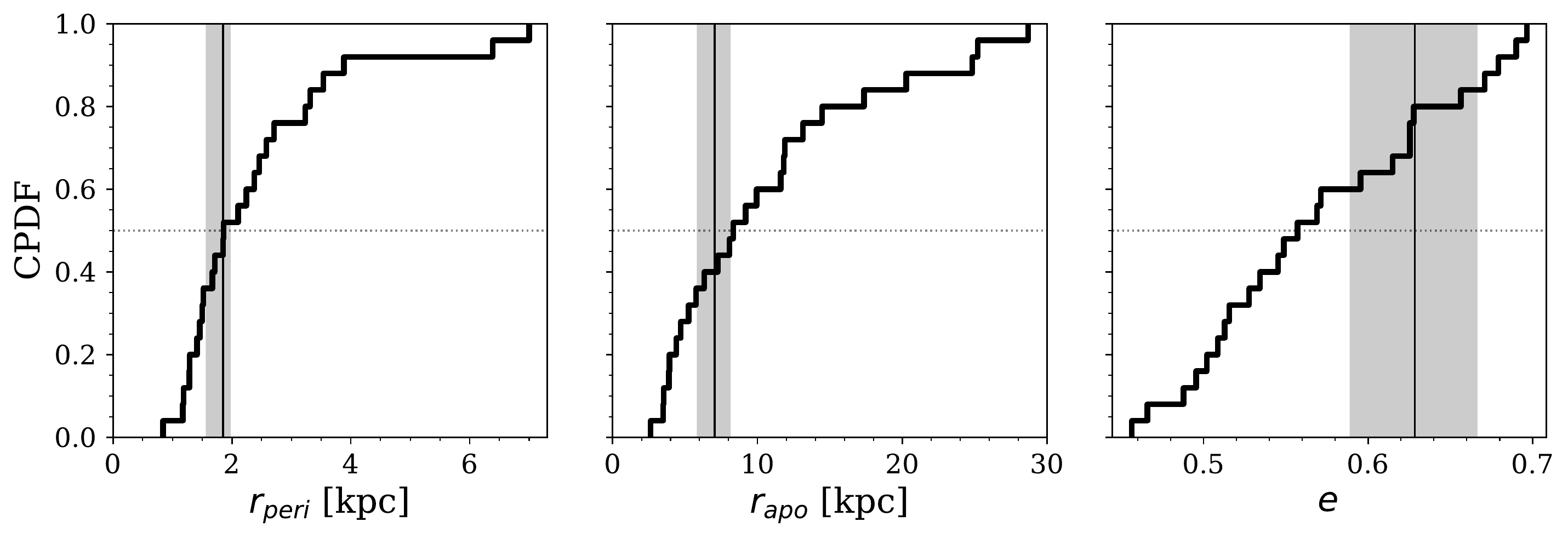}
    \caption{Distribution of median GC orbital characteristics. From left to right the panels show the cumulative PDF of the median pericentre, apocentre, and eccentricity of the GC systems respectively. The MW values and uncertainties are shown by the vertical line and shading. The orbital characteristics of MW GCs are typically found in the simulations.}
\label{fig:orbits}
\end{figure*}

To simplify the calculation of the orbits in the simulations, $\rperi$ and $\rapo$ are obtained following \citet{Mackereth19}, assuming the potential is spherically symmetric and finding the roots of the implicit equation 
\begin{equation}
   L^2 + 2r^2 [\Phi(r) - E] = 0 
\end{equation}
for the galactocentric radius $r$, where $L$ is the magnitude of the angular momentum, $\Phi$ is the gravitational potential and $E$ is the total GC energy. The eccentricity is then calculated as
\begin{equation}
    e = \frac{ \rapo - \rperi }{ \rapo + \rperi } .
\end{equation}
For the MW GCs we obtained the orbital parameters from the catalogue by \citet{Baumgardt19}, where the orbits were integrated assuming Model I for the MW potential from \citet{Irrgang13}. Figure~\ref{fig:orbits} shows the distribution of median orbital characteristics across the 25 galaxies and compares them to the MW GC system. The simulations show a broad distribution of orbital pericentres and apocentres. The median MW orbits are typical in the simulations, except for its slightly elevated median eccentricity. 

\begin{figure*}
    \includegraphics[width=0.67\textwidth]{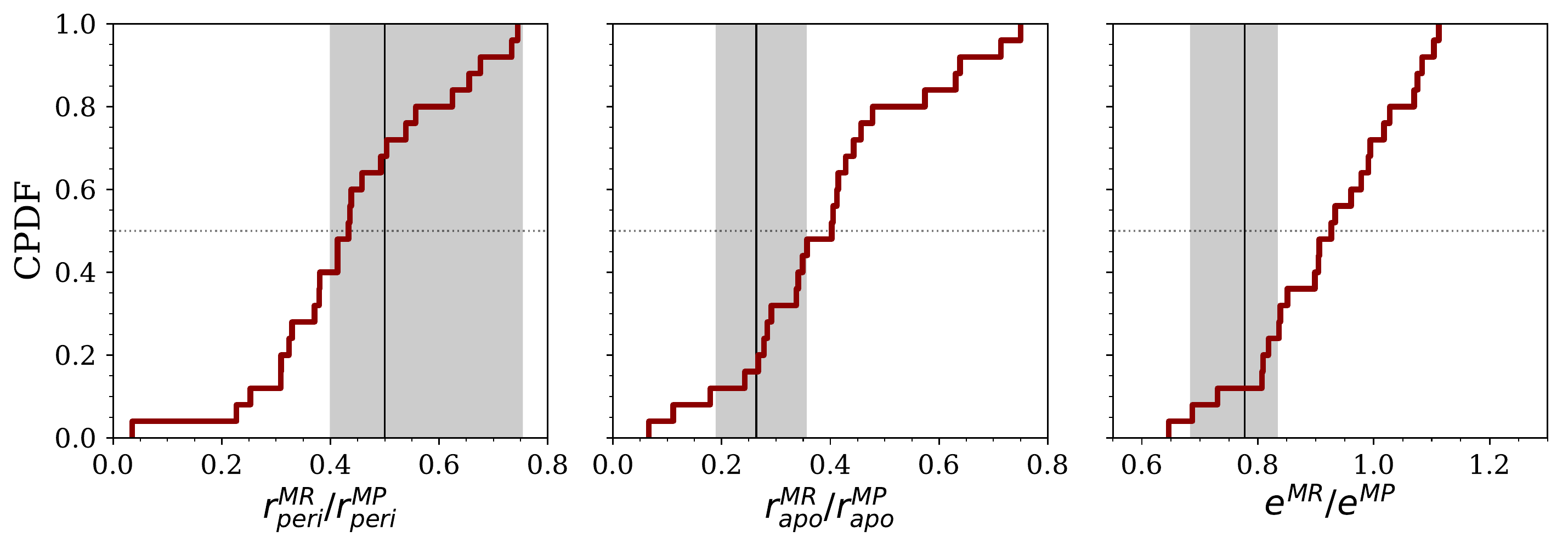}
    \includegraphics[width=0.67\textwidth]{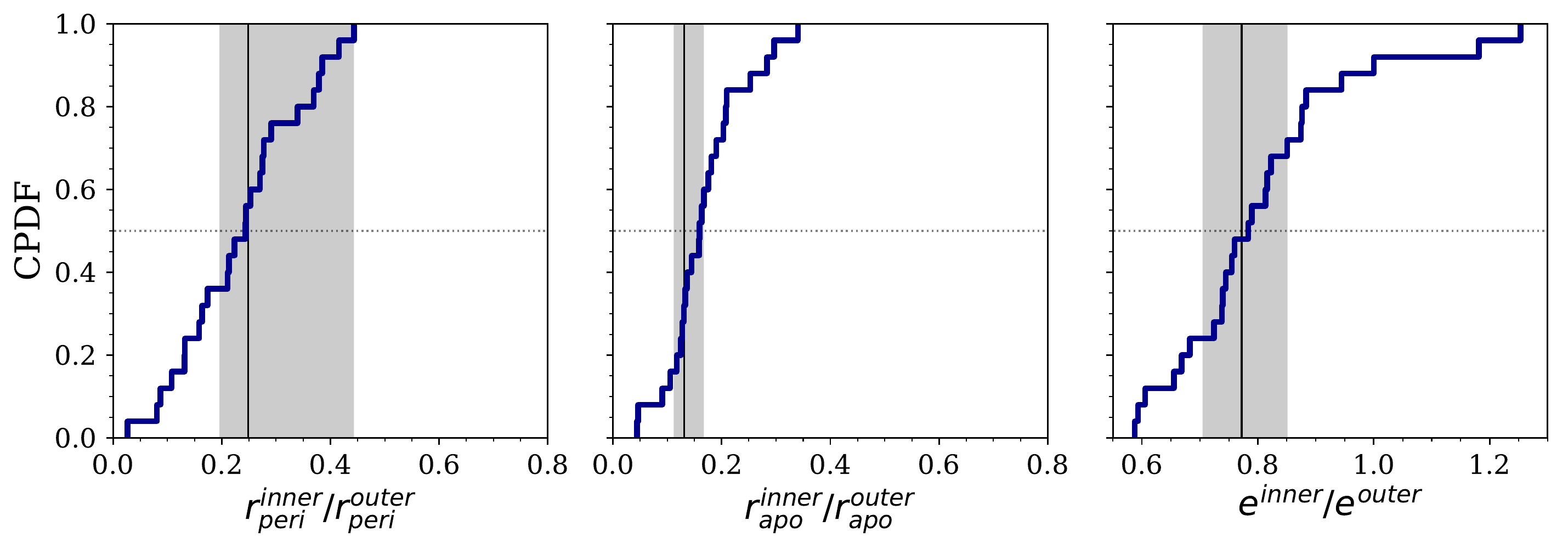}
    \caption{Comparison of the distribution of orbital characteristics of GC subpopulations in metallicity (top row) and galactocentric radius (bottom row). From left to right the panels show the cumulative PDF of the ratio of median pericenter, apocenter, and eccentricity of the two GC populations, respectively. The MW values and uncertainties are shown by the vertical line and shading. The median metal-poor or outer GC \emph{always} orbits at larger radii and typically with higher eccentricities than the median metal-rich or inner GC.}
\label{fig:orbits_pops}
\end{figure*}

We can further compare the orbits of metal-poor versus metal-rich and inner versus outer GC populations. Figure~\ref{fig:orbits_pops} shows the distribution of the ratio of the median orbital parameters of the subpopulations. There are clear systematic differences in the relative orbits of the subpopulations split by metallicity and galactocentric radius. In the simulations, the median metal-poor GC \emph{always} orbits at a larger distance than the median metal-rich GC. In general, the metal-poor as well as the outer GCs have more eccentric orbits than $\sim70$ per cent of the metal-rich and the inner GCs. Figure~\ref{fig:orbits_pops} also shows that in the MW, the ratio between the eccentricities of metal-rich and metal-poor GCs is lower than in about 85~per~cent of the simulations. Moreover, the ratio between the apocentre radii of metal-rich and metal-poor GCs in the MW is also relatively small (smaller than in about 80~per~cent of the simulations). These features could be tracers of the fraction of GCs that formed ex-situ and inherited the eccentric orbital motion of their host satellite. In Section~\ref{sec:tracing_assembly_with_GCs}, we show that there is a strong correlation between the relative eccentricity of metal-poor and metal-rich GCs and the redshift of the last major merger.

\subsection{Integrals of motion}
\label{sec:integrals}

Integrals of motion are functions of the phase-space coordinates that remain constant along the orbit and are independent of time \citep[e.g.,][]{BinneyTremaine}. They provide a more robust way of describing the GC kinematics by removing the time-dependence of the instantaneous phase-space coordinates. The set of integrals of motion for a given problem is defined based on the spatial symmetry of the potential and its variability in time. Here we use those quantities which are conserved under the fewest restrictions, namely the magnitude of the angular momentum vector $L$ (conserved in the absence of external torques), the $z$-component of the angular momentum vector $\Lz$ (an integral of motion in axisymmetric potentials), and the Hamiltonian or total energy $E$ (constant in a static potential if forces are conservative). To obtain the potential energies of the MW GCs we use {\sc galpy} \citep{galpy}, assuming Model I for the MW potential from \citet{Irrgang13} for consistency with the orbits calculated by \citet{Baumgardt19}.
 
\begin{figure*}
    \includegraphics[width=0.75\textwidth]{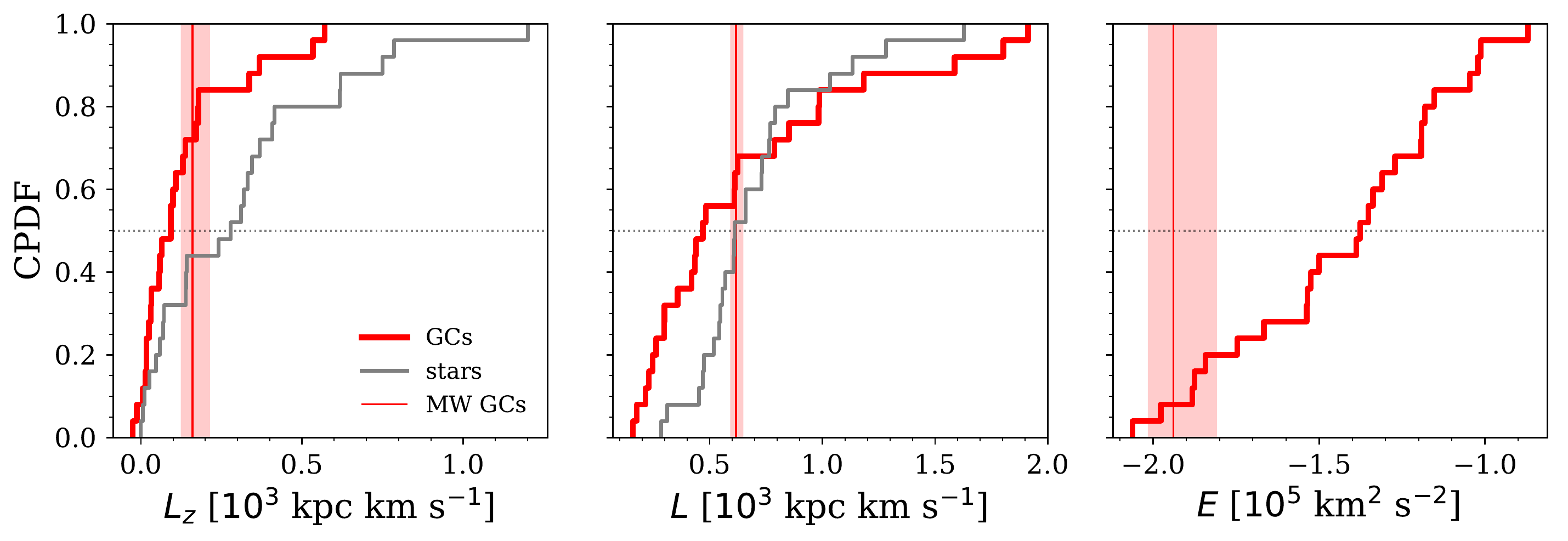}
    \caption{Distribution of the integrals of motion of the GC systems and the stars in the simulations. The left, middle, and right panels show the cumulative PDF of the median $z$-component of angular momentum, magnitude of the total angular momentum, and total energy of the GC systems of the 25 simulations respectively. The grey lines show the angular momentum distribution for all the stars bound to each galaxy. The observed values for the MW GC system and the associated uncertainties are shown by the vertical line and shading. With few exceptions, GCs have prograde orbits  and lower vertical angular momenta than the stars. The MW GC system has a significantly larger median binding energy than the average E-MOSAICS galaxy.}
\label{fig:integrals}
\end{figure*}

Figure~\ref{fig:integrals} shows the distributions of the median GC integrals of motion, as well as the median angular momenta of the stars. The total angular momentum distributions of the GCs and the stars are similar, but differ in that $\Lz$ for GCs is lower than for stars, indicating that their rotation is not strictly aligned. The MW GCs are fairly typical in terms of angular momentum but lie near the high binding energy tail of the simulations. This may signify that the MW's in-situ GCs formed earlier than is typical for $L^*$ galaxies. Alternatively, it may be explained by an underestimation of binding energies in the simulations due to the slightly low stellar masses ($\sim 0.2$ dex) of EAGLE galaxies with $\M200 \sim 10^{12}\Msun$ compared to observations \citep[see fig. 8 in][]{Schaye15}. By comparing instead the relative energies of the GCs subpopulations we can remove this systematic and investigate the origin of this feature in the MW. 

\begin{figure*}
    \includegraphics[width=0.67\textwidth]{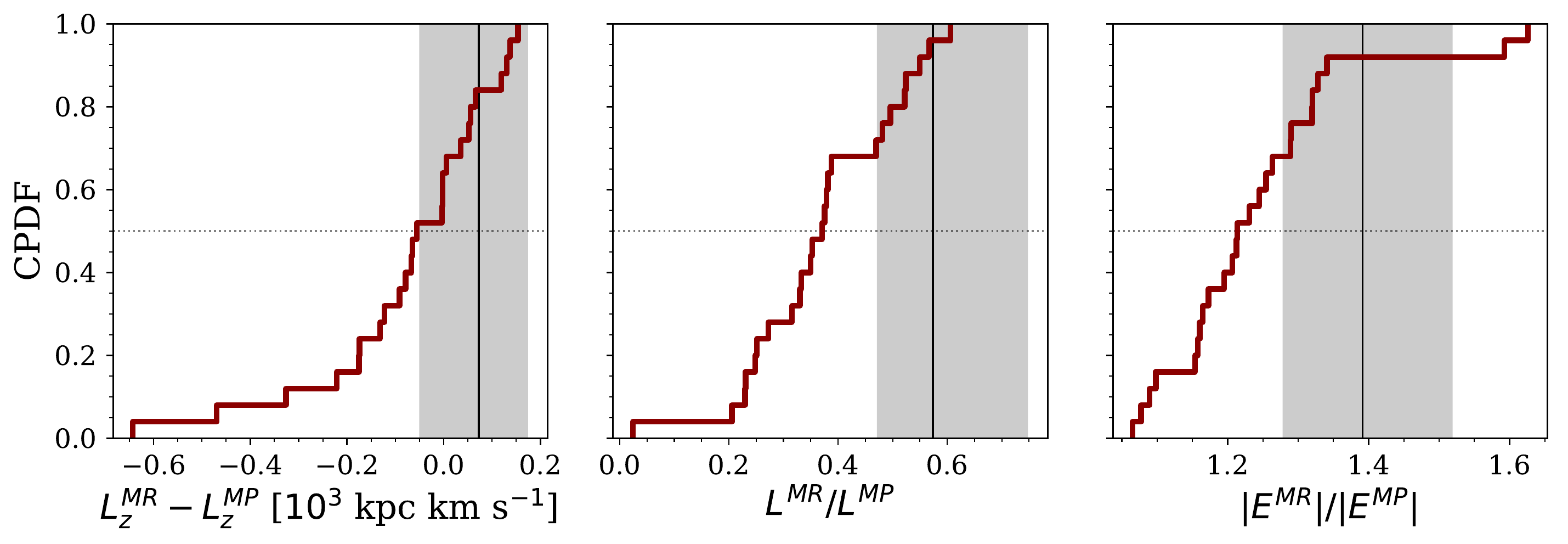}
    \includegraphics[width=0.67\textwidth]{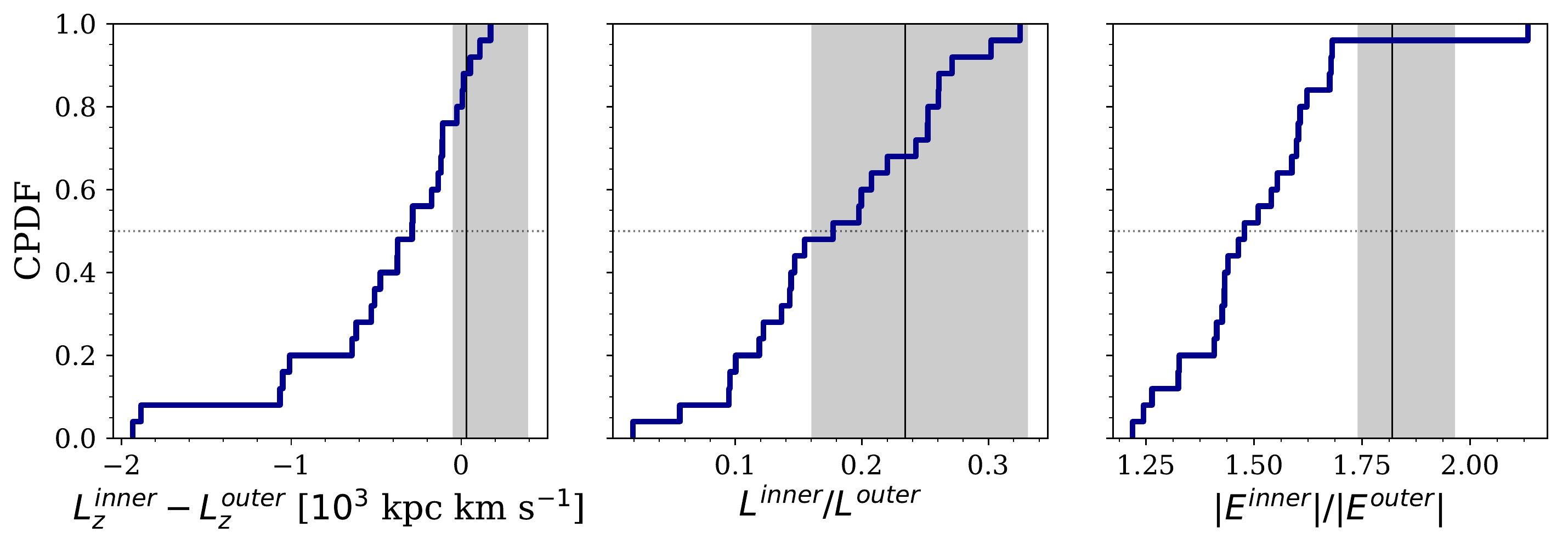}
    \caption{Comparison of the integrals of motion of metallicity (top row) and galactocentric radius (bottom row) GC subpopulations. The left, middle, and right panels show the cumulative PDF of the difference of median $L_z$, ratio of median $L$, and ratio of median $|E|$ of the two subpopulations respectively. The observed values for the MW GC system and the associated uncertainties are shown by the vertical line and shading. The MW is atypical: its metal-rich (and inner) GCs are on average more tightly bound and have larger angular momenta relative to its metal-poor (and outer) GCs.}
\label{fig:integrals_pops}
\end{figure*}

Figure~\ref{fig:integrals_pops} shows the distribution of the difference in median $\Lz$, ratio of median $L$, and ratio of median $|E|$ of the metal-rich/metal-poor and inner/outer GC subpopulations. In the simulations, the majority of galaxies have a metal-rich (and inner) GC component with lower angular momentum and higher binding energy than the metal-poor (and outer) GCs. The similarity between the distributions of inner/outer and metal-rich/metal-poor subpopulations in the simulations is not entirely surprising, since on average 78 per cent of the inner GCs are metal-rich, and 61 per cent of the outer GCs are metal-poor. The MW GC system is atypical in this respect, as it lies in the tail of the binding energy ratio distributions, with its metal-rich (and inner) GCs significantly more bound than 90 per cent of the simulations. As we show in Section~\ref{sec:tracing_assembly_with_GCs}, this is a signature of the relatively early assembly of the MW disc and its lack of late major mergers. The feature is explained by the efficacy with which massive satellites deliver clusters to the inner galaxy through a combination of dynamical friction and more resilience to the early tidal stripping of their tightly bound GCs. 

\section{Kinematics of accreted versus in-situ GCs}
\label{sec:origin}

Simulations provide the unique advantage of tracking the galaxy where each GC formed. We now consider the kinematic signatures of clusters that formed within their present-day galaxy host or within satellites that were later accreted. For each cluster in the simulations, we assign an 'in-situ' or 'accreted' label based on whether the star particle hosting the cluster formed from a gas particle that was bound the main progenitor or to another galaxy. This classification can be ambiguous in cases where the cluster formed from a gas particle that was accreted during the time interval between two simulation snapshots, but this only corresponds for a small fraction of the GCs, for which we assume that the GC was accreted \citep[for details see][]{emosaicsI}. 

\begin{figure*}
    \includegraphics[width=0.78\textwidth]{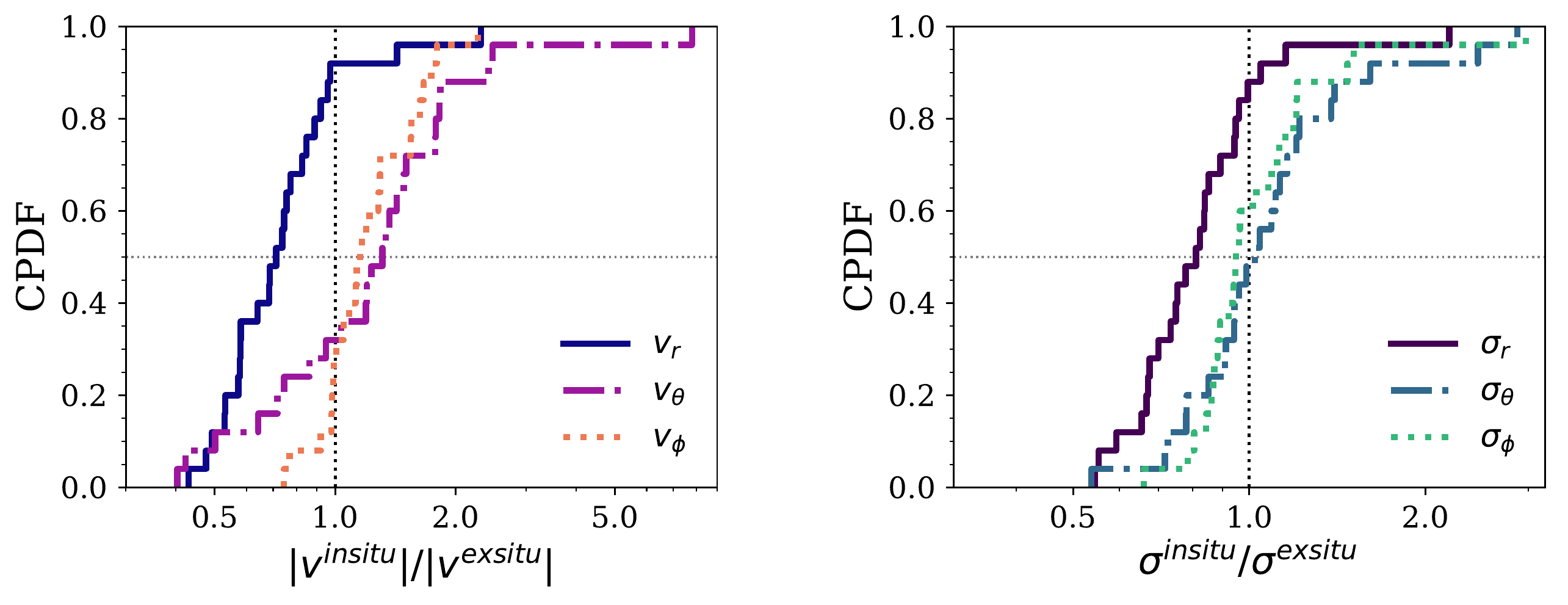}
    \caption{Comparison of the median 3D velocities of in-situ and accreted GC populations. Left: cumulative PDF of the ratio between the median velocity components of in-situ and accreted GCs. Right: same for the velocity dispersions. In-situ GCs typically rotate faster than the accreted population, while in about one third of the simulations the accreted GCs dominate the rotation as a result of recent mergers. Accreted GCs have larger radial velocities and dispersions in more than 80 per cent of the galaxies.}
\label{fig:velocities_acc}
\end{figure*}

Figure~\ref{fig:velocities_acc} shows the distribution of relative velocities for in-situ versus accreted clusters. The median azimuthal velocities of in-situ clusters are larger than in  accreted clusters in about 70 per cent of the galaxies. Perhaps surprisingly, accreted GCs in the remaining $\sim30$ per cent of the simulations dominate the rotation velocity of the system. These galaxies all had recent mergers, and the majority are undergoing mergers at $z=0$. In some cases the recently accreted satellites fall in along a trajectory aligned with the rotation of the disc, while in others they carry enough orbital angular momentum to change the direction of the total angular momentum of the system (which is used to define the $z$-axis of the galaxy for particles within $30\kpc$ of the centre). The only clear discriminator between the in-situ and accreted populations is the radial component, with accreted GCs having larger radial velocities and dispersions in $\sim85$ per cent of the galaxies. The elevated dispersions are a result of the fact that accreted GCs are typically brought in by several satellite accretion events with different orbits, and this leads to a broad radial velocity distribution compared to that of the in-situ GCs (which inherit the circular orbits of the gas disc). 

\begin{figure*}
    \includegraphics[width=0.75\textwidth]{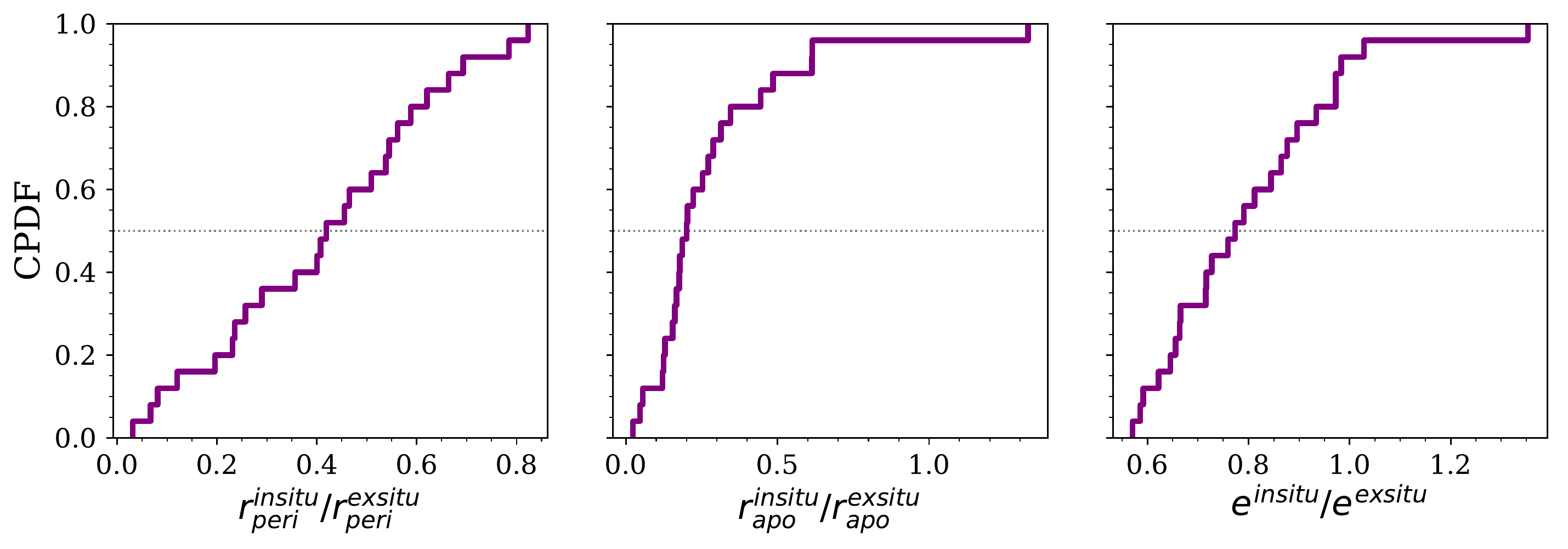}
    \caption{Comparison of the orbits of in-situ and accreted GC populations. From left to right, the panels shows the cumulative PDF of the ratio between the median pericentre, apocentre, and eccentricity of in-situ and accreted GCs respectively. In-situ and accreted GC populations split clearly in orbital space, with in-situ clusters having predominantly smaller median pericentres, apocentres, and eccentricities.}
\label{fig:orbits_acc}
\end{figure*}

\begin{figure*}
    \includegraphics[width=0.75\textwidth]{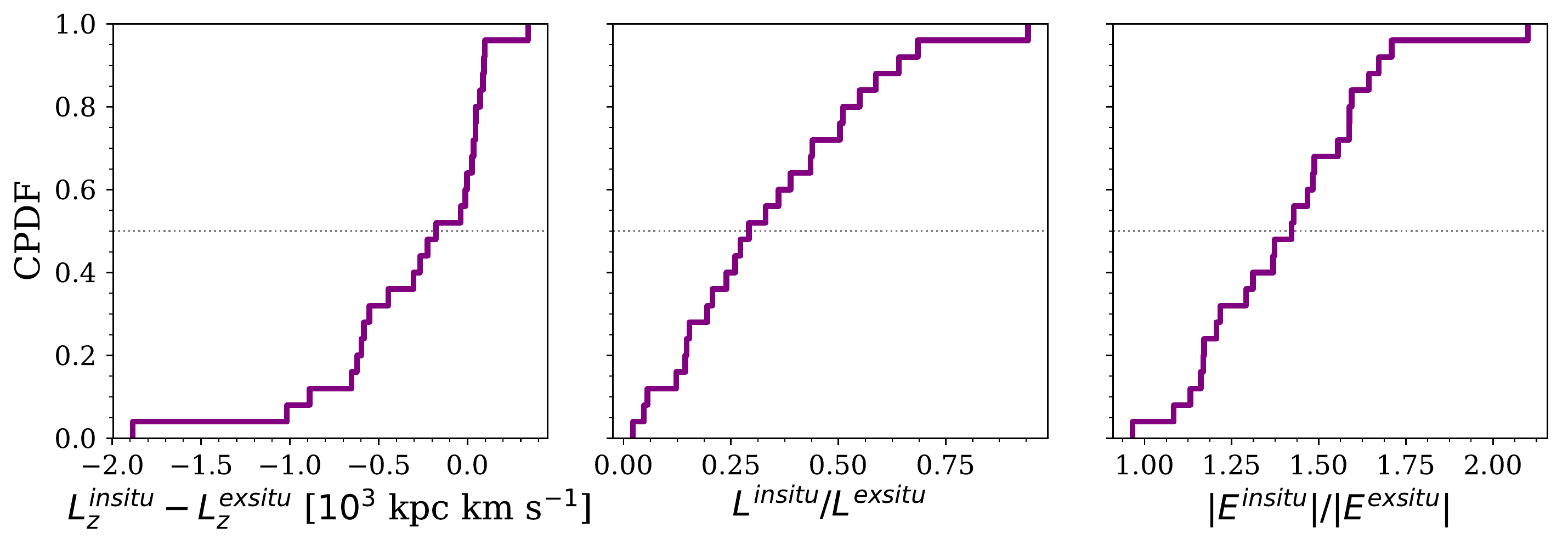}
    \caption{Comparison of the integrals of motion of in-situ and accreted GC populations. From left to right the panels show the cumulative PDF of the median difference of $\Lz$, ratio of $L$, and ratio of $|E|$ of in-situ and accreted GCs respectively. In-situ GCs in nearly all MW-mass galaxies have on average lower total angular momentum and higher binding energy than accreted GCs.}
\label{fig:integrals_accreted}
\end{figure*}

Figure~\ref{fig:orbits_acc} shows the ratio of the orbital parameters of in-situ and accreted GCs. In more than 90 per cent of the galaxies, in-situ clusters orbit at smaller galactocentric distances and with lower eccentricities than accreted clusters. In Figure~\ref{fig:integrals_accreted} we show the distribution of the relative angular momentum and binding energy of in-situ and accreted GC populations. In the vast majority of galaxies the in-situ GCs have lower median angular momentum and higher binding energy than accreted GCs. This is not surprising since accreted GCs orbit at larger radii on average (Figure~\ref{fig:orbits_acc}), and therefore have a larger maximum range of $L$ values. This is the same trend observed in Section~\ref{sec:integrals} for metal-rich versus metal-poor or inner versus outer clusters, and indicates that differences in the kinematics of metallicity or galactocentric radius subpopulations can, on average, be traced directly to their origins. 

Summarizing, we find that GC origin imprints a strong signature in the distribution of relative eccentricities, apocentres, angular momenta, and binding energies. Across the simulations, accreted GCs \emph{on average} orbit at larger distances (median $r= 21.8\kpc$, median $[\rm{Fe/H}]= -1.40$) and have lower metallicities than in-situ GCs (median $r = 4.6\kpc$, median $[\rm{Fe/H}] = -0.85$). These trends translate directly to the relative distributions of GC subpopulations distinguished by radius and metallicity shown in Figures~\ref{fig:orbits_pops} and \ref{fig:integrals_pops}. The distributions of orbits and integrals of motion of the metallicity and radial GC subpopulations should therefore be excellent tracers of the relative importance of in-situ and ex-situ galaxy growth. Of course, these trends apply only to averages across entire populations, and neither the metallicity nor the galactocentric radius of an individual cluster is enough to establish its origin (Reina-Campos et al. in prep.).


\section{Tracing galaxy and halo assembly histories using GC kinematics}
\label{sec:tracing_assembly_with_GCs}

In this section, we apply a general statistical approach to investigate the physical origin of the MW GC kinematic features found in Section~\ref{sec:comparison}.
We adopt and a priori `agnostic' approach, in which we exploit the wealth of information that can be extracted from the 6D phase space distribution of GCs to understand how much of the present-day properties of the galaxy, its dark matter halo, and their assembly history is traced by the GC system kinematics. This is done by performing an unbiased search for statistical correlations between each of the properties describing the assembly of the simulated galaxies, and each of the kinematic tracers. The procedure is summarised as follows.

\begin{enumerate}

\item Following \citet{emosaicsII}, a relevant set of galaxy and halo quantities is selected to comprehensively characterise the diversity of mass distributions, environments, and assembly histories of the DM and stellar components of the 25 MW-mass simulated galaxies from E-MOSAICS. This set of metrics is described in Section~\ref{sec:metrics}.

\item A comprehensive set of GC kinematic tracers is constructed based on the 3D velocities and positions of the GCs in each simulated galaxy at $z=0$. This is done using statistical descriptors (median, inter-quartile range, skewness, and kurtosis) of the distributions of each tracer. The set of tracers is described in Section~\ref{sec:tracers} and includes all the kinematic features that were shown to be sensitive to the details of the assembly histories in Sections~\ref{sec:comparison}. The simulated GC systems are divided into a total of 7 kinematic samples: one for the entire GC system, one for each of the metallicity and the galactocentric radius subpopulations as defined in Section~\ref{sec:simulations}, one for the relative statistics of the metal-rich and metal-poor subpopulations, and one for the relative statistics of the inner and outer subpopulations. The relative statistics are obtained by calculating the ratios of each of the 4 statistics (median, inter-quartile range, skewness, and kurtosis) for the metal-rich/metal-poor, and inner/outer subpopulations. For statistics that are not positive-definite we use the difference instead of the ratio. 

\item A search is performed for statistically significant  correlations between each of the $N \times M$ combinations possible between $N$ GC system kinematic tracers and the entire set of $M$ assembly metrics. The Spearman rank-correlation test is used to assess whether the relationship between each pair of tracer and assembly metric can be described by a monotonic function. All correlations with Spearman $p < 0.05$ (accounting for the effect of multiple comparisons, see Appendix~\ref{sec:statistics}) are selected as statistically significant. We then fit linear regression models to the relationship between each kinematic tracer (as the independent variable) and each assembly metric (as the dependent variable). Out of this set of linear models we select those with the most predictive power (according to their Pearson linear correlation coefficients) for each of the halo and galaxy assembly metrics. The details of the method are described in Appendix~\ref{sec:statistics}. The search for correlations is performed separately for each of the 7 kinematics samples defined above.

\item The observed kinematics of the MW GC system and its subpopulations are used to make quantitative predictions (including their statistical uncertainties) using the selected linear models for several relevant aspects of the the formation and assembly history of the Galaxy.

\end{enumerate}

\subsection{Quantifying galaxy assembly and GC system kinematics}

\subsubsection{Galaxy and dark matter halo assembly metrics}
\label{sec:metrics}

In this work we use the set of assembly metrics from \citet{emosaicsII}. We briefly describe these metrics here and refer the reader to Section 4.2 of \citet{emosaicsII} for a detailed discussion. The assembly metrics are divided into four groups: quantities describing the present-day mass distribution of the galaxy and its dark matter halo, properties describing the time-scales of halo and stellar mass growth, quantities describing the topology of the merger tree, and lastly, quantities describing the in-situ/accreted origin of stars and GCs.

The mass distribution of the galaxy is described using the virial mass $\M200$, maximum circular velocity $\vmax$, galactocentric radius at which the circular velocity reaches its maximum, $\Rvmax$, and NFW profile \citep{nfw97} concentration parameter, $\cnfw$. The mass growth history of the DM halo is characterised using the lookback time when the galaxy reached 25, 50, 75 and 100 per cent\footnote{The maximum mass can in some cases occur at $z>0$ due to the temporary overestimation of $\M200$ during mergers. This effect leads to a maximum discrepancy of about 30 per cent (although typically only a few percent) compared to the $z=0$ mass.} of its total mass ($\tau_{25}$, $\tau_{50}$, $\tau_{75}$, and $\tmax$ respectively), the time when the galaxy main progenitor formed half of its stellar mass $\ta$, and the time when all the progenitors together formed half of their stellar mass $\tf$. To quantify the importance of in-situ versus ex-situ growth using the formation and assembly time-scales we use 
\begin{equation}
    \delta_t \equiv 1 - \ta/\tf ,
    \label{eq:delta_t}
\end{equation}
with $\delta_t > 0.1$ indicating significant growth of the stellar component through mergers \citep{Qu17}. 

The merger tree of each galaxy is described using merger time-scales and demographics. The time-scales consist of the lookback time of the last major merger $\tmm$ (where a major merger is defined by a stellar mass ratio greater than $1/4$), the time when the last merger (of any mass ratio) occurred $\tam$, and the ratio of the merger time-scales for major versus all mergers
\begin{equation}
    \rt \equiv \frac{\tH - \tmm}{\tH - \tam} ,
\label{eq:timescales_ratio}
\end{equation}
where $\tH$ is the Hubble time. The major merger time-scale is also expressed alternatively in terms of the redshift, expansion parameter, time since the Big Bang, and their logarithms, to ensure that the best linear predictor is found.  The demographics of the  merger tree are characterised by considering the total number of branches connecting to the main branch $\Nbr$ (i.e., the total number of mergers experienced by the main progenitor), the number of branches connecting to the main branch at $z>2$ $\Nbrz$ (i.e., the number of $z>2$ mergers), the ratio of the number of mergers at high redshift over all mergers $r_{z>2} \equiv \Nbrz/\Nbr$, the total number of progenitors (or `leaves') $\Nleaf$, and the number of major $\Nmajor$ (stellar mass ratio $>1/4$), minor $\Nminor$ (mass ratio between $1/100$ and $1/4$), small $\Nsmall$ (mass ratio between $1/100$ and $1/20$), medium $\Nmedium$ (mass ratio between $1/20$ and $1/4$), and tiny $\Ntiny$ (mass ratio $<1/100$) mergers. In addition, the relative importance of major mergers is quantified using the ratio of the number of major mergers to all other mergers, 
\begin{equation}
    \rmm \equiv \frac{\Nmajor}{\Nminor + \Ntiny} .
\label{eq:merger_ratio}
\end{equation}
Since the resolution of the simulations limits the minimum resolved mass of a galaxy to $M_* \geq 4.5\times10^6\Msun$, mergers below this mass scale are unresolved and therefore considered smooth mass accretion. In the remainder of the paper we refer to resolved mergers simply as `mergers'. Lastly, the origin of stars and GCs is quantified using the fraction of mass in GCs and stars formed ex-situ, $\fexgcs$ and $\fexstars$ respectively.\footnote{The fraction of ex-situ clusters is defined relative to the total number of GCs with mass $>10^5\Msun$ regardless of  metallicity, maintaining the general metallicity selection of $-2.5<\feh<-0.5$ mentioned in Section~\ref{sec:simulations}.}

\subsubsection{GC system kinematics tracers}
\label{sec:tracers}

The set of kinematic tracers of galaxy assembly used here is selected to include the typical quantities used in dynamical studies complemented by several additional physically motivated properties. These are the $z$-component of the angular momentum vector $\Lz$, the magnitude of the angular momentum $L$, the kinetic energy $\Ekin$ per unit mass, the total energy per unit mass $E = \Ekin + \Epot$, where $\Epot$ is the potential energy per unit mass, and the orbital characteristics (pericentre and apocentre radius, and eccentricity). We also include quantities that describe the instantaneous kinematics of the GC system, namely the median 3D velocities, the tangential velocity $\vt \equiv \sqrt{\vtheta^2 + \vphi^2}$, and the velocity anisotropy parameter $\beta$. To reduce the dimensionality of the kinematic data, the distribution of the GC system associated to each simulated galaxy is described using the four statistics for each of the kinematic tracers listed above: the median, inter-quartile range, skewness, and kurtosis. The inter-quartile range is a measure of the width of the distribution, while the skewness quantifies its deviation from symmetry around the median, and the kurtosis measures the weight of the `wings' relative to the the central peak.

\subsection{Correlations between galaxy assembly and GC system kinematics}
\label{sec:correlations}

Following the procedure outlined in Section~\ref{sec:tracing_assembly_with_GCs}, we now search for correlations between each GC kinematic tracer (defined in  Section~\ref{sec:tracers}) and each of the galaxy assembly metrics (listed in Section~\ref{sec:metrics}). The search is first performed using the statistical descriptors (median, inter-quartile range, skewness, and kurtosis) of each of the kinematics across the entire GC population of each simulated galaxy without using additional metallicity or spatial information. For this we follow the statistical method described in detail in Appendix~\ref{sec:statistics}. In short, we assess whether a monotonic function can describe the relationship between each pair of variables by performing Spearman rank-order correlation tests using the kinematic tracer as the independent variable and the assembly metric as the dependent variable. After correcting the threshold $p$-value used to determine statistical significance for the effect of multiple comparisons (see Appendix~\ref{sec:statistics} for details), we select only those correlations with Spearman $p<0.05$. We then perform linear regression fits to each of the correlated pairs and calculate the linear correlation coefficient, or Pearson $r$, which indicates the fraction of the variation in the data that is explained by a linear model. Only those with $|r| > 0.7$ are selected, and in a few interesting cases the requirement is relaxed to $|r| > 0.6$. A total of 10 correlations are found which satisfy the two criteria: statistical significance (Spearman $p < 0.05$), and linear correlation coefficient $|r| > 0.7$. To mitigate biases due to the underproduction of stellar mass in the EAGLE model (see Section~\ref{sec:integrals}), we avoid whenever possible using the correlations found with kinematic tracers that are most affected by the galaxy potential, such as the width of the total energy distribution. 

\begin{figure}
    \includegraphics[width=\columnwidth]{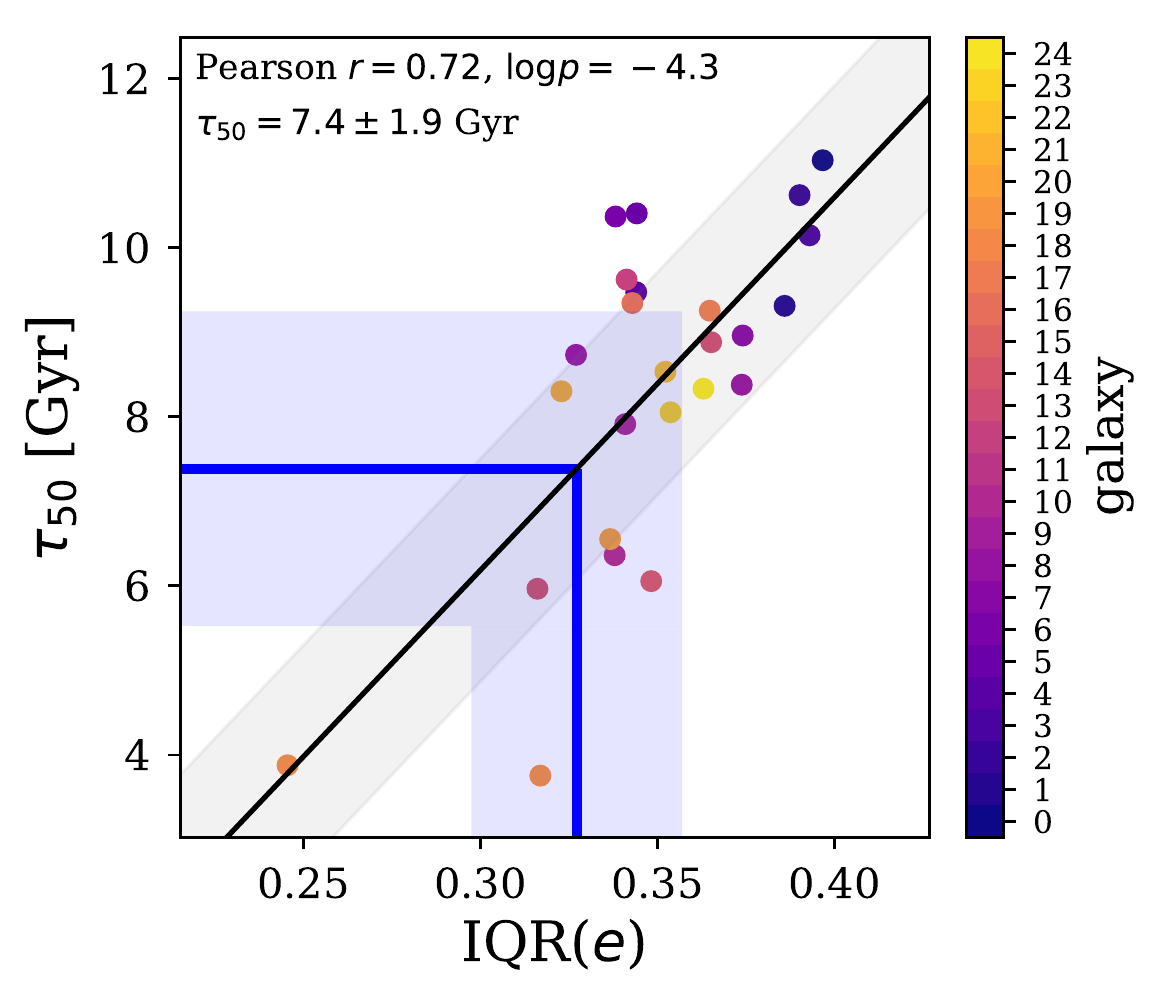}
    \caption{Example of a correlation between a 3D kinematic tracer of the entire GC system ($x$-axis) and a galaxy assembly metric ($y$-axis). The figure shows the half-mass assembly lookback time of the DM halo, $\t50$ versus the inter-quartile range of the distribution of GC orbital eccentricity.  The solid black lines and shading show the best fit linear regression, and the legend shows the Pearson correlation coefficient and $p$-value. The blue lines and shading show the predictions and uncertainties for the MW based on the observed GC kinematics. As a result of stripping during infall, galaxies that assembled half their halo mass earlier have a larger spread in the distribution of GC eccentricity compared to galaxies that assembled later. }
\label{fig:predictions_all}
\end{figure}

Several unexpected signatures of galaxy assembly are present in the kinematic data. Figure~\ref{fig:predictions_all} shows an example of an interesting correlation. The inter-quartile range of the orbital eccentricity correlates with the halo mass growth time-scale, with larger eccentricity spreads found in galaxies that reached half of their total halo mass earlier. Haloes that assemble earlier have an earlier end to their major merger epoch \citep[see table A2 in][]{emosaicsII}. Therefore, the eccentricities of GCs brought in by massive satellites are initially clustered at the time of accretion and slowly drift apart as a result of dynamical friction and tidal stripping. Table~\ref{tab:correlations_all} lists all the correlations selected for the entire GC populations.

\begin{figure*}
    \includegraphics[width=\textwidth]{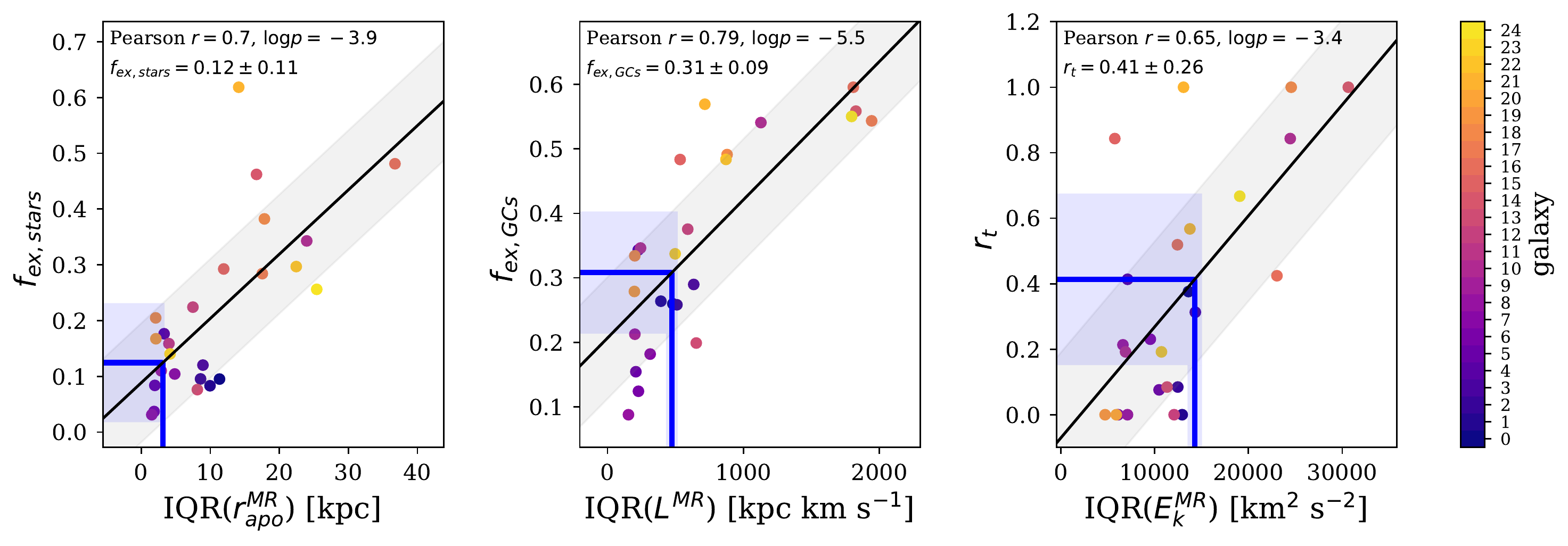}
    \caption{Examples of correlations between 3D kinematic tracers of the \emph{metal-rich} GC population and galaxy assembly metrics. Left: fraction of all stars formed ex-situ $\fexstars$ versus inter-quartile range of the orbital apocentres of metal-rich GCs. Middle: fraction of all GCs formed ex-situ versus inter-quartile range of the magnitude of the GC angular momenta. Right: ratio of major merger to overall merger time-scales $\rt$ versus inter-quartile range of the GC kinetic energy distribution. The symbols, lines, and legend follow the convention of Figure~\ref{fig:predictions_all}.  The accreted fractions of stars and GCs correlate with the inter-quartile ranges of the apocentre and angular momentum distribution of metal-rich GCs respectively. This indicates the GCs accreted from each massive satellite increasingly broaden the orbit distribution. The duration of the major merger epoch relative to all mergers correlates with the spread in the metal-rich GC kinetic energies, reflecting how major mergers bring metal-rich GCs to the inner galaxy more effectively than minor mergers. }
\label{fig:predictions_highmet}
\end{figure*}

As shown in Section~\ref{sec:subpopulations}, the kinematics of  metal-poor and outer GC subpopulations can be significantly different from the kinematics of metal-rich and inner clusters, and this could potentially provide a direct connection to the origin of the GCs  (Section~\ref{sec:origin}) and ultimately the assembly history of the host galaxy. Following this idea, we  repeat the correlation analysis for each of the subpopulations split by metallicity and galactocentric radius as defined in Section~\ref{sec:subpopulations}. Using only the metal-rich population we find an additional 10 significant correlations with Pearson $|r| > 0.7$. Figure~\ref{fig:predictions_highmet} shows a number of interesting correlations for the metal-rich GC population. For instance, the fraction of accreted stars and GCs correlates strongly with the width of the distribution of orbital apocentres  and total angular momenta, respectively (left and middle panels). This indicates, as qualitatively expected, that metal-rich accreted stars and GCs (which originate from relatively massive progenitors as a result of the mass-metallicity relation) have a dominant contribution to broadening the high angular momentum and apocentre tail of the distributions (as these tend to be larger for accreted satellites). Furthermore, the inter-quartile range of the binding energy distribution correlates with the ratio of the merger time-scales $\rt$, such that larger values of $\rt$ (i.e., a later end of the major merger epoch relative to all mergers) result in a larger spread in kinetic energy (right panel). A similarly strong correlation is found with the skewness of the energy distribution, and both are explained by the fact that massive satellites bring higher metallicity GCs and sink to the centre of the galaxy more efficiently than low mass ones, such that their clusters become more tightly bound over a shorter period of time. This reduces the accumulation of GCs at low binding (or kinetic) energies quickly after the last major merger. Table~\ref{tab:correlations_highmet} lists all the selected correlations for the metal-rich GC population.   

\begin{figure*}
    \includegraphics[width=0.75\textwidth]{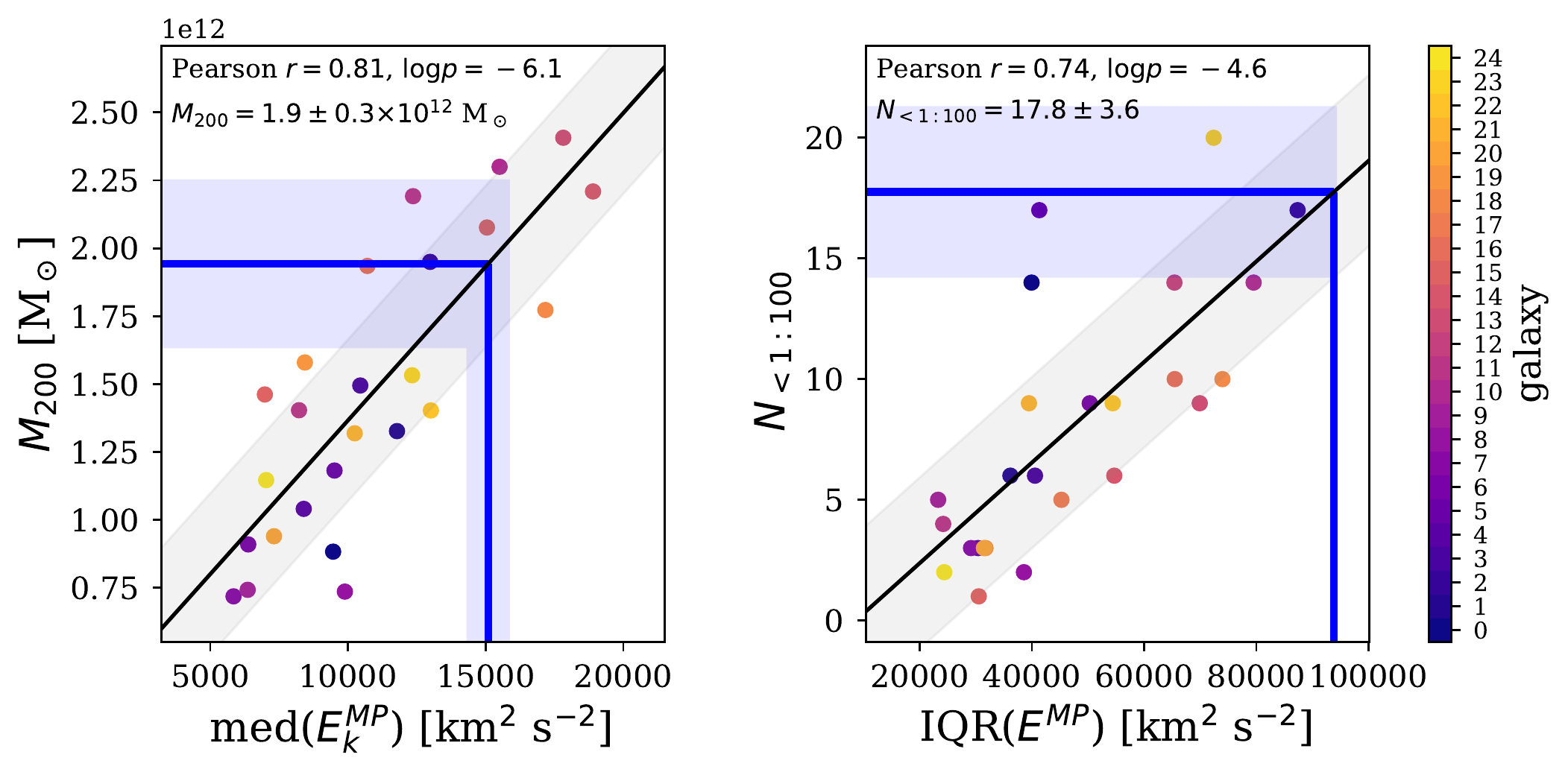}
    \caption{Examples of correlations between 3D kinematic tracers of the \emph{metal-poor} GC population and galaxy assembly metrics. Left: virial mass $\M200$ versus median GC kinetic energy. Right: number of tiny mergers $\Ntiny$ (with mass ratios $<1/100$)  versus inter-quartile range of GC orbital energy. The symbols, lines, and legend follow the convention of Figure~\ref{fig:predictions_all}. The halo virial mass correlates with the median of the metal-poor GC kinetic energy distribution as expected from dynamical equilibrium. The number of tiny mergers correlates with the width of the energy distribution, showing that metal-poor GCs brought in by the lowest-mass satellites are a sensitive probe of these accretion events.}
\label{fig:predictions_lowmet}
\end{figure*}

Selecting only the metal-poor GCs results in an additional 9 strong correlations (Pearson $|r| >0.7$). Figure~\ref{fig:predictions_lowmet} shows two examples. Since metal-poor clusters tend to orbit at larger radii, they provide tight constraints on the total mass distribution, quantified by $\M200$, through a correlation  with the median of the kinetic energy distribution as expected from virial equilibrium. Interestingly, even the number of tiny ($< 1/100$ mass ratio) mergers leave a signature in the metal-poor GC kinematics as found in the correlation with the width of the orbital energy distribution. This result should be interpreted with caution since tiny mergers include satellites with $M_* < 2\times10^7\Msun$, which are resolved with fewer than 100 baryonic particles in the simulations, making the structure of their stellar component prone to numerical artefacts. An obvious interpretation of the correlation between $\Ntiny$ and the spread in GC energies would be an underlying correlation between the number of tiny mergers and virial mass set by hierarchical mass growth. However, these quantities are poorly correlated (Pearson $r = 0.36$). Instead, this might be evidence of a direct imprint of low mass mergers in the GC kinematics. If true, this correlation then indicates that the present-day kinematics of the metal-poor population (which was in part accreted from low-mass satellites) retains memory of the orbital energy of each individual accretion event, even for satellites with less than 1 per cent of the stellar mass of the galaxy. This confirms our expectation that GC tracers should be more sensitive to low-mass accretion events compared to stellar halo tracers. As a result of the increase in the number of GCs normalised by the galaxy stellar mass in dwarf galaxies \citep{Peng08, Georgiev10}, low mass mergers contribute more GCs per unit accreted stellar mass compared to major mergers. Table~\ref{tab:correlations_lowmet} lists all the selected correlations for the metal-poor GC population.   

\begin{figure*}
    \includegraphics[width=\textwidth]{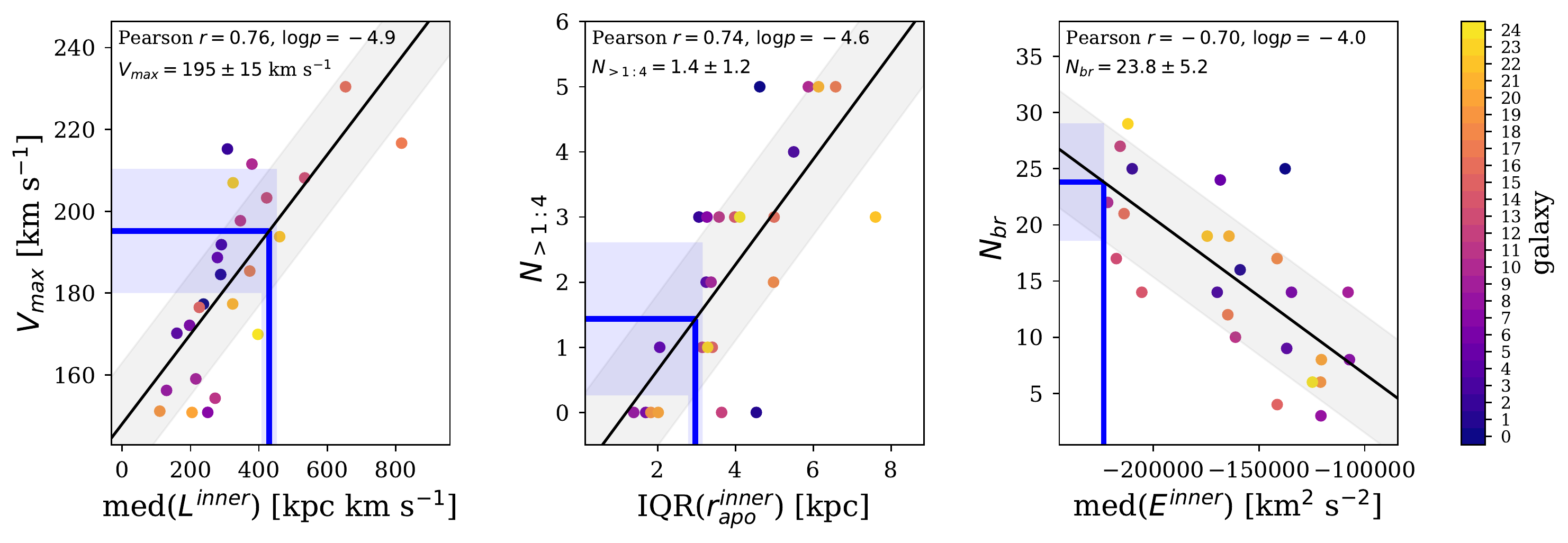}
    \caption{Examples of correlations between 3D kinematic tracers of \emph{inner} GCs and galaxy assembly metrics. Left: maximum circular velocity $\vmax$ versus median GC angular momentum. Middle: total number of major mergers $\Nmajor$ (with mass ratios $>1/4$) versus inter-quartile range of GC orbital apocentre. Right: total number of mergers (with galaxies of mass $M_* \ge 4.5\times10^6\Msun$) experienced by the main progenitor $\Nbr$ versus median orbital energy. The symbols, lines, and legend follow the convention of Figure~\ref{fig:predictions_all}. The maximum circular velocity correlates with the median angular momentum of inner GCs because these are typically born with highly circular orbits when formed in-situ. The number of major mergers correlates with the width of the apocentre distribution due to the dynamical heating effect of massive satellites as they sink to the centre of the halo. The total number of mergers correlates with the median GC orbital energy and this follows from the correlation between number of mergers and mass probed by the GC energies. }
\label{fig:predictions_disk}
\end{figure*}

Selecting only the inner population ($r < 8\kpc$) results in an additional 11 significant correlations with Pearson $|r| > 0.7$, demonstrating that the kinematics of GCs in the inner galaxy encode significant amounts of information relating to its assembly history. Figure~\ref{fig:predictions_disk} shows three relevant examples. The left panel shows that the median angular momentum of GCs in the inner galaxy is a good predictor of the maximum circular velocity of the galaxy. This is likely a result of the dynamical dominance of the baryonic component in the central region of the DM halo, where $\vmax$ is typically found in $L^*$ galaxies. Inner GCs are typically formed in-situ with highly circular orbits \citep{Pfeffer20}, making them ideal tracers of the circular velocity. The scatter in the relation between $\vmax$ and median inner GC angular momentum ($\sim 15\kms$) is similar  to the scatter in the prediction of $\vmax$ using $\M200$. The second panel of Figure~\ref{fig:predictions_disk} shows that the width of the apocentre distribution of inner clusters traces the total number of major mergers experienced by the main progenitor. This suggests that massive satellites contribute significantly to heating and broadening the distribution of orbits of GCs in the inner galaxy because they are more effective (as a result of dynamical friction) at delivering their clusters to the centre of the galaxy \citep{Pfeffer20}. We do not find a correlation between $\M200$ or $\vmax$ and $\Nmajor$, which shows that the kinematics-based prediction is not trivial. The third panel of Figure~\ref{fig:predictions_disk} shows, as expected from the causal relation between the total number of resolved progenitors and halo virial mass, that the number of mergers (or `branches')  correlates with the median energy of inner GCs. This tight relation likely follows from the similarly strong correlation between  $\Nbr$ (or $\Nleaf$) and $\vmax$ set by hierarchical structure formation in \LCDM.  Table~\ref{tab:correlations_disk} lists all the selected correlations for the inner GC population.

\begin{figure}
    \includegraphics[width=\columnwidth]{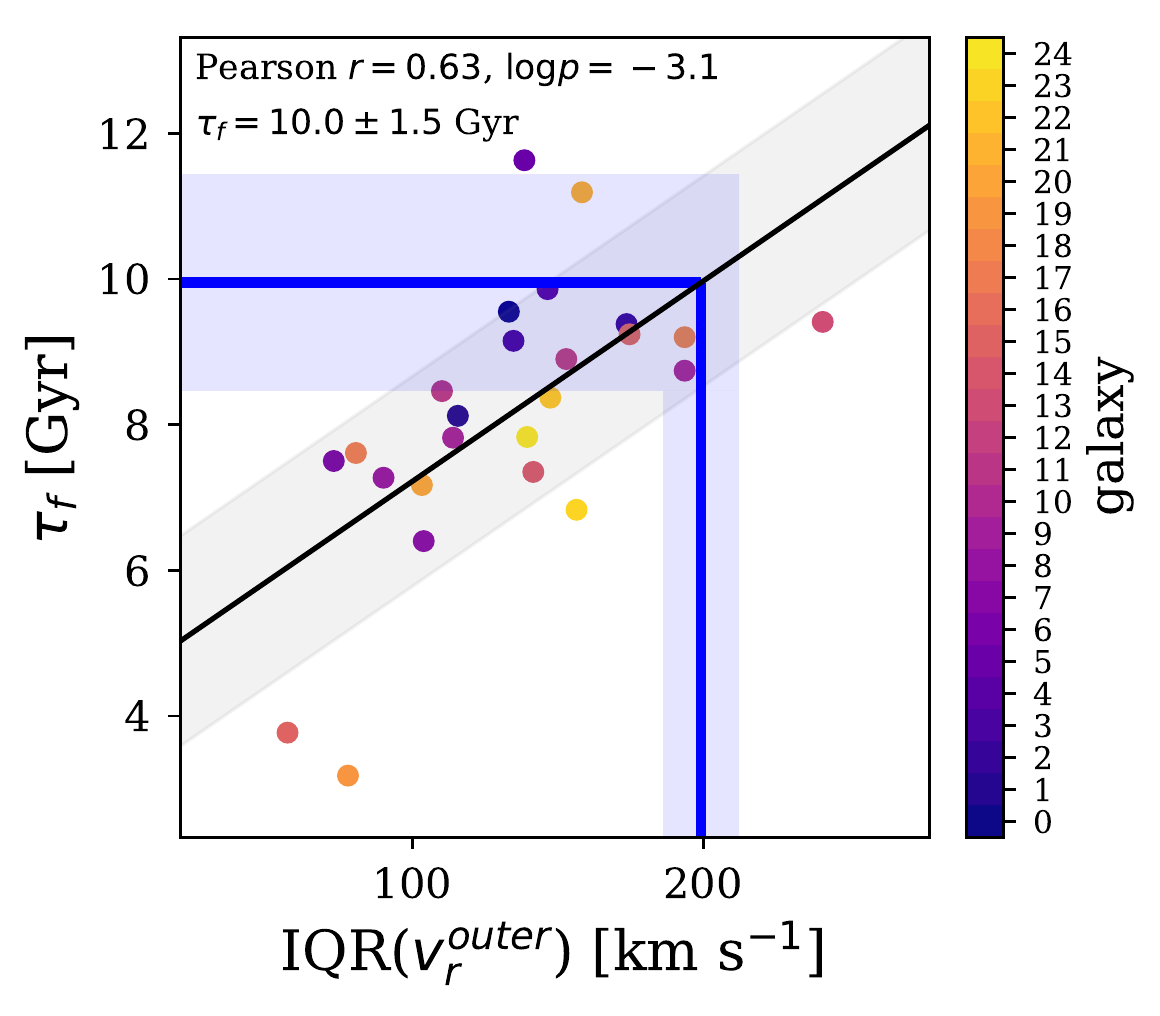}
    \caption{Example of a correlation between 3D kinematic tracers of \emph{outer} ($r>8\kpc$) GCs and galaxy assembly metrics, showing the relation between the median stellar age of the galaxy $\tf$ and the inter-quartile range of the radial velocities of the outer GCs. The symbols, lines, and legend follow the convention of Figure~\ref{fig:predictions_all}. An earlier assembly of the stellar component through a larger number of $z>2$ mergers increases the variety of accreted GC orbits and the width of the radial velocity distribution of GCs in the outer galaxy.}
\label{fig:predictions_halo}
\end{figure}

Using only the outer clusters results in an additional 5 significant correlations with Pearson $|r|>0.7$. Figure~\ref{fig:predictions_halo} shows an interesting example. Using only the 3D velocities, and requiring no knowledge of the potential, it is possible to probe the median formation lookback time of the stellar component $\tf$. Specifically, earlier median stellar formation epochs result in a broader distribution of radial velocities in the outer clusters. Insight into the origin of this correlation is provided by an additional correlation between the width of the radial velocity distribution and the number of high redshift mergers $\Nbrz$. An increased number of early mergers results in faster growth of the stellar component by accretion at $z>2$, as well as a larger variety of accreted GC orbits (because they fall in from many different directions). In addition, more recent star formation leads to more clusters forming in circular orbits and reduces the spread of radial velocities in relatively younger galaxies. Table~\ref{tab:correlations_halo} lists all the selected correlations for the outer GC population.   

\begin{figure*}
    \includegraphics[width=0.31\textwidth]{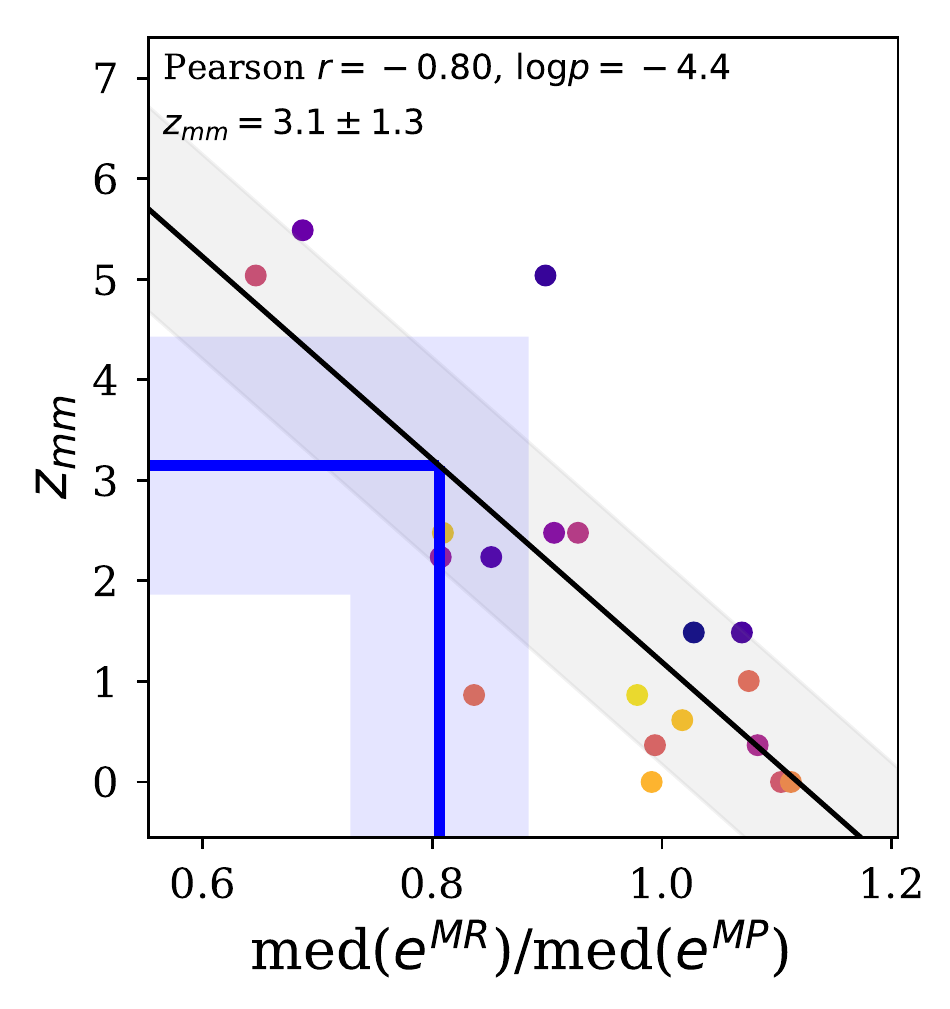}
    \includegraphics[width=0.685\textwidth]{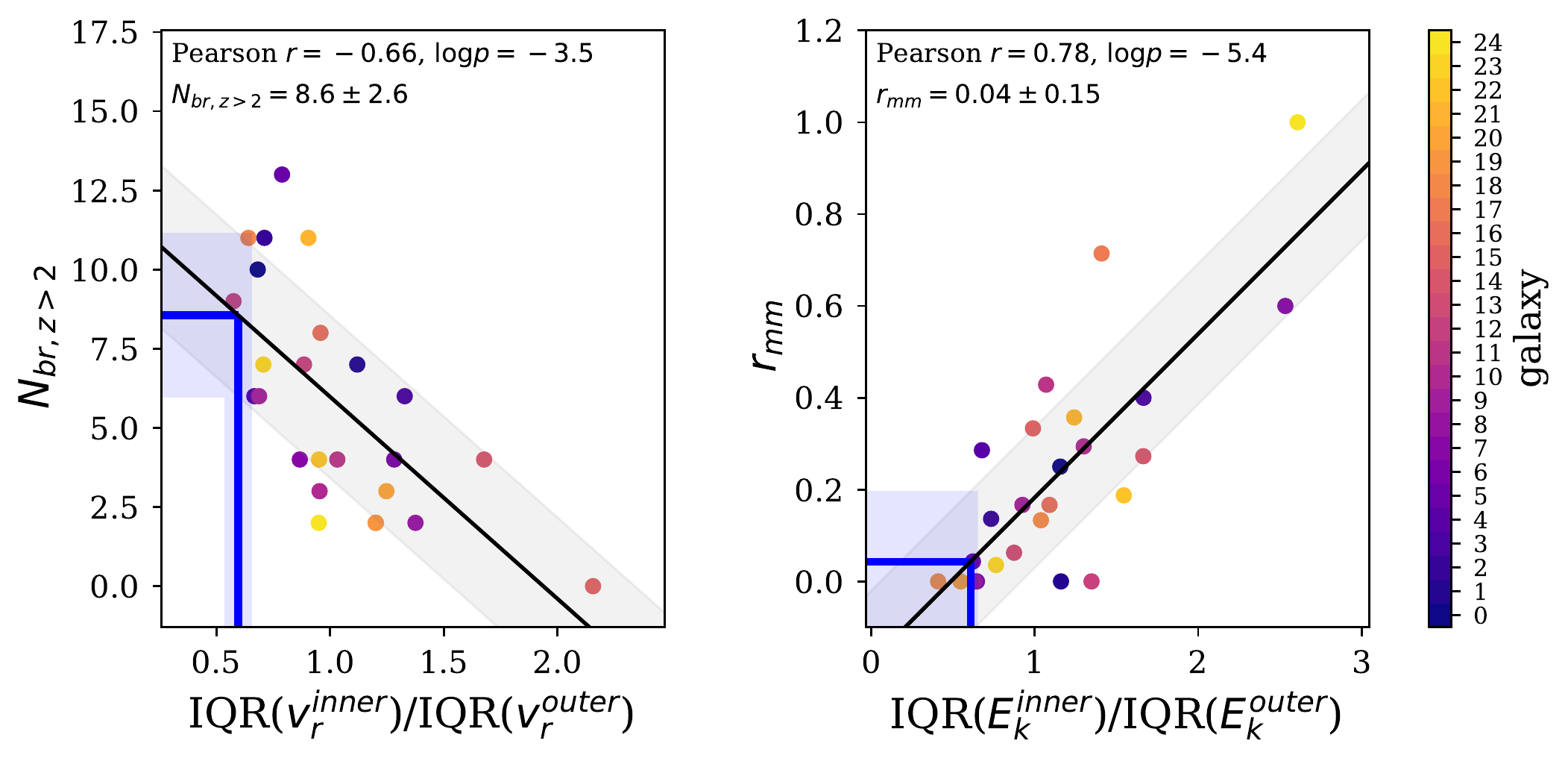}
    \caption{Examples of correlations between \emph{relative} kinematic tracers of GC \emph{metallicity} (left panel) and \emph{radial} (middle and right panels) subpopulations and galaxy assembly metrics. Left: redshift of the last major merger $\zmm$ versus ratio of median eccentricity of metal-rich and metal-poor GC populations. Middle: total number of mergers at $z>2$  $\Nbrz$ versus ratio of the inter-quartile ranges of radial velocities of inner and outer GCs. Right: ratio of the number of major mergers to the number of non-major mergers $\rmm$ versus ratio of the inter-quartile ranges of kinetic energies of inner and outer GCs. The symbols, lines, and legend follow the convention of Figure~\ref{fig:predictions_all}. The redshift of the last major merger anti-correlates with the ratio of median orbital eccentricities of the metal-rich and metal-poor GCs. In the kinematics of inner versus outer populations we find that the number of high-redshift mergers anti-correlates with the relative width of the radial velocity distributions, while the number of major mergers relative to all other mergers correlates with the ratio of their kinetic energy inter-quartile ranges. All these relations are evidence of the dynamical heating of GCs in the inner galaxy due to the accretion of massive satellites, while the outer clusters are more efficiently heated by many recent minor mergers.}
\label{fig:predictions_sub}
\end{figure*}

In the comparison of the kinematics of metallicity and radial subpopulations in Sections~\ref{sec:orbits} and~\ref{sec:integrals} we found clear differences which indicate that these could be sensitive tracers of galaxy formation and assembly. In the next step, we repeat the correlation analysis using as tracers the \emph{relative} medians, inter-quartile ranges, skewness, and kurtosis of the metal-rich versus metal-poor subpopulations. This results in 2 additional correlations with Pearson $|r|>0.7$. The left panel of Figure~\ref{fig:predictions_sub} shows a strong anti-correlation between the ratio of the median orbital eccentricity of the metal-rich and metal-poor populations and the redshift of the last major merger. This correlation is caused by the dynamical heating (i.e., increase in eccentricity) of the orbits of GCs in the inner galaxy (which are on average more metal-rich) following a major merger, in addition to the accretion of kinematically hot metal-rich GCs from the massive satellite itself. The metallicity of massive satellites will also be larger for more recent mergers, therefore contributing a proportionally larger fraction of its dynamically hot, metal-rich clusters to the host galaxy. Earlier occurrence of the last major merger allows for a longer period of accretion of low-mass satellites until $z=0$, which increases the number of accreted GCs and therefore the median eccentricity of metal-poor clusters. 

Lastly, using the relative kinematic tracers of the inner and outer GC subpopulations results in 4 correlations with Pearson $|r|>0.7$. The middle panel of Figure~\ref{fig:predictions_sub} shows that the total number of $z>2$ mergers is predicted by the ratio of the inter-quartile ranges of radial velocity of outer and inner GC populations, such that smaller spreads in inner GC radial velocities relative to the outer GCs are a signature of a larger number of high-redshift mergers. This is a direct consequence of the large contribution of accreted GCs to the outer GC population. The simulations show that an increase in the number of $z>2$ mergers broadens the distribution of radial velocities of accreted clusters in the outer galaxy, while radial dispersion in the inner galaxy is mostly independent of early accretion and dominated by the in-situ component. The right panel of Figure~\ref{fig:predictions_sub} shows another consequence of orbit heating due to massive mergers. The ratio of the number of major mergers to the number of non-major mergers $\rmm$ increases with the ratio of inter-quartile ranges of kinetic energy of inner and outer GCs. Once again, this is due to the ability of massive satellites to deliver GCs to the central galaxy, broadening the distribution of kinetic energy of the inner GCs compared to those in the outer halo. Table~\ref{tab:correlations_sub} lists all the selected correlations for the relative tracers of the GC subpopulations selected by metallicity and galactocentric radius.     

\subsection{Predicting the assembly history of the MW from observed GC kinematics}
\label{sec:predictions}

In total, we have identified 51 strong (Pearson $|r|>0.7$), statistically significant correlations between GC kinematics and galaxy assembly histories. Figures~\ref{fig:predictions_all}--\ref{fig:predictions_sub} show the predictions obtained using the linear models and the observed kinematics of the MW GC system for the correlations discussed in Section~\ref{sec:correlations}. Table~\ref{tab:predictions3D} presents a summary of the most significant and reliable correlations selected from the full list in Appendix~ \ref{sec:correlation_tables}. The table also presents the quantitative predictions obtained from these correlations for a total of 18 assembly metrics using the observed 3D velocities, orbits, and integrals of motion of the MW GCs. More than half of these predictions do not require a-priori knowledge of the potential of the galaxy, but only the instantaneous 3D velocities and positions. In addition to the direct inferences using the GC kinematics, we can indirectly derive additional constraints by combining the predictions (albeit with larger uncertainties). 

Using the predictions for $\tf = 10.0\pm1.5\Gyr$ and $\delta_t = 0.13\pm0.12$ obtained from the radial velocities of outer GCs and from the kinetic energies of metal-rich GCs, respectively (see Table~\ref{tab:predictions3D}), equation~(\ref{eq:delta_t}) gives a constraint on the lookback time at which half of the stellar mass of the main progenitor was in place, $\ta = 8.7_{-1.5}^{+1.7}\Gyr$.  Furthermore, using the predictions for the lookback time of the last major merger $\tmm = 11.6_{-1.6}^{+0.7}\Gyr$, the ratio of the merger time-scales $\rt = 0.41\pm0.26$, and equation~(\ref{eq:timescales_ratio}) we obtain the lookback time of the last resolved merger (with $M_* > 4.5\times10^6\Msun$), $8.7^{+1.8}_{-3.4}\Gyr$. Despite the large uncertainty, the 1$\sigma$ lower limit on this estimate indicates that the merger epoch of the MW ended earlier than 96 per cent of the simulations (see Section~\ref{sec:discussion}).

The predictions for the formation and assembly of the MW based on the GC system kinematics can now be placed in the broader context of the entire population of $L^*$ galaxies. Table~\ref{tab:predictions3D} also lists the relative location of each prediction within the distributions for $L^*$ galaxies as sampled by the 25 E-MOSAICS simulations. Overall, these distributions show that the MW is very particular in several aspects related to the early formation of its DM halo and stellar component, along with the significant fraction of its growth due to low mass mergers at early times. 

\begin{table*}
	\centering
	\caption{Summary of GC kinematic tracers and their predictions for the assembly history of the MW and its DM halo. From left to right, the columns list the galaxy and halo assembly metric, the corresponding GC kinematics tracer, the Pearson $r$ correlation coefficient of the linear model, the prediction of the model using the GC system kinematics of the MW, and its percentile placement within the distribution of the 25 E-MOSAICS simulations. For each assembly metric we list here only the most correlated tracer while avoiding tracers that could be biased by the slight underproduction of stellar mass in $L^*$ galaxies in EAGLE (Section~\ref{sec:correlations}). The last column lists the predictions obtained by \citet{kraken} using the age-metallicity relation of MW GCs. Units of kinematic tracers are: radii $[\rm{kpc}]$, energies $[{\rm km}^2 {\rm s}^{-2}]$, velocities $[\kmsns]$, and angular momenta $[\rm{kpc}\kms]$. }
	\label{tab:predictions3D}
	\begin{tabular}{lccccccc} 
		\hline 
         Assembly metric & Kinematic tracer & GC population &  \parbox[c]{1.5cm}{Correlation coefficient} & \parbox[c]{1.2cm}{Pearson $\log p$} & MW prediction &
         Percentile & \parbox[c]{1.4cm}{Prediction based on GC age/metallicity}  \\
		\hline \hline
		$\M200$ [$10^{12}\Msun$] & med($\Ekin$)  & metal-poor & 0.81  & -6.1 & $1.94\pm0.31$ & 76 & -- \\
        $\vmax$ [$\kmsns$]       & med($L$)      & inner      & 0.76  & -4.9 & $195\pm15$    & 68 & $180\pm17$  \\
        $\tau_{25}$  [Gyr]       & med($\rperi$) & metal-poor & -0.80 & -5.8 & $11.2\pm0.9$  & 72 & $11.5\pm0.8$ \\
        $\t50$ [Gyr]             & IQR($e$)      &  all       & 0.72  & -4.3 & $7.4\pm1.9$   & 24 & $9.4\pm1.4$ \\
        $\tf$ [Gyr]              & IQR($\vr$)    & outer      & 0.63  & -3.1 & $10.0\pm1.5$  & 92 & $10.1\pm1.4$ \\
        $\delta_t$               & IQR($\Ekin$)  & metal-rich & 0.60  & -2.9 & $0.13\pm0.12$ & 80 & -- \\
        \hline
        $\Nbr$                   & med($\Etot$)  & inner      & -0.70 & -4.0 & $23.8\pm5.2$  & 80 & $15.1\pm3.3$ \\
        $\Nbrz$ & IQR($\vr^{\rm inner}$)$/$IQR($\vr^{\rm outer}$) & inner/outer  & -0.66 & -3.5 & $8.6\pm2.6$ & 76 & $9.2\pm1.9$ \\
        $\Nleaf$                 & med($\Etot$)  & inner      & -0.78 & -5.5 & $40.8\pm7.2$  & 88 & $24.1\pm10.2$\\
        $\rbl$                   & IQR($L$)      & metal-rich & -0.73 & -4.5 & $0.64\pm0.09$ & 56 & -- \\
        $\Nmajor$                & IQR($\rapo$)  & inner      & 0.74  & -4.6 & $1.4\pm1.2$   & 40 & -- \\
        $\Nsmall$                & med($\Ekin$)  & metal-poor & 0.70  & -4.1 & $4.1\pm1.3$   & 80 & -- \\
        $\Ntiny$                 & IQR($\Etot$)  & metal-poor & 0.74  & -4.6 & $17.8\pm3.6$  & 96 & $7.9\pm2.2$ \\
        \hline
        $\zmm$ & med($e^{\rm MR}$)/med($e^{\rm MP}$) & metal-rich/metal-poor & -0.80 & -4.4 & $3.1\pm1.3$ & 84 & -- \\
        $\rmm$ & IQR($\Ekin^{\rm inner}$)$/$IQR($\Ekin^{\rm outer}$) & inner/outer & 0.78 & -5.4 & $0.04\pm0.15$ & 28 & -- \\
        $\rt$                    & IQR($\Ekin$)  & metal-rich & 0.65 & -3.4 & $0.41\pm0.26$ & 60 & -- \\
        \hline
        $\fexstars$              & IQR($\rapo$)  & metal-rich & 0.70 & -3.9 & $0.12\pm0.11$ & 40 & -- \\
        $\fexgcs$                & IQR($L$)      & metal-rich & 0.79 & -5.5 & $0.31\pm0.09$ & 44 & --  \\
		\hline 
	\end{tabular}
\end{table*}

\subsection{Comparison with constraints on the assembly history of the MW and $L^*$ galaxies}
\label{sec:discussion}

One way to verify the reliability of the constraints derived here is by comparing them with the results of other independent methods. Using the distribution of GCs in age-metallicity space in the E-MOSAICS simulations as well as observed ages and metallicities of the MW GCs, \citet{kraken} derived several quantitative constraints on the assembly history of the MW.  The last column of Table~\ref{tab:predictions3D} shows a comparison with the predictions obtained by \citet{kraken} using only the GC ages and metallicities. Of the 18 assembly metrics listed in Table~\ref{tab:predictions3D}, 10 are uniquely probed by GC kinematics, while 8 overlap with those derived using ages and metallicities. In general, the predictions agree remarkably well within their uncertainties, with a few exceptions. We find tension in those predictions for which the kinematic tracer values fall outside the range of the simulations, which are technically extrapolations, and should be treated with caution because they indicate that the MW is not represented in the models. There is slight tension in the values of $\Nbr$ (1.4$\sigma$), and $\Nleaf$ (1.3$\sigma$), and a  significant discrepancy in $\Ntiny$ (2.3$\sigma$). Furthermore, since all these involve either the median or the width of the GC energy distribution (and therefore total galaxy stellar --and DM-- mass), we expect systematic effects in our method from the known underestimation of stellar-to-halo mass ratios in haloes with $\M200 \sim 10^{12}\Msun$ in EAGLE \citep[see Section 5.2 in][]{Schaye15}. This effect can lead to different biases in the predictions depending on which tracer is used. This is particularly evident in the predictions for $\Nleaf$. In this case, our method predicts the number of progenitors based on the GC energies in galaxies with lower-than-observed potentials, causing an overestimation of the MW prediction due to its deeper disc potential. This bias explains why kinematic predictions using the energies produce larger $\vmax$ (in one of the discarded correlations) and larger numbers of progenitors and tiny mergers compared to the age-metallicity relation. We confirmed this by examining predictions for $\Nleaf$ from tracers that are independent of the potential (albeit less precise), such as the median angular momentum of inner clusters, for which we obtain $\Nleaf = 30.6\pm8.9$. This prediction agrees within the error bars with the age-metallicity result from \citet{kraken}. In the following discussion we therefore adopt the more accurate predictions $\Nbr=15.1\pm3.3$, $\Nleaf=24.1\pm10.2$, and $\Ntiny=7.9\pm2.2$ from \citet{kraken}. Their estimate of the total number of mergers is particularly robust because it combines predictions using three different tracers, and two of them are completely independent (the age-metallicity slope, and the number of GCs). Using these results, we estimate that each of the $\sim 15$ progenitors of the Galaxy experienced on average $0.6\pm0.2$ prior mergers. To avoid systematics in the remaining predictions listed in  Table~\ref{tab:predictions3D}, we verified that the value of each MW kinematic tracer lies within the range covered by the 25 simulations.

In addition to the overall correlations in age-metallicity space, \citet{kraken} also estimated the properties of the most recent mergers experienced by the MW using the average stellar mass growth histories of all EAGLE galaxies. From the stellar masses of the progenitors they estimate that roughly 43 per cent of the MW GCs formed ex-situ. Using direct predictions based on the GC system kinematics we provide a partially independent constraint (based on the same set of simulations), of $\fexgcs = 31\pm9$ per cent. However, this number includes clusters with $\feh>-0.5$, which are known to be overabundant in E-MOSAICS due to numerical underdisruption \citep{emosaicsI}. Correcting for the total number of GCs in the simulations, a factor of $\sim 5$ at $\feh>-0.5$ \citep[see figure D1 of][]{emosaicsII}, our estimate increases to $37\pm11$ per cent. Recently, \citet{Kruijssen20} combined the available observational constraints on individual MW GCs to obtain an accreted fraction in the range $35-50$ per cent. Including the correction for underdisruption, our prediction is consistent with this result. 

Our analysis predicts the total mass $\M200 = (1.94\pm0.31) \times 10^{12}\Msun$ of the MW DM halo using the kinetic energy distribution of metal-poor GCs. The virial mass of the MW has been estimated using several different methods requiring a variety assumptions. Systematics in these assumptions are difficult to account for and  produce a spread between results that is larger than the individual error bars of each study, with most values in the range $0.5 - 2.5\times10^{12}\Msun$ \citep{Callingham19}. Our prediction is on the higher end but consistent with many of these estimates. In addition to the $\sim 15$ per cent uncertainty in our prediction, an additional  systematic effect due to the underprediction of the stellar masses of $L^*$ galaxies in EAGLE should also be included. Metal-poor GCs are located near the disc (at a median galactocentric distance of $\approx 13\kpc$ in the simulations), where the potential is underestimated in E-MOSAICS. As a result, their kinetic energies could also be underestimated, biasing our halo mass estimate towards artificially high values.  

The lookback times at which the DM halo reached 25 and 50 per cent of its current mass, $\tau_{25} = 11.2\pm0.9\Gyr$ and $\tau_{50} = 7.4\pm1.9\Gyr$ were predicted using the pericentres and eccentricities of the metal-poor and the entire GC populations respectively. The first quarter of the mass of the halo was assembled earlier than 72 per cent of the E-MOSAICS galaxies, while half the mass was in place earlier than only 24 per cent of the sample. For an average halo with $\M200=10^{12}\Msun$, the corresponding predictions of the Extended Press-Schechter formalism are $[\tau_{25},\tau_{50}] = [10.8, 8.5] \Gyr$ \citep{Correa15}, which could indicate that although the earliest period of mass growth in the MW DM halo was faster than average, the following stage was relatively slow. Note however, that the predictions are consistent with the mean within the uncertainties.

Our results predict that half of the stars in the Galaxy had formed at a lookback time $\tf = 10.0\pm1.5\Gyr$. This is in excellent agreement with the direct measurement of the star formation history derived by \citet{Snaith14}, who estimate that half of the stellar mass of the Galaxy had formed $10.5\pm1.5\Gyr$ ago. Using studies of the evolution of the progenitors of $L^*$ galaxies, we can place the MW in the context of the distribution of galaxies of the same present-day stellar mass. \citet{Pacifici16} used the star formation histories of a large sample of low-redshift galaxies to estimate a mean half-mass formation lookback time $\sim 7.7\Gyr$ for all MW-mass galaxies, and $\sim 6.9\Gyr$ for those that are still star-forming at $z=0$. This confirms that across all its progenitors, the MW stars formed much earlier than the average $L^*$ galaxy, and could be a result of the fast early DM halo growth combined with an active merger epoch at $z>2$ (as predicted by the above-average value of $\Nbrz$).

The indirect prediction (through $\tf$ and $\delta_t$) for the lookback time at which the main progenitor had formed half of its stellar mass, $\ta = 8.7_{-1.5}^{+1.7}\Gyr$, is in excellent agreement with the estimate using the slope of the MW GC age-metallicity relation from \citet{kraken}, $8.6_{-2.2}^{+1.3}\Gyr$. This lends more confidence to our conclusion that the MW grew significantly by mergers at $z>2$. \citet{Papovich15} obtained the stellar mass as a function of redshift for the population of high-redshift progenitors of MW-mass galaxies, estimating that on average half of the stars in the main progenitor were assembled $\sim 7.5\Gyr$ ago. \citet{Behroozi19} combined the observed evolution of the galaxy population with large-volume $N$-body cosmological simulations to constrain the growth of galaxies as a function of mass, redshift, and colour.  They find that the average MW-mass galaxy assembled half of its stars $8.3\Gyr$ ago. The difference between the two studies above points to systematics in the inferences. Despite the relatively large uncertainty, the GC kinematics predicts that the MW assembled half of its mass earlier than both estimates above. The predicted  early stellar mass growth is also consistent with the finding of \citet{Mackereth18} that the MW disc alpha-element abundances are only present in about 5 per cent of MW-mass galaxies in the EAGLE Ref-L100N1504 simulation, and that they originate from an early period of fast gas accretion and star formation. \citet{Hughes20} reach a similar conclusion using observational constraints on the contribution of disrupted GCs to the MW bulge.

Our results also predict the redshift of the last major merger, $\zmm = 3.1\pm1.3$. This agrees with several studies which obtain a lower limit of $z\sim2$ \citep{Wyse01, Hammer07, Stewart08, Shen10, Bland-HawthornGerhard16}, as well as with the more stringent limit from the GC age-metallicity relation, $z\ga4$ \citep{kraken, Kruijssen20}. The GC kinematics predict that the Galaxy experienced $\Nbr \sim 24$ mergers (with stellar masses $M_* > 4.5\times10^6\Msun$), which is considerably larger than the estimate of $15.1\pm3.3$ obtained by \citet{kraken} as a result of systematics in the stellar masses of $L^*$ galaxies in EAGLE. Here we adopt the value from \citet{kraken} because it is a more accurate estimate because it combines three different tracers, where two are independent (see discussion at the beginning of this section). The total number of $\sim 15$ mergers experienced by the MW is about twice larger than the 7 accretion events for which evidence has been found so far in the kinematics and chemistry of halo stars and GCs, namely Sagittarius \citep{Ibata94}, the progenitor of the Helmi streams \citep{Helmi99,Koppelman19}, \emph{Gaia}-Enceladus \citep{Helmi18, sausage}, Kraken \citep{kraken,Kruijssen20,Massari19}, Sequoia \citep{sequoia}, and Thamnos 1 and 2 \citep{thamnos}. This suggests that up to $\sim 8$ hidden structures may remain yet to be discovered (at or least the subset corresponding to late accretion events whose dynamical structure has not been dispersed).  

The kinematics of the MW GC system predict the lookback time of the last resolved merger $\tam = 8.7^{+1.8}_{-3.4}\Gyr$. The most recent observed accretion event, that of the Sagittarius dSph galaxy, took place  $>5-7\Gyr$ ago \citep[as inferred from its star formation history;][]{deBoer15}. This would place it within the large uncertainty in our prediction. Our result is also consistent with a recent analysis combining the orbits, ages, and metallicities of GCs associated with Sagittarius  \citep{Kruijssen20}, who find that it was accreted $6.7^{+2.0}_{-1.1}\Gyr$ ago. 

A total of $\sim 55$ known accreted GCs can be associated to the 5 known GC-bearing accretion events \citep{Kruijssen20}. In addition to these, \citet{Massari19} find 11 clusters with high orbital energies and a broad distribution of angular momenta which have not been associated to any known progenitors. Using the predicted merger demographics we can infer how many clusters were contributed by putative undiscovered satellites. Our result for the total number of mergers $\sim 23$ is likely an overestimate due to systematics in $\Nbr$ (see discussion above).  Using the more accurate estimate of $\Nbr = 15.1\pm3.3$ from \citet{kraken} requires that the $\sim 10$ remaining  accreted satellites host at least the 11 high energy GCs, and possibly a few more undiscovered ones. Thamnos 1 and 2 appear to be remnants of two satellites without any associated GCs \citep{thamnos}, implying that the remaining 8 predicted satellites which have not yet been detected hosted at least 11 clusters, between 1 and 2 per galaxy on average. This small average number of associated GCs complicates the identification of these satellites through the clustering of GCs in position-velocity-age-metallicity space, as had been done previously.

After applying the correction for GC underdisruption discussed above, the estimate for the fraction of the 157 MW GCs that were accreted from satellites is $\fexgcs = 37\pm11$ per cent, in agreement with \citet{Kruijssen20}. This corresponds to about 58 clusters and is considerably larger than the estimate of 40 GCs obtained by \citet{MackeyGilmore04}. This difference is due to the inclusion in our sample of the GCs that overlap with the main progenitor branch in age-metallicity space \citep[see][]{kraken}. Our prediction is below the ex-situ fraction of 59 per cent obtained by \citet{Massari19} based on the kinematics of GCs relative to disc and halo stars. More recently, \citet{Forbes20b} estimated, by associating the ambiguous GCs to the 5 known mergers, that 55 per cent were formed ex-situ, still larger than our predicted range. The discrepancy with the independent estimates by \citet{Massari19} and \citet{Forbes20b} might be due to systematics in the kinematics prediction, or alternatively, it may be resolved if the GCs that remain hidden in the MW bulge (which are not excluded in the simulations) are mostly of in-situ origin. 

Using semi-empirical galaxy growth histories, \citet{Behroozi19} estimate the fraction of ex-situ stars as a function of halo mass. For $\M200 \approx 10^{12}\Msun$ about $68$ per cent of galaxies accreted $\approx 12-14$ per cent of their stars. Our prediction, $\sim 12$ per cent, falls within the standard deviation of the population. This suggests that although the Galaxy had considerable early growth via many mergers with low-mass galaxies at $z>2$, these mergers were not massive enough to contribute a large fraction of its total stellar content.

It is interesting that the picture of early growth and assembly of the DM halo of the Galaxy through many low-mass mergers that emerges from our analysis also agrees with the conclusions of \citet{Carlesi20}, who analysed the assembly histories of galaxies that form in simulations constrained to reproduce the large-scale environment of the Local Group. They found that the cosmological environment of the MW produces galaxies which assemble $\sim 0.5\Gyr$ earlier, and have their last major merger\footnote{Defined in \citet{Carlesi20} by a halo mass ratio $>1/10$.} $\sim 1.5\Gyr$ earlier than galaxies in random environments. They also find that the environment of the Local Group causes the last major merger to occur more often within the first half of cosmic history. 

The results of this statistical analysis reveal a detailed story in which the Galaxy assembled very rapidly, with one quarter of its DM halo mass already in place $\sim 11\Gyr$ ago (72nd percentile of the 25 E-MOSAICS simulations), when it grew quickly through many mergers with low mass galaxies. Its stellar mass grew even more rapidly during the same period, with 50 per cent of stars already formed across all its progenitors $10.0\pm1.5\Gyr$ ago (92nd percentile). The stellar component of the main progenitor also assembled relatively early, with half of its mass in place $8.7_{-1.5}^{+1.7}\Gyr$ ago; 73rd percentile). The rapid buildup of the stars was partly the result of accretion from a total of $15.1\pm3.3$ mergers with galaxies of masses $M_* > 4.5\times10^6\Msun$ (52nd percentile), out of which $9.2\pm1.9$ took place at $z>2$ \citep[76th percentile;][]{kraken}. The Galaxy experienced only $1.4\pm1.2$ major mergers in its entire history (40th percentile), with the last one taking place at $z = 3.1\pm1.3$ (84th percentile), much earlier than the median $L^*$ galaxy, for which this occurs at $z\approx1.5$ in E-MOSAICS. The period of major mergers spanned about 40 per cent of the entire merger epoch, implying that the last resolved merger occurred $8.7^{+1.8}_{-3.4}\Gyr$ ago (96nd percentile for the $1\sigma$ lower limit), and that out of the total of $\sim 15$ mergers, only about 6 took place at $z<2$. The vast majority of the MW's mergers involved satellites with mass ratios $<1/4$. Out of these, $\sim 6$ were the most significant, with mass ratios $>1/100$ (64th percentile), while $\sim 8$ had less than 1 per cent of the mass of the main progenitor (52nd percentile). On average, each of the MW progenitors experienced $<2.3$ mergers prior to accretion onto the Galaxy. About 88 per cent of the stars formed within the main progenitor, and 12 per cent were accreted from satellites during its early merger phase at $z\la1$ (40th percentile). A larger fraction of its GCs, about 37 per cent or $\sim 58$ objects, were brought in by accreted satellites (44th percentile).

\section{Limitations and Caveats}
\label{sec:limitations}

In this section we discuss the limitations of the simulations and their implications for the results of the analysis presented in this work.

The lack of a resolved cold and dense ISM in the simulations results in underdisruption of GCs with $\feh>-1.0$, and leads to a GC excess of a factor of $\sim2.5$ at $-1.0<\feh<-0.5$ with respect to the observed metallicity distribution of the MW and M31 (see discussion in Section~\ref{sec:simulations}). To quantify the impact of this excess on the kinematic distributions of GC systems in the simulations, we remove a random subset of 60 per cent of the GCs with $-1.0<\feh<-0.5$. Appendix~\ref{sec:limitations_appendix} shows the comparison of the fiducial orbits with the orbits of the GC sample corrected for underdisruption. For both the median and IQR of the orbits, a two-sample Kolmogorov-Smirnov \citep{Smirnov39} test fails to reject at high significance the null hypothesis that both distributions are identical ($p \geq 0.41$). We therefore expect that GC underdisruption will have a minimal impact on our analysis, including the MW predictions.

The baryonic mass resolution of the E-MOSAICS simulations, $m_{\rm gas} = 2.25\times10^5\Msun$ means that galaxies with stellar masses $\Mstar \la 2\times10^7\Msun$ will be resolved with fewer than $\sim100$ star particles. The lack of resolution in the lowest-mass GC-forming progenitors could therefore artificially exclude their GCs from our analysis, introducing a bias in the comparison to the MW system. However, it should be noted that most GCs in the local Universe are hosted by MW-mass galaxies, and only $\sim 5$ per cent seem to inhabit galaxies with $\Mstar < 10^7\Msun$ \citep{Harris16}. Nevertheless, to assess the impact of resolution, we repeat the kinematic analysis after artificially removing GCs formed in the lowest mass galaxies in the simulations. Appendix~\ref{sec:limitations_appendix} shows the results for the distribution of GC orbits across the 25 simulations. To approximate the effect of removing low-mass progenitors, we removed all GCs with $\feh<-1.5$. This corresponds to the average metallicity of GCs formed in galaxies with $\Mstar\la10^8\Msun$ at $z\approx2$ \citep[see][figure 9]{emosaicsII}, the median GC formation time in E-MOSAICS \citep{Keller20}. This cut removes most GCs formed in galaxies with masses up to $\sim5$ times the resolution limit, or about 21 per cent of all the GCs in the  simulations. The results for an additional intermediate cut at $\feh<-1.8$, comparable to $\Mstar\la10^{7.6}\Msun$, are also shown. The K-S test fails to reject the null hypothesis that the orbit distributions of the fiducial and reduced samples are identical ($p \geq 0.24$).  This indicates that the statistics used to obtain the MW predictions in Section~\ref{sec:predictions} (i.e. the median and IQR), are robust to the removal of GCs formed in the lowest-mass galaxies. We therefore do not expect that GCs formed in under-resolved galaxies would significantly alter our results.

The metallicities of dwarf galaxies at $z=0$ in EAGLE are $\sim 0.5$ dex above those of Local Group dwarf galaxies \citep{Schaye15}. However, this discrepancy seems to be absent at $z\ga2$ \citep{DeRossi17}. In E-MOSAICS, GC metallicities follow the observed mass-metallicity relation of their birth galaxies at $z\simeq2-3$, the main epoch of GC formation in the simulations \citep[][figure 11]{Keller20}. This indicates that the GC metallicities are in general better reproduced than the stellar metallicities. Furthermore, the apparent discrepancy may disappear when the observational methods to infer metallicities are applied to simulations \citep{Nelson18}. To understand the impact of possible systematics in the GC metallicities on our predictions, we return to the full correlation analysis and select the best tracers of the assembly metrics in Table~\ref{tab:predictions3D} that do not use any metallicity information. Only two metrics, $\tau_{25}$, and $\delta_t$, cannot be predicted without GC metallicities. As before, we avoid tracers that could be biased due to the undermassive discs in EAGLE. The correlations and predictions are listed in Appendix~\ref{sec:limitations_appendix}, along with a comparison with the fiducial predictions using the full GC information. All of the  metallicity-independent predictions show excellent agreement with the results of the full analysis shown in Table~\ref{tab:predictions3D}. Given that our prediction for $\tau_{25}$ agrees with the \citet{kraken} value obtained using only GC ages, this further confirms the robustness of our results.  

The threshold metallicity $\feh = -1.2$ used to separate the metal-poor and metal-rich GC populations throughout our analysis is rather arbitrary and could have a potential impact on the MW assembly predictions. We have already shown above (and in Appendix~\ref{sec:limitations_appendix}) that most of the predictions are consistent with those made without using the GC metallicities. However, to explore the effect of the threshold, we repeat the correlation analysis while varying its value by $\pm0.4$ dex, i.e. $\feh = [-1.6, -0.8]$, and compare the results to the fiducial MW predictions in Appendix~\ref{sec:limitations_appendix}. Dividing the populations at $\feh = -0.8$ generally reduces the statistical significance and strength of the correlations, but the MW predictions remain consistent within the uncertainties. Using $\feh = -1.2$ also reduces the correlation strengths but to a smaller extent. In addition, some correlations seem to remain strong ($r\ga0.5$) regardless of the threshold value (i.e. $\M200$ and $\Nsmall$). Interestingly, none of the correlations become stronger than the fiducial case (with a threshold at $\feh=-1.2$). Some tracers strongly prefer a lower threshold (e.g. $\tau_{25}$), while others prefer a higher metallicity threshold (e.g. $\fexstars$).

Lastly, there is significant uncertainty in the stellar mass - halo mass (SMHM) relation of galaxies below $\sim 10^{11}\Msun$ \citep{Behroozi19}. The EAGLE model has been shown to reproduce the SMHM relation at $z=0$ obtained by \citet{Behroozi13} using empirical constraints. However, observational constraints on the SMHM relation begin to diverge below $\sim 10^{11}\Msun$, as shown in Fig. 34 of \citet{Behroozi19}. For example, the \citet{Behroozi13} SMHM relation agrees with E-MOSAICS, but is $\sim 0.5$ dex above the \citet{Moster18} and \citet{Behroozi19} relations at $\Mhalo=10^{10}\Msun$. This offset between the different published relations likely points to systematics in the observational data (e.g. the low-mass stellar mass function and the cosmic SFR density evolution), as well as to the lack of complementary constraints on the SMHM relation (i.e. low-mass galaxy clustering.) Testing for the impact of the SMHM relation on GC kinematics would require running entirely new simulations with different stellar mass functions, which requires re-tuning of the subgrid physics. Since this is beyond the scope of this work, we perform a simple test instead. Assuming that stellar masses in the simulations are on average $\sim 50$ per cent larger in haloes with $\Mstar<10^{11}\Msun$, we artificially remove from the sample 50 per cent of those GCs that were likely born in these galaxies. To do this we used the mean stellar metallicity at $z\approx2$ \citep[the median GC formation redshift in E-MOSAICS;][]{Keller20} as a function of birth galaxy stellar mass from figure 34 of \citet{emosaicsII}. For GCs formed in galaxies with $\Mstar \la 10^9\Msun$, this method estimates their metallicities to be $\feh \la -1.0$. The effect on the orbital distributions of removing half of all GCs with $\feh \la -1.0$ is shown in Appendix~\ref{sec:limitations_appendix}. The effect is negligible for the median orbits, and slightly larger for the inter-quartile range. In all cases, the K-S test is unable to reject the null hypothesis that the modified and the fiducial distributions are identical ($p \geq 0.65$).

\section{Discussion and conclusions}
\label{sec:conclusions} 

In this paper, we present a detailed comparison of the kinematics of the MW GC system with the predictions of a cosmologically representative set of hydrodynamical galaxy formation simulations that include a subgrid model for the formation and evolution of star clusters. The E-MOSAICS galaxies and their GC populations have been shown to reproduce many observables \citep{emosaicsI,emosaicsII}. This makes the selection of 25 simulated galaxies based exclusively on present-day halo mass ideally suited to probe the distribution of GC kinematics that arise from differences in the formation and assembly of MW-mass galaxies. We compared the distributions of 3D velocities, orbital characteristics, and integrals of motion of GC populations (with metallicity in the range  $-2.5<\feh<-0.5$) across the 25 simulations with the MW GC system.  In addition, to gain insight into the signatures of clusters with different origins, we compared the relative distributions of subpopulations selected based on metallicity and galactocentric radius, i.e., metal-rich ($\feh>-1.2$) versus metal-poor ($\feh<-1.2$) and inner ($r<8\kpc$) versus outer ($r>8\kpc$) GCs. We find that GCs generally follow the kinematics of field stars in the simulations, with the largest difference in the azimuthal velocities where, although GCs most commonly have prograde rotation, they do so at slower speeds than the stars. In addition, GCs are typically more radially anisotropic than stars.

Although the MW GC population fits well within the distribution we find for $L^*$ galaxies in the simulations, the kinematics of its GC system are not typical in several aspects.  This is evident for example in the degree of prograde rotation, which is larger in the MW than in 80 per cent of the simulations, hinting at the lack of destructive mergers since the formation of the inner GCs. The velocity dispersions are also significantly higher in the MW GCs, placing them in above the 80th percentile of the distribution, hinting at an elevated number of minor accretion events. 

When comparing the median velocities of the GC subpopulations, we find that the rotation signal is dominated by the metal-rich GCs, which typically (in $\sim65$ per cent of cases) rotate faster than the metal-poor population.  The distribution of relative velocities of inner and outer clusters is surprisingly broad in the simulations, with many cases where the outer GCs rotate faster than the inner GCs due to a massive accretion event that dominates the angular momentum of the galaxy. In the MW, the metallicity subpopulations are more distinct kinematically. The fast rotation and low dispersion of its metal-rich GCs relative to the metal-poor population places it in the 80th-90th percentile of $L^*$ galaxies. This is caused by a relative lack of disc dynamical heating from late massive mergers.

The MW GC system is fairly typical with respect to the distribution of median orbital pericentre, apocentre, and eccentricity in the simulations. Both in the simulations and in the MW, metal-rich and metal-poor (or inner and outer) populations are clearly split in orbital parameters, with metal-rich (or inner) GCs typically at smaller apocentres and eccentricities than metal-poor (or outer) GCs. This is because the metal-poor (or outer) populations follow, \emph{on average}, a similar distribution of orbits compared to accreted GCs, and metal-rich (or inner) GCs track \emph{on average} the distribution of the in-situ population which was initially dynamically cold (Section~\ref{sec:origin}). 

The integrals of motion reveal additional insights. While field stars and GCs in the simulations have similar distributions of total angular momentum, the GCs have a smaller disc-aligned component ($\Lz$) than stars. The MW GCs have rather typical total angular momenta, but larger median $\Lz$ and binding energy than $\sim70$ and $\sim90$ per cent of galaxies respectively. The MW GC subpopulations separate clearly in this space, where the relative separation in median $\Lz$ and binding energy of inner and outer (or metal-rich and metal-poor) GCs is larger than in $>80$ per cent of the simulations. This indicates that the metal-rich (or inner) component of the MW is very compact and has a relatively high rotational support, and is consistent with the absence of late massive mergers that would have otherwise destroyed the disc (and the satellite would also have had less time to sink to the centre of the galaxy). Indeed, the statistical analysis in Section~\ref{sec:tracing_assembly_with_GCs} confirms that major mergers ended relatively early in the MW's history. 

To obtain quantitative constraints on the trends observed in the kinematics, we performed a blind search for statistical correlations between the kinematics and a comprehensive set of DM halo and galaxy assembly metrics with the goal of characterising the assembly history and formation environment that produced the MW GC system. This search was further expanded using the kinematics of subpopulations selected by metallicity and galactocentric radius. The analysis found several dozen significant correlations, with many of the metrics correlated with more than one kinematic tracer. Overall, many of the strong correlations are explained by the physics of infall, accretion, stripping, dynamical friction. We find that the kinematics of metal-rich/metal-poor and inner/outer subpopulations trace \emph{on average} the evolution of the orbits of in-situ and accreted GCs, which were born initially separated in angular momenta and binding energy. The number of mergers, their masses, and their time-scales subsequently modify the orbits of these GC subpopulations in very distinct ways. These are driven by the relative efficiency with which massive satellites sink to the centre of the galaxy (compared to low-mass ones) due to dynamical friction, and produce several of the observed correlations. For example, due to the large differences in satellite infall orbits, a larger number of mergers with mass ratios $<1/100$ produces a relative increase in the width of the distribution of orbital energy of the metal-poor GCs. Major (as well as early) mergers heat the orbits of inner GCs more effectively than minor (and also late) mergers, causing the ratio of median eccentricity of the metal-poor and metal-rich GCs to decrease in galaxies with more recent major mergers. Lastly, some of the observed relations result from correlations between assembly features that arise naturally in hierarchical structure formation, such as the one between the total number of mergers and the virial mass.

From the results of the search we selected the strongest, most significant and robust correlations and used them to predict 18 different aspects of the assembly history of the Galaxy and its DM halo with their associated uncertainties. In many cases the results probe new, unexplored aspects of the history of the Galaxy while in other cases they confirm and even enhance the precision of existing constraints from other methods. In particular, the known increase in GC specific frequency towards lower mass galaxies seems to make the GC system kinematics sensitive to even the lowest-mass accretion events. These predictions can be compared to the population of $L^*$ galaxies using the distribution of assembly histories of the 25 E-MOSAICS simulations. Below we summarise our quantitative constraints on the assembly of the Galaxy\footnote{As discussed in Section~\ref{sec:metrics}, we define as mergers only those that involve galaxies with stellar masses $M_* > 4.5\times10^6\Msun$ due to the limited resolution of the simulations. Below this mass, mergers are counted as smooth mass accretion.}:

\begin{itemize}

    \item The MW assembled very quickly, with half of its present-day stellar mass already formed across all its progenitors $10.0\pm1.5\Gyr$ ago, and earlier than 92 per cent of the simulated $L^*$ galaxies in E-MOSAICS. 
    
    \item The fast growth of the stellar component was caused by a quick assembly of the initial 25 per cent of the total mass of its host DM halo $11.2\pm0.9\Gyr$ ago (earlier than 72 per cent of the E-MOSAICS haloes). The following stage of halo mass growth was relatively slow, reaching 50 per cent $7.4\pm1.9\Gyr$ ago (earlier than just 24 per cent of the E-MOSAICS haloes). 
    
    \item The MW main progenitor assembled its stellar mass through a combination of in-situ star formation and mergers relatively early, with half of its stellar mass already in place $8.7_{-1.5}^{+1.7}\Gyr$ ago (73rd percentile). This early growth was partially driven by a relatively large number of $9.2\pm1.9$ mergers at $z>2$ (76th percentile).
    
    \item Compared to the average galaxy of its mass, the MW had an atypically low number of major mergers, $1.4\pm1.2$, lower than 60 per cent of the 25 $L^*$ galaxies in E-MOSAICS.
    
    \item The relative eccentricities of metal-rich and metal-poor GCs constrain the redshift of the last major merger. We predict it took place at $z = 3.1\pm1.3$. This is consistent with earlier lower limits and much earlier that the median, $z\approx 1$, placing it in the 84th percentile of galaxies of the same mass in E-MOSAICS.  
    
    \item The Galaxy had a quiescent late merger history, with only $5.9\pm2.4$ mergers occurring at $z<2$ (28th percentile). Despite the large uncertainty, the merger epoch of the MW is predicted to have ended significantly earlier than the average $L^*$ galaxy, with the last merger occurring $8.7^{+1.8}_{-3.4}\Gyr$ ago (where the lower limit is earlier than in 96 per cent of the simulations). 
    
    \item Due to the MW's relatively quiescent late ($z<2$) merger history, satellite accretion did not contribute a large overall fraction of the its stars and GCs, $12\pm11$ and $37\pm11$ per cent respectively. These fractions are fairly typical in $L^*$ galaxies (40th and 44th percentile for stars and GCs respectively).  
    
    \item The Galaxy experienced a total of $15.1\pm3.3$ mergers throughout its entire history (52nd percentile). After the single major merger, the two most massive events had mass ratios in the range 1:20--1:4, and the other $\sim 4$ mergers had smaller mass ratios in the range 1:100--1:20. Most of the MW's mergers, or about 8, involved relatively tiny galaxies with mass ratios $<$1:100. While $\Nmedium$ was atypically low (28th percentile), $\Nsmall$ was relatively high (80th percentile), and $\Ntiny$ is near the average (52nd percentile) compared to galaxies of the same mass. 
    
    \item Each of the $\sim 15$ galaxies that merged into the main progenitor to assemble the MW experienced fewer than 2 prior mergers on average.
    
\end{itemize}  

These predictions agree in general with the body of existing observational and theoretical constraints on the assembly of the MW and its halo. The constraints paint a picture of rapid early growth of the DM halo and the galaxy through accretion of many subhaloes. In the hierarchical assembly that characterises \LCDM, this is the natural result of DM haloes formed in overdense environments. Recent studies find that the MW lives in a region of $8\Mpc$ radius that is 2.5$\sigma$ overdense with respect to the mean matter density \citep{Neuzil19}. Constrained hydrodynamical simulations that reproduce the high density large-scale environment of the Local Group indeed predict a relatively early assembly of the Galaxy \citep{Carlesi20}, which confirms this scenario.

Many aspects of the formation of the MW remain uncertain. To further reconstruct the details of the merger history of the Galaxy, in a series of recently submitted papers we combine the kinematics with the ages and metallicities of individual GCs to identify their progenitors and improve the constraints on the timing and mass of merger events \citep{Pfeffer20, Kruijssen20}. Together, the results of analyses using GC ages, metallicities, and kinematics are beginning to demonstrate the potential of GC as excellent tracers of galaxy formation and assembly.  This will be essential in understanding the formation histories of galaxies across the entire mass range out to distances of several megaparsecs with existing facilities, and out to cosmological distances using the upcoming generation of 30-metre class ground based observatories. We will explore the extension of our method to line-of-sight kinematics and other GC-based diagnostics in future work.

\section*{Acknowledgements}

The authors would like to thank the anonymous referee for a critical review that resulted in important improvements to the quality of the manuscript. This work made use of the software packages {\sc numpy} \citep{numpy}, {\sc scipy} \citep{scipy}, {\sc pandas} \citep{pandas}, {\sc matplotlib} \citep{matplotlib} and {\sc pynbody} \citep{pynbody}. STG, JMDK, and MRC gratefully acknowledge funding from the European Research Council (ERC) under the European Union's Horizon 2020 research and innovation programme via the ERC Starting Grant MUSTANG (grant agreement number 714907). JMDK gratefully acknowledges funding from the Deutsche Forschungsgemeinschaft (DFG, German Research Foundation) through an Emmy Noether Research Group (grant number KR4801/1-1) and the DFG Sachbeihilfe (grant number KR4801/2-1). JP and NB gratefully acknowledge funding from the ERC under the European Union's Horizon 2020 research and innovation programme via the ERC Consolidator Grant Multi-Pop (grant agreement number 646928, PI Bastian). MRC is supported by a Fellowship from the International Max Planck Research School for Astronomy and Cosmic Physics at the University of Heidelberg (IMPRS-HD). BWK acknowledges funding in the form of a Postdoctoral Fellowship from the Alexander von Humboldt Stiftung. NB and RAC are Royal Society University Research Fellows. This work used the DiRAC Data Centric system at Durham University, operated by the Institute for Computational Cosmology on behalf of the STFC DiRAC HPC Facility (www.dirac.ac.uk). This equipment was funded by BIS National E-infrastructure capital grant ST/K00042X/1, STFC capital grants ST/H008519/1 and ST/K00087X/1, STFC DiRAC Operations grant ST/K003267/1 and Durham University. DiRAC is part of the National E-Infrastructure. This study also made use of high performance computing facilities at Liverpool John Moores University, partly funded by the Royal Society and LJMUs Faculty of Engineering and Technology.

\section*{Data availability}

The data underlying this article will be shared on reasonable request to the corresponding author.


\bibliographystyle{mnras}
\bibliography{merged}


\appendix

\section{Statistical methods}
\label{sec:statistics}

Here we describe the method we apply to search systematically for significant correlations between GC system kinematic tracers and galaxy and halo assembly metrics in the simulations. The list of kinematic tracers calculated for the GC system of each the 25 E-MOSAICS galaxies (described in Section~\ref{sec:metrics}), together with the assembly metrics for each galaxy (described in Section~\ref{sec:tracers}), define a $N \times M = 47 \times 32$ grid of possible correlations between the 47 tracers (as independent variables) and the 32 assembly metrics (as dependent variables) considered here. For each set of 25 data points (corresponding to the assembly metrics versus the kinematic tracers of the GC population for each of the 25 simulations) in this grid we first apply a statistical test to establish the correlation coefficients and $p$-values (the probability that the observed correlation is purely random) of each pair of tracer/assembly variables. For this purpose we choose the Spearman rank-correlation test because it makes no assumptions about the linearity of the relationship between the variables, but only tests for their rank-ordering.  This procedure yields a grid with $47\times32 = 1504$ entries. We then proceed to select as significant all correlations with a Spearman $p$-value $\peff < \pref$, with $\pref = 0.05$ which sets our significance threshold at 95 per cent confidence that the correlation did not arise randomly.   

Given the large number of variables pairs, $N \times M$, in this data set, to avoid selecting spurious correlations we correct the raw $p$-values by calculating an effective threshold that accounts for the total number of independent pairs of variables in our grid search. This is done using the Holm-Bonferroni method \citep{Holm79} which essentially adjusts for the probability of finding spurious correlations when performing multiple comparisons. For instance, searching among 100 possible correlations, we expect 5 spurious ones to appear statistically significant for a threshold $p$-value of 0.05. To eliminate these, the method scales the $p$-values by the total size of the search grid, 
$\peff = \pref /\Ncorr$. Since this correction assumes that all the variable pairs are uncorrelated, it must be adjusted to include only the list of remaining pairs as we step through the rank-ordered list of $p$-values obtained in the search above,
\begin{equation}
    \peff = \frac{\pref}{\Ncorr + 1 - i} ,    
\label{eq:peff}
\end{equation}
where $i$ is the rank (in increasing order) of the initial $p$-values. This ensures that as we test for correlation in a particular pair of variables, only the number of remaining pairs to be tested in the list scales the effective significance threshold for this pair. 

To set the value $\Ncorr$ we must consider the total number of independent variable pairs. Following the discussion in Appendix B of \citet{emosaicsII}, this number corresponds to the number of kinematic tracers that are independent per galaxy assembly metric. After dropping $\vt$, $\Lz$, $|L|$, $\beta$, $E$, $\Ekin$, $\rperi$, $\rapo$, and $e$ because they correlate with the 3D velocities and dispersions, we obtain $\Ncorr = 14$. Using equation~\ref{eq:peff} we obtain a range of effective $p$-values between $3\times10^{-3}$ and $5\times10^{-2}$. After obtaining the effective $p$-value for each pair of tracer/metric variables, those with $p < \peff$ are selected as statistically significant. 

Out of the entire list of significant correlations we then make a final selection based on which have Pearson $r$-values (describing how well the variation in the data is explained by a linear model) that exceed a threshold correlation coefficient $|r| > 0.7$. Since many assembly metrics are found to correlate with more than one kinematic tracer, we select for each metric only the tracer with the strongest linear correlation coefficient. In a few interesting cases of correlations below the threshold, we relax it to $|r| > 0.6$.

\section{Summary of correlations}
\label{sec:correlation_tables}

This section lists all the significant and strong (Pearson $|r|>0.7$) correlations found between GC kinematic tracers and galaxy assembly metrics. Additional relevant correlations with lower Pearson $r$ are also listed. Table~\ref{tab:correlations_all} lists the correlations for the entire GC system. Tables~\ref{tab:correlations_highmet} and \ref{tab:correlations_lowmet} list the correlations for the metal-rich and metal-poor GC population, respectively. Tables~\ref{tab:correlations_disk} and \ref{tab:correlations_halo} list the correlations for the inner and outer GC populations, respectively. Table~\ref{tab:correlations_sub} lists the correlations for the relative tracers of metallicity and galactocentric radius subpopulations.

\begin{table*}
\centering
\caption{Summary of correlations between kinematic tracers for the entire GC system and galaxy assembly metrics. From left to right, the columns list the galaxy and DM halo assembly metric, the GC kinematics tracer it correlates with, the Spearman $p$-value, the Pearson $r$ correlation coefficient, the Pearson $p$-value, the slope and intercept of the linear regression, and the scatter of the data around the regression line. We list only all of the strongest correlations (Pearson $|r|>0.7$) in addition to other selected correlations with lower correlation coefficients listed in Table~\ref{tab:predictions3D}. }
\label{tab:correlations_all}
\begin{tabular}{llcccccc}
\toprule
            Assembly metric ($y$) &                        Tracer ($x$) & Spearman $\log p$ & Pearson $r$ & Pearson $\log p$ & Slope (${\rm d}y/{\rm d}x$) &   Intercept ($y_0$) &             Scatter \\
\midrule
         $M_{200}~[\rm{M}_\odot]$ &          IQR($E$) [km$^2$ s$^{-2}$] &             -4.20 &        0.75 &            -4.76 &          $1.82\times10^{7}$ & $6.86\times10^{11}$ & $3.40\times10^{11}$ \\
         $M_{200}~[\rm{M}_\odot]$ &  med($E_{\rm k}$) [km$^2$ s$^{-2}$] &             -4.26 &        0.74 &            -4.58 &          $1.04\times10^{8}$ & $4.28\times10^{11}$ & $3.46\times10^{11}$ \\
   $V_{\rm max}~[\rm{km~s}^{-1}]$ &          IQR($E$) [km$^2$ s$^{-2}$] &             -4.53 &        0.75 &            -4.77 &         $8.13\times10^{-4}$ &  $1.49\times10^{2}$ &               15.16 \\
   $V_{\rm max}~[\rm{km~s}^{-1}]$ &  med($E_{\rm k}$) [km$^2$ s$^{-2}$] &             -6.16 &        0.76 &            -4.92 &         $4.76\times10^{-3}$ &  $1.36\times10^{2}$ &               14.93 \\
          $\tau_{50}~[{\rm Gyr}]$ &                            IQR($e$) &             -2.97 &        0.72 &            -4.28 &                       44.11 &               -7.04 &                1.32 \\
 $\log(1+\tau_{\rm mm}/\rm{Gyr})$ &                            IQR($e$) &             -1.50 &        0.71 &            -3.17 &                        8.28 &               -2.05 &                0.27 \\
                   $N_{\rm leaf}$ &  IQR($E_{\rm k}$) [km$^2$ s$^{-2}$] &             -3.66 &        0.72 &            -4.35 &         $1.47\times10^{-3}$ &                6.26 &                8.01 \\
                 $f_{\rm ex,GCs}$ &                              S($E$) &             -3.90 &       -0.78 &            -5.30 &                       -0.17 &                0.51 &                0.10 \\
                 $f_{\rm ex,GCs}$ &                              K($E$) &             -3.01 &       -0.71 &            -4.19 &                       -0.06 &                0.40 &                0.11 \\
                 $f_{\rm ex,GCs}$ &            med($r_{\rm apo}$) [kpc] &             -4.64 &        0.75 &            -4.75 &                        0.02 &                0.19 &                0.10 \\
\bottomrule
\end{tabular}
\end{table*}

\begin{table*}
\centering
\caption{Summary of correlations between kinematic tracers for the \emph{metal-rich} GC population and galaxy assembly metrics. Columns follow the format of Table~\ref{tab:correlations_all}. }
\label{tab:correlations_highmet}
\begin{tabular}{llcccccc}
\toprule
          Assembly metric ($y$) &                        Tracer ($x$) & Spearman $\log p$ & Pearson $r$ & Pearson $\log p$ & Slope (${\rm d}y/{\rm d}x$) &    Intercept ($y_0$) &             Scatter \\
\midrule
       $M_{200}~[\rm{M}_\odot]$ &          med($E$) [km$^2$ s$^{-2}$] &             -4.19 &       -0.72 &            -4.29 &         $-1.09\times10^{7}$ & $-1.81\times10^{11}$ & $3.56\times10^{11}$ \\
 $V_{\rm max}~[\rm{km~s}^{-1}]$ &          med($E$) [km$^2$ s$^{-2}$] &             -4.59 &       -0.72 &            -4.25 &        $-4.84\times10^{-4}$ &   $1.11\times10^{2}$ &               15.94 \\
 $V_{\rm max}~[\rm{km~s}^{-1}]$ &          IQR($E$) [km$^2$ s$^{-2}$] &             -4.18 &        0.74 &            -4.62 &         $9.37\times10^{-4}$ &   $1.54\times10^{2}$ &               15.38 \\
 $V_{\rm max}~[\rm{km~s}^{-1}]$ &  med($E_{\rm k}$) [km$^2$ s$^{-2}$] &             -5.71 &        0.74 &            -4.66 &         $4.59\times10^{-3}$ &   $1.40\times10^{2}$ &               15.31 \\
                   $N_{\rm br}$ &          med($E$) [km$^2$ s$^{-2}$] &             -3.76 &       -0.71 &            -4.09 &        $-1.53\times10^{-4}$ &                -7.83 &                5.19 \\
                   $r_{\rm bl}$ &          IQR($L$) [kpc km s$^{-1}$] &             -3.33 &       -0.73 &            -4.51 &        $-1.71\times10^{-4}$ &                 0.72 &                0.09 \\
               $f_{\rm ex,GCs}$ &          IQR($L$) [kpc km s$^{-1}$] &             -5.14 &        0.79 &            -5.51 &         $2.14\times10^{-4}$ &                 0.21 &                0.09 \\
               $f_{\rm ex,GCs}$ &           IQR($r_{\rm peri}$) [kpc] &             -4.34 &        0.71 &            -4.08 &                        0.06 &                 0.20 &                0.11 \\
               $f_{\rm ex,GCs}$ &            med($r_{\rm apo}$) [kpc] &             -4.36 &        0.76 &            -4.97 &                        0.02 &                 0.20 &                0.10 \\
               $f_{\rm ex,GCs}$ &            IQR($r_{\rm apo}$) [kpc] &             -5.59 &        0.81 &            -5.98 &                        0.01 &                 0.21 &                0.09 \\
                     $\delta_t$ &  IQR($E_{\rm k}$) [km$^2$ s$^{-2}$] &             -1.37 &        0.60 &            -2.86 &         $1.37\times10^{-5}$ &                -0.07 &                0.12 \\
                        $r_{t}$ &  IQR($E_{\rm k}$) [km$^2$ s$^{-2}$] &             -2.95 &        0.65 &            -3.42 &         $3.39\times10^{-5}$ &                -0.07 &                0.26 \\
\bottomrule
\end{tabular}
\end{table*}

\begin{table*}
\centering
\caption{Summary of correlations between kinematic tracers for the \emph{metal-poor} GC population and galaxy assembly metrics. Columns follow the format of Table~\ref{tab:correlations_all}. }
\label{tab:correlations_lowmet}
\begin{tabular}{llcccccc}
\toprule
          Assembly metric ($y$) &                        Tracer ($x$) & Spearman $\log p$ & Pearson $r$ & Pearson $\log p$ & Slope (${\rm d}y/{\rm d}x$) &   Intercept ($y_0$) &             Scatter \\
\midrule
       $M_{200}~[\rm{M}_\odot]$ &          IQR($E$) [km$^2$ s$^{-2}$] &             -4.06 &        0.72 &            -4.34 &          $1.98\times10^{7}$ & $5.30\times10^{11}$ & $3.54\times10^{11}$ \\
       $M_{200}~[\rm{M}_\odot]$ &  med($E_{\rm k}$) [km$^2$ s$^{-2}$] &             -5.78 &        0.81 &            -6.11 &          $1.13\times10^{8}$ & $2.34\times10^{11}$ & $2.98\times10^{11}$ \\
 $V_{\rm max}~[\rm{km~s}^{-1}]$ &          IQR($E$) [km$^2$ s$^{-2}$] &             -6.58 &        0.79 &            -5.48 &         $9.62\times10^{-4}$ &  $1.39\times10^{2}$ &               14.14 \\
 $V_{\rm max}~[\rm{km~s}^{-1}]$ &  med($E_{\rm k}$) [km$^2$ s$^{-2}$] &             -5.98 &        0.74 &            -4.57 &         $4.57\times10^{-3}$ &  $1.34\times10^{2}$ &               15.46 \\
 $V_{\rm max}~[\rm{km~s}^{-1}]$ &  IQR($E_{\rm k}$) [km$^2$ s$^{-2}$] &             -5.32 &        0.81 &            -5.92 &         $3.45\times10^{-3}$ &  $1.32\times10^{2}$ &               13.55 \\
        $\tau_{25}~[{\rm Gyr}]$ &           med($r_{\rm peri}$) [kpc] &             -2.79 &       -0.80 &            -5.80 &                       -0.23 &               11.66 &                0.92 \\
                 $N_{\rm leaf}$ &  IQR($E_{\rm k}$) [km$^2$ s$^{-2}$] &             -3.49 &        0.73 &            -4.48 &         $1.59\times10^{-3}$ &                1.98 &                7.91 \\
                   $N_{<1:100}$ &          IQR($E$) [km$^2$ s$^{-2}$] &             -5.53 &        0.74 &            -4.60 &         $2.08\times10^{-4}$ &               -1.79 &                3.55 \\
               $N_{1:100-1:20}$ &  med($E_{\rm k}$) [km$^2$ s$^{-2}$] &             -3.70 &        0.70 &            -4.08 &         $3.51\times10^{-4}$ &               -1.18 &                1.30 \\
\bottomrule
\end{tabular}
\end{table*}

\begin{table*}
\centering
\caption{Summary of correlations between kinematic tracers for the \emph{inner} GC population and galaxy assembly metrics. Columns follow the format of Table~\ref{tab:correlations_all}. }
\label{tab:correlations_disk}
\begin{tabular}{llcccccc}
\toprule
          Assembly metric ($y$) &                        Tracer ($x$) & Spearman $\log p$ & Pearson $r$ & Pearson $\log p$ & Slope (${\rm d}y/{\rm d}x$) &    Intercept ($y_0$) &             Scatter \\
\midrule
       $M_{200}~[\rm{M}_\odot]$ &          med($E$) [km$^2$ s$^{-2}$] &             -6.73 &       -0.85 &            -7.12 &         $-1.17\times10^{7}$ & $-4.36\times10^{11}$ & $2.70\times10^{11}$ \\
 $V_{\rm max}~[\rm{km~s}^{-1}]$ &            med($v_t$) [km s$^{-1}$] &             -4.75 &        0.70 &            -4.04 &                        0.54 &   $1.23\times10^{2}$ &               16.27 \\
 $V_{\rm max}~[\rm{km~s}^{-1}]$ &        IQR($L_z$) [kpc km s$^{-1}$] &             -5.08 &        0.75 &            -4.88 &                        0.11 &   $1.50\times10^{2}$ &               15.00 \\
 $V_{\rm max}~[\rm{km~s}^{-1}]$ &          med($L$) [kpc km s$^{-1}$] &             -5.59 &        0.76 &            -4.89 &                        0.11 &   $1.48\times10^{2}$ &               14.98 \\
 $V_{\rm max}~[\rm{km~s}^{-1}]$ &          IQR($L$) [kpc km s$^{-1}$] &             -4.00 &        0.74 &            -4.67 &                        0.13 &   $1.41\times10^{2}$ &               15.31 \\
 $V_{\rm max}~[\rm{km~s}^{-1}]$ &          med($E$) [km$^2$ s$^{-2}$] &             -8.75 &       -0.87 &            -7.82 &        $-5.37\times10^{-4}$ &                97.04 &               11.24 \\
 $V_{\rm max}~[\rm{km~s}^{-1}]$ &          IQR($E$) [km$^2$ s$^{-2}$] &             -6.20 &        0.76 &            -4.95 &         $2.32\times10^{-3}$ &   $1.39\times10^{2}$ &               14.89 \\
 $V_{\rm max}~[\rm{km~s}^{-1}]$ &  med($E_{\rm k}$) [km$^2$ s$^{-2}$] &             -5.08 &        0.73 &            -4.41 &         $2.57\times10^{-3}$ &   $1.52\times10^{2}$ &               15.70 \\
                   $N_{\rm br}$ &          med($E$) [km$^2$ s$^{-2}$] &             -4.22 &       -0.70 &            -4.03 &        $-1.39\times10^{-4}$ &                -7.15 &                5.23 \\
                 $N_{\rm leaf}$ &          med($E$) [km$^2$ s$^{-2}$] &             -5.49 &       -0.78 &            -5.47 &        $-2.46\times10^{-4}$ &               -14.02 &                7.19 \\
                     $N_{>1:4}$ &            IQR($r_{\rm apo}$) [kpc] &             -4.33 &        0.74 &            -4.56 &                        0.81 &                -0.97 &                1.16 \\
\bottomrule
\end{tabular}
\end{table*}

\begin{table*}
\centering
\caption{Summary of correlations between kinematic tracers for the \emph{outer} GC population and galaxy assembly metrics. Columns follow the format of Table~\ref{tab:correlations_all}. }
\label{tab:correlations_halo}
\begin{tabular}{llcccccc}
\toprule
          Assembly metric ($y$) &                Tracer ($x$) & Spearman $\log p$ & Pearson $r$ & Pearson $\log p$ & Slope (${\rm d}y/{\rm d}x$) &   Intercept ($y_0$) &             Scatter \\
\midrule
       $M_{200}~[\rm{M}_\odot]$ &  med($E$) [km$^2$ s$^{-2}$] &             -4.11 &       -0.77 &            -5.13 &         $-1.32\times10^{7}$ & $1.79\times10^{10}$ & $3.28\times10^{11}$ \\
       $M_{200}~[\rm{M}_\odot]$ &  IQR($E$) [km$^2$ s$^{-2}$] &             -4.79 &        0.78 &            -5.43 &          $2.73\times10^{7}$ & $4.28\times10^{11}$ & $3.18\times10^{11}$ \\
 $V_{\rm max}~[\rm{km~s}^{-1}]$ &  med($E$) [km$^2$ s$^{-2}$] &             -4.43 &       -0.75 &            -4.80 &        $-5.75\times10^{-4}$ &  $1.21\times10^{2}$ &               15.11 \\
 $V_{\rm max}~[\rm{km~s}^{-1}]$ &  IQR($E$) [km$^2$ s$^{-2}$] &             -5.50 &        0.74 &            -4.65 &         $1.15\times10^{-3}$ &  $1.40\times10^{2}$ &               15.33 \\
                 $N_{\rm leaf}$ &  med($E$) [km$^2$ s$^{-2}$] &             -3.93 &       -0.70 &            -4.05 &        $-2.74\times10^{-4}$ &               -4.26 &                8.25 \\
     $\tau_{\rm f}~[{\rm Gyr}]$ &    IQR($v_r$) [km s$^{-1}$] &             -3.30 &        0.63 &            -3.09 &                        0.03 &                4.47 &                1.44 \\
\bottomrule
\end{tabular}
\end{table*}

\begin{table*}
\centering
\caption{Summary of correlations between the relative kinematic tracers for the \emph{metallicity} and \emph{radial} GC subpopulations and galaxy assembly metrics. Columns follow the format of Table~\ref{tab:correlations_all}. }
\label{tab:correlations_sub}
\begin{tabular}{llcccccc}
\toprule
            Assembly metric ($y$) &                        Tracer ($x$) & Spearman $\log p$ & Pearson $r$ & Pearson $\log p$ & Slope (${\rm d}y/{\rm d}x$) & Intercept ($y_0$) & Scatter \\
\midrule
$z_{\rm mm}$ &      med($e^{\rm MR}$) / med($e^{\rm MP}$) &             -3.76 &       -0.80 &            -4.37 &                      -10.08 &             11.27 &    1.02 \\
$r_{\rm mm}$ &  IQR($r_{\rm apo}^{\rm MR}$) / IQR($r_{\rm apo}^{\rm MP}$).  &             -4.92 &        0.74 &            -4.66 &                        0.84 &             -0.02 &    0.16 \\
\midrule
$z_{\rm mm}$ &  K($E_{\rm k}^{\rm inner}$) - K($E_{\rm k}^{\rm outer}$) &             -2.69 &        0.76 &            -3.75 &                        0.42 &              1.03 &    1.10 \\
$a_{\rm mm}$ &  med($e^{\rm inner}$) / med($e^{\rm outer}$) &             -5.09 &        0.79 &            -4.20 &                        1.58 &             -0.76 &    0.17 \\
$r_{\rm mm}$ &  IQR($E_{\rm k}^{\rm inner}$) / IQR($E_{\rm k}^{\rm outer}$)  &             -3.65 &        0.78 &            -5.41 &                        0.36 &             -0.17 &    0.15 \\
$N_{{\rm br},z>2}$ &  IQR($|v_r^{\rm inner}|$) / IQR($|v_r^{\rm outer}|$) &             -3.64 &       -0.66 &            -3.49 &                       -6.38 &             12.36 &    2.57 \\
\bottomrule
\end{tabular}
\end{table*}

\section{Impact of simulation limitations}
\label{sec:limitations_appendix}

\begin{figure*}
    \includegraphics[width=0.70\textwidth]{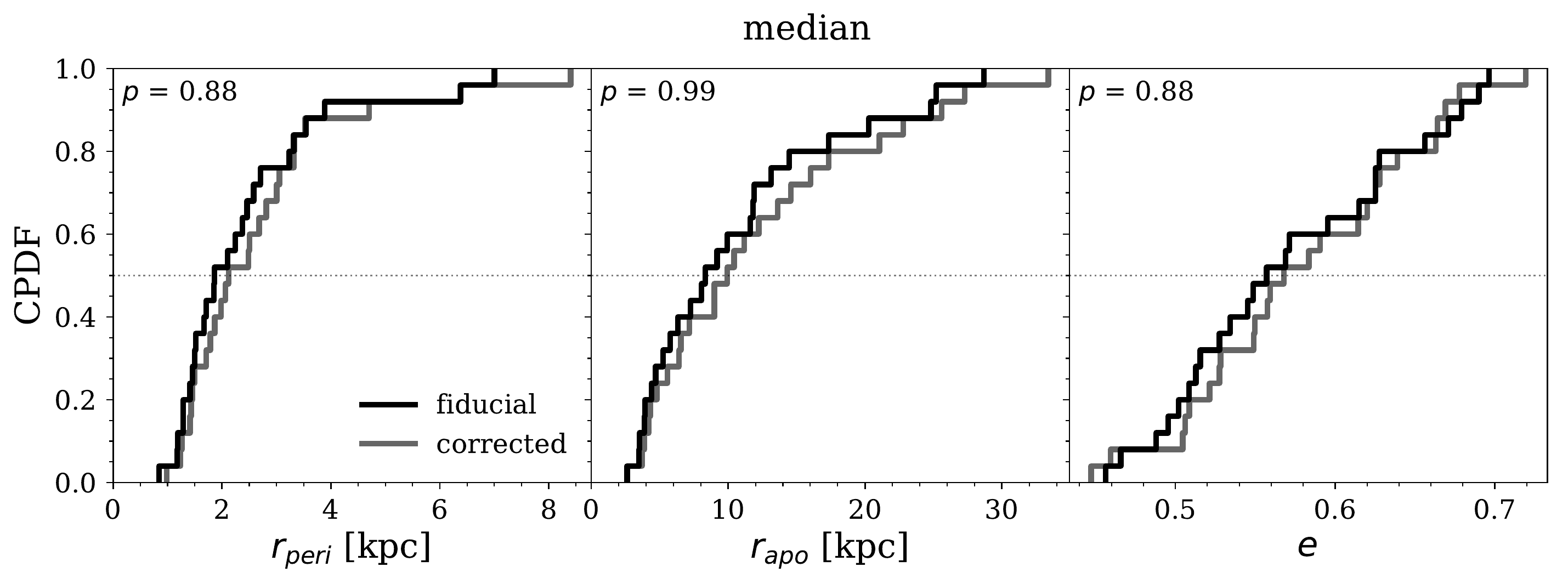}
    \includegraphics[width=0.70\textwidth]{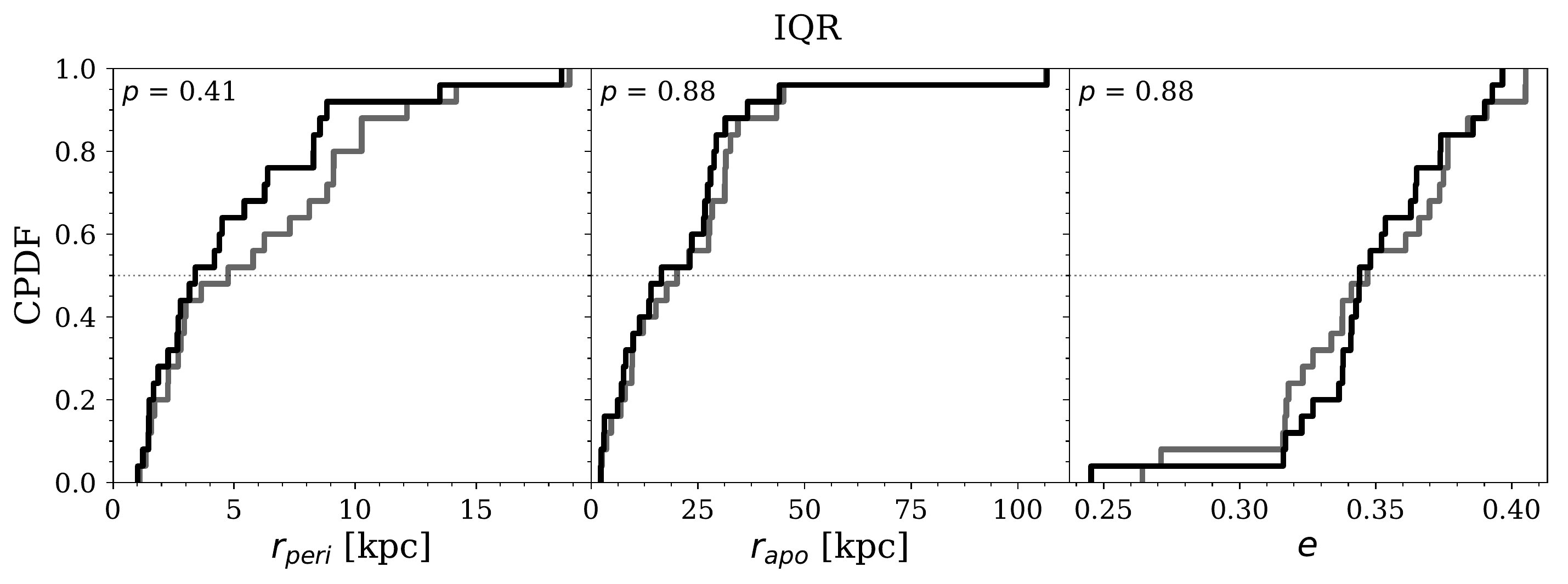}
    \caption{Impact of GC underdisruption in E-MOSAICS on the median (top), and inter-quartile range (bottom) of the orbital parameters distributions of simulated GCs. To correct for the excess of GCs, we artificially remove 60 per cent of all GCs with $-1.0<\feh-0.5$. This corresponds to the excess of metal-rich GCs in the simulations compared to the MW and M31 (see Section~\ref{sec:simulations}). The orbital distribution of the corrected sample is shown in grey, while the fiducial is shown in black. The K-S test $p$-values are indicated in each panel.}
\label{fig:underdisruption_effect}
\end{figure*}

\begin{figure*}
    \includegraphics[width=0.70\textwidth]{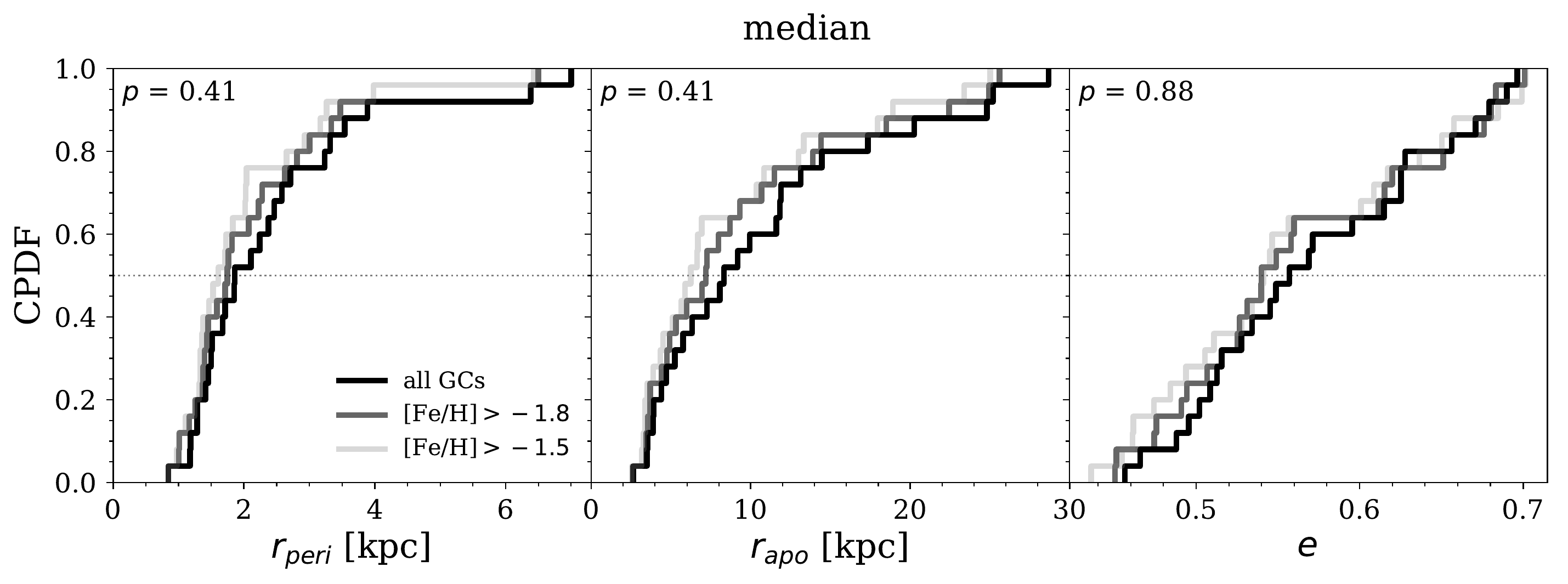}
    \includegraphics[width=0.70\textwidth]{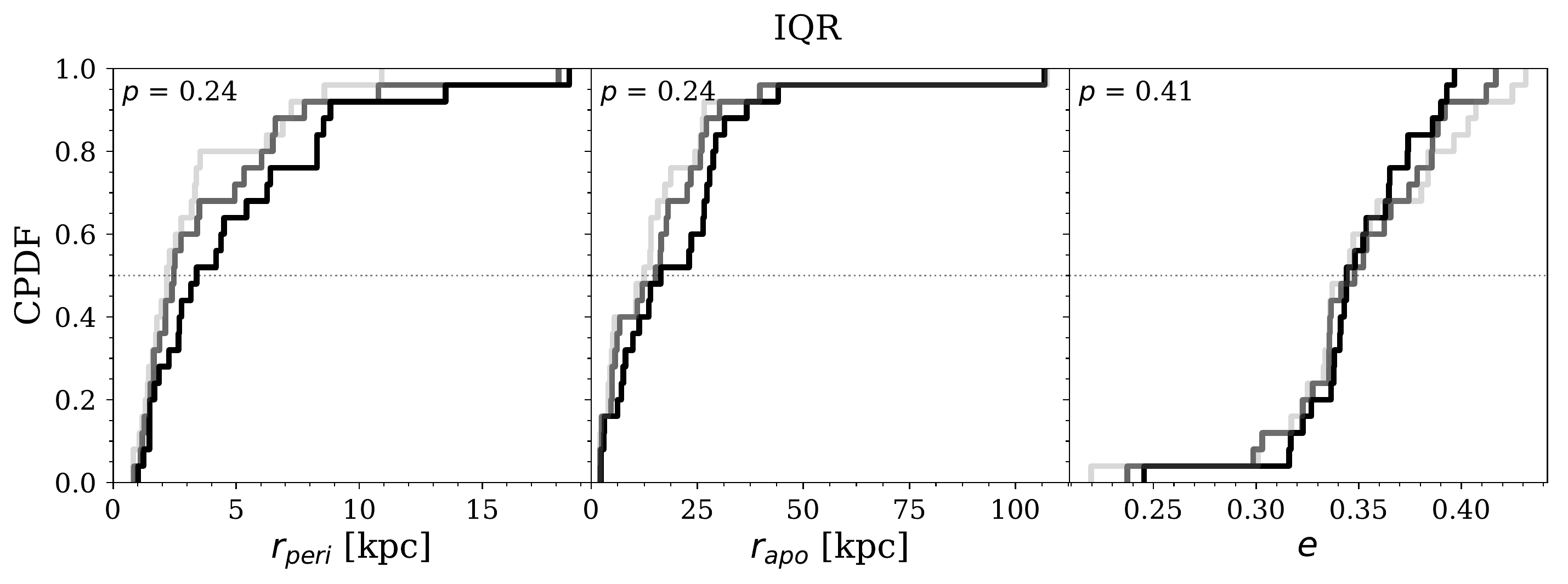}
    \caption{Test of the effect of mass resolution on the kinematics of simulated GCs for the median (top), and inter-quartile range (bottom) of the orbital parameters. To emulate the lack of GCs formed in poorly resolved galaxies, we artificially remove the GCs contributed by the lowest-mass galaxies, with masses about 5 times larger than the formal resolution ($\Mstar \approx 2\times10^7\Msun$ or $\sim 100$ star particles). This is done using a metallicity cut based on the evolution of the mean stellar masses and metallicities of the simulated galaxies (see Section~\ref{sec:limitations}. The orbital distribution after removing all GCs with metallicities $\feh < -1.5$ and ages (equivalent to GC progenitors with $\Mstar\la10^8\Msun$) is shown in light grey. For comparison, the distributions using all GCs in the simulations is shown in black, and an intermediate cut for $\Mstar\la10^{7.6}\Msun$ is shown in dark grey. The K-S test $p$-values for the most stringent cut are indicated in each panel.}
\label{fig:resolution_effect}
\end{figure*}

\begin{figure*}
    \includegraphics[width=0.70\textwidth]{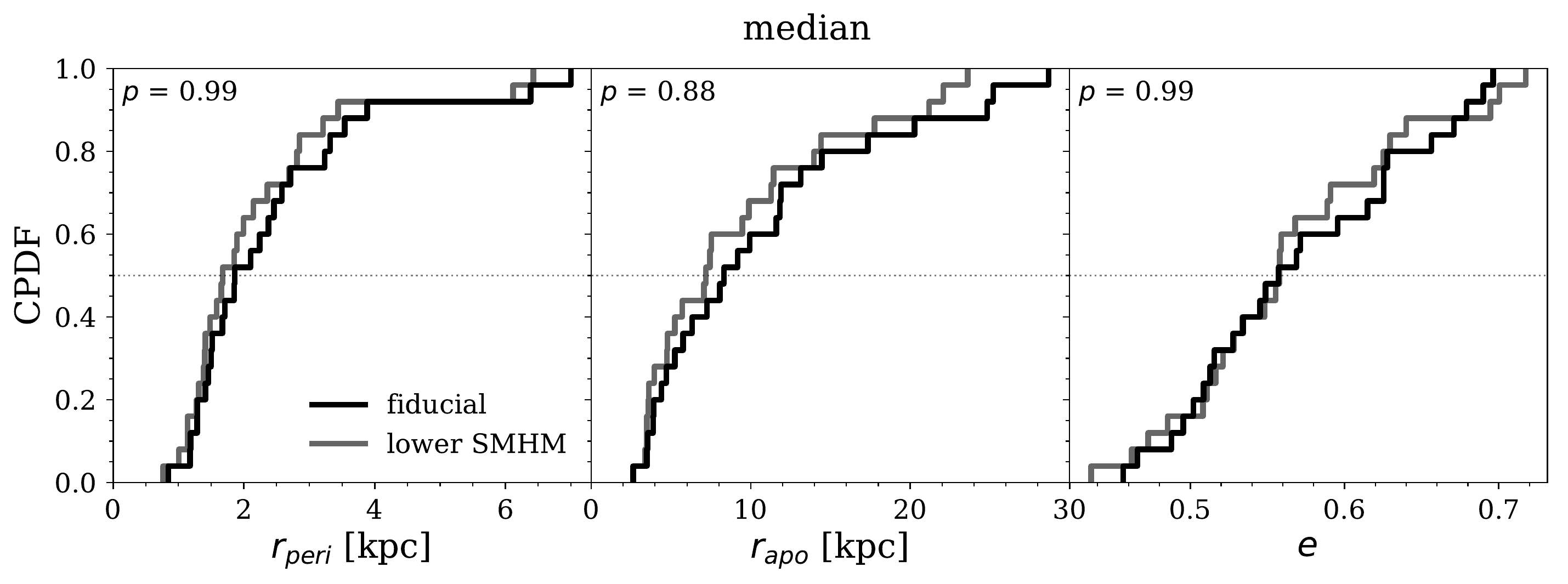}
    \includegraphics[width=0.70\textwidth]{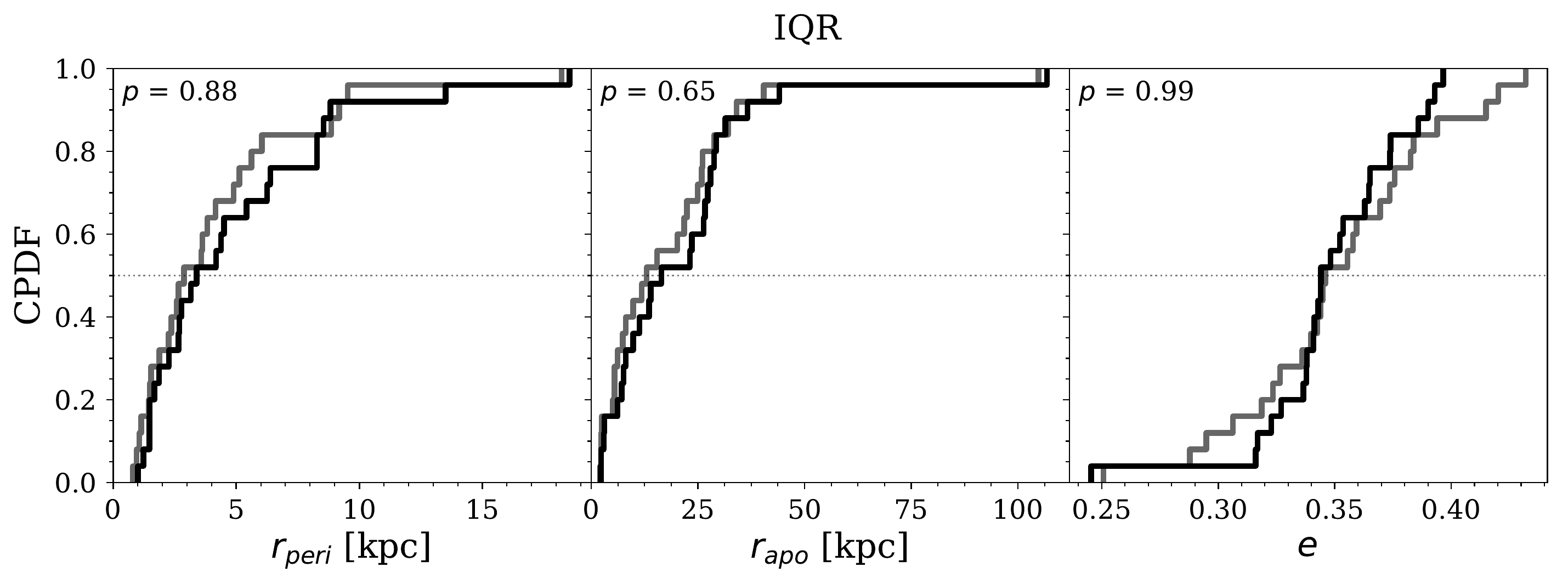}
    \caption{Effect of a systematic downward shift in the low-mass SMHM relation on the median (top), and inter-quartile range (bottom) of the orbital parameters of simulated GCs. To emulate the $-0.3$ dex shift in the stellar masses, we artificially remove 50 per cent of the GCs born in galaxies with $\Mstar\la10^9\Msun$. This is done using a metallicity cut based on the evolution of the mean stellar masses and metallicities of the simulated galaxies (see Section~\ref{sec:limitations}). The orbital distribution of the modified SMHM sample is shown in grey, while the fiducial is shown in black. The K-S test $p$-values are indicated in each panel.}
\label{fig:smhm_effect}
\end{figure*}

Figure~\ref{fig:underdisruption_effect} shows the effect of GC underdisruption in the simulations on the orbital distributions. Table~\ref{tab:predictions_nomet} lists the results of using metallicity-independent tracers to check the robustness of the predictions in Table~\ref{tab:predictions3D} that rely on GC metallicity subpopulations. Tables~\ref{tab:predictions_threshold0.8} and \ref{tab:predictions_threshold1.6} show the effect of varying by $\pm0.4$ dex the metallicity threshold that separates the metal-poor and metal-rich GC subpopulations. Figure~\ref{fig:resolution_effect} shows the impact on the orbital distributions of an artificial resolution cut to emulate the absence of GCs formed in the lowest-mass under-resolved galaxies. Figure~\ref{fig:smhm_effect} shows the effect of artificially reducing the GC populations in progenitors with $\Mstar\la10^9\Msun$ to mimic the effect of a steeper SMHM relation in the simulations. 

\begin{table*}
	\centering
	\caption{Alternative GC kinematic tracers that do not require metallicity information, and their predictions for the assembly history of the MW and its DM halo. From left to right, the columns list the galaxy and halo assembly metric, the corresponding GC kinematics tracer, the Pearson $r$ correlation coefficient of the linear model, and the prediction of the model using the MW GC system kinematics. For each assembly metric we select only the most correlated tracer while avoiding tracers that could be biased by the slight underproduction of stellar mass in $L^*$ galaxies in EAGLE (Section~\ref{sec:correlations}). The last column reproduces those predictions from Table~\ref{tab:predictions3D} which use GC metallicities. }
	\label{tab:predictions_nomet}
	\begin{tabular}{lcccccc} 
		\hline 
         Assembly metric & Kinematic tracer & GC population &  \parbox[c]{1.5cm}{Correlation coefficient} & \parbox[c]{1.2cm}{Pearson $\log p$} & 
         \parbox[c]{2.7cm}{MW prediction using\\ metallicity-independent tracers} & \parbox[c]{2.2cm}{Prediction using all GC information}  \\
		\hline \hline
		$\M200$ [$10^{12}\Msun$] & med($\Ekin$)  & all        & 0.74  & -4.6 & $2.03\pm0.35$ & $1.94\pm0.31$ \\
        \hline
        $\rbl$                   & med($\rapo$)  & inner      & -0.64 & -3.3 & $0.67\pm0.10$ & $0.64\pm0.09$ \\
        $\Nsmall$                & med($\Ekin$)  & all        & 0.69  & -3.9 & $4.5\pm1.3$   & $4.1\pm1.3$ \\
        $\Ntiny$                 & IQR($\Ekin$)  & outer      & 0.64  & -3.3 & $21.3\pm4.4$  & $17.8\pm3.6$ \\
        \hline
        $\zmm$ & med($e^{\rm inner}$)/med($e^{\rm outer}$) & inner/outer & -0.73 & -3.4 & $2.1\pm1.3$   & $3.1\pm1.3$ \\
        $\rt$                    & IQR($\Ekin$)  & inner      & 0.63  & -3.2 & $0.49\pm0.27$ & $0.41\pm0.26$  \\
        \hline
        $\fexstars$              & med($\rapo$)  & all        & 0.62  & -3.0 & $0.17\pm0.12$ & $0.12\pm0.11$ \\
        $\fexgcs$                & med($\rapo$)  & all        & 0.75  & -4.7 & $0.30\pm0.10$ & $0.31\pm0.09$ \\
		\hline 
	\end{tabular}
\end{table*}

\begin{table*}
	\centering
	\caption{Effect of a higher metallicity threshold, $\feh=-0.8$, for dividing the metal-poor and metal-rich GC populations on the GC kinematic tracers and their predictions for the MW assembly history. From left to right, the columns list the galaxy and halo assembly metric, the corresponding GC metallicity-dependent kinematics tracer, the Pearson $r$ correlation coefficient of the linear model, and the prediction of the model using the MW GC system kinematics. The last column reproduces the predictions from Table~\ref{tab:predictions3D} for the fiducial threshold. }
	\label{tab:predictions_threshold0.8}
	\begin{tabular}{lcccccc} 
		\hline 
         Assembly metric & Kinematic tracer & GC population &  \parbox[c]{1.5cm}{Correlation coefficient} & \parbox[c]{1.2cm}{Pearson $\log p$} & 
         \parbox[c]{2.7cm}{MW prediction using $\feh=-0.8$ \\ threshold} & \parbox[c]{2.0cm}{Prediction using fiducial threshold}  \\
		\hline \hline
		$\M200$ [$10^{12}\Msun$] & med($\Ekin$)  & metal-poor & 0.72  & -4.3 & $1.99\pm0.36$ & $1.94\pm0.31$ \\
		$\tau_{25}$  [Gyr]       & med($\rperi$) & metal-poor & -0.26 & -0.7 & $10.7\pm1.5$ & $11.2\pm0.9$ \\
        $\delta_t$               & IQR($\Ekin$)  & metal-rich & 0.35 & -1.1 & $0.11\pm0.14$ & $0.13\pm0.12$ \\
        \hline
        $\rbl$                   & IQR($L$)      & metal-rich & -0.31 & -0.9 & $0.61\pm0.12$ & $0.64\pm0.09$ \\
        $\Nsmall$                & med($\Ekin$)  & metal-poor & 0.67  & -3.6 & $4.4\pm1.4$   & $4.1\pm1.3$ \\
        $\Ntiny$                 & IQR($\Etot$)  & metal-poor & 0.49  & -1.9 & $15.1\pm4.6$  & $17.8\pm3.6$ \\
        \hline
        $\zmm$ & med($e^{\rm MR}$)/med($e^{\rm MP}$) & metal-rich/metal-poor & -0.50 & -1.6 & $3.9\pm1.5$ & $3.1\pm1.3$ \\
        $\rt$                    & IQR($\Ekin$)  & metal-rich      & 0.47  & -1.8 & $0.37\pm0.30$ & $0.41\pm0.26$  \\
        \hline
        $\fexstars$              & IQR($\rapo$)  & metal-rich  & 0.67 & -3.7 & $0.15\pm0.11$ & $0.12\pm0.11$ \\
        $\fexgcs$                & IQR($L$)      & metal-rich  & 0.48 & -1.8 & $0.34\pm0.13$ & $0.31\pm0.09$ \\
		\hline 
	\end{tabular}
\end{table*}

\begin{table*}
	\centering
	\caption{Effect of a lower metallicity threshold, $\feh=-1.6$, for dividing the metal-poor and metal-rich GC populations on the GC kinematic tracers and their predictions for the MW assembly history. From left to right, the columns list the galaxy and halo assembly metric, the corresponding GC metallicity-dependent kinematics tracer, the Pearson $r$ correlation coefficient of the linear model, and the prediction of the model using the MW GC system kinematics. The last column reproduces the predictions from Table~\ref{tab:predictions3D} for the fiducial threshold. }
	\label{tab:predictions_threshold1.6}
	\begin{tabular}{lcccccc} 
		\hline 
         Assembly metric & Kinematic tracer & GC population &  \parbox[c]{1.5cm}{Correlation coefficient} & \parbox[c]{1.2cm}{Pearson $\log p$} & 
         \parbox[c]{2.7cm}{MW prediction using $\feh=-1.6$ \\ threshold} & 
         \parbox[c]{2.2cm}{Prediction using fiducial threshold}  \\
		\hline \hline
		$\M200$ [$10^{12}\Msun$] & med($\Ekin$)  & metal-poor & 0.57  & -2.6 & $1.76\pm0.42$ & $1.94\pm0.31$ \\
		$\tau_{25}$  [Gyr]       & med($\rperi$) & metal-poor & -0.78 & -5.4 & $11.2\pm1.0$  & $11.2\pm0.9$ \\
		
        $\delta_t$               & IQR($\Ekin$)  & metal-rich & 0.58  & -2.6 & $0.19\pm0.12$ & $0.13\pm0.12$ \\
        \hline
        $\rbl$                   & IQR($L$)      & metal-rich & -0.47 & -1.7 & $0.62\pm0.12$ & $0.64\pm0.09$ \\
        $\Nsmall$                & med($\Ekin$)  & metal-poor & 0.66  & -3.5 & $3.9\pm1.4$   & $4.1\pm1.3$ \\
        $\Ntiny$                 & IQR($\Etot$)  & metal-poor & 0.54  & -2.2 & $14.5\pm4.4$  & $17.8\pm3.6$ \\
        \hline
        $\zmm$ & med($e^{\rm MR}$)/med($e^{\rm MP}$) & metal-rich/metal-poor & -0.67 & -2.7 & $2.2\pm1.4$ & $3.1\pm1.3$ \\
        $\rt$                    & IQR($\Ekin$)  & metal-rich & 0.62  & -3.0 & $0.57\pm0.27$ & $0.41\pm0.26$  \\
        \hline
        $\fexstars$              & IQR($\rapo$)  & metal-rich & 0.48 & -1.8 & $0.21\pm0.13$ & $0.12\pm0.11$ \\
        $\fexgcs$                & IQR($L$)      & metal-rich & 0.62 & -3.0 & $0.34\pm0.12$ & $0.31\pm0.09$ \\
		\hline 
	\end{tabular}
\end{table*}

\bsp	
\label{lastpage}
\end{document}